\documentclass[superscriptaddress,aps,twocolumn,longbibliography]{revtex4-1}
\usepackage{amsmath,amssymb}
\usepackage{graphicx}
\usepackage{dcolumn}
\usepackage{enumitem}
\usepackage{float}
\usepackage{bm,dsfont}
\usepackage{multirow}
\usepackage{simpler-wick}
\usepackage{cuted}
\usepackage{color}
\usepackage{hyperref}
\hypersetup{
colorlinks = true,
linkcolor = [rgb]{0.70,0.13,0.13},
citecolor = [rgb]{0.13,0.55,0.13},
urlcolor = [rgb]{0.25, 0.41, 0.88}}
\newcommand{\bra}[1]{{\langle #1|}}
\newcommand{\ket}[1]{{|#1 \rangle}}
\newcommand{\braket}[2]{{\langle #1|#2\rangle}}

\newcommand{\ii}{\mathrm{i}}
\newcommand{\id}{\mathds{1}}
\newcommand{\U}{\mathrm{U}}
\newcommand{\SU}{\mathrm{SU}}

\newcommand{\dsE}{\mathbb{E}}
\newcommand{\dsZ}{\mathbb{Z}}
\newcommand{\dsR}{\mathbb{R}}

\newcommand{\dsH}{\mathbb{H}}

\newcommand{\scE}{\mathcal{E}}

\newcommand{\scO}{\mathcal{O}}

\newcommand{\scX}{\mathcal{X}}
\newcommand{\Tr}{\operatorname{Tr}}

\newcommand{\vect}[1]{{\bm{#1}}}

\newcommand{\mat}[1]{\left[\begin{matrix}#1\end{matrix}\right]}

\newcommand{\R}[2]{{R_{\scriptscriptstyle #1}^{\scriptscriptstyle #2}}}
\newcommand{\dia}[3]{\raisebox{#3pt}{\includegraphics[height=#2pt]{dia_#1}}}
\newcommand{\eq}[1]{\begin{equation}#1\end{equation}}
\newcommand{\eqs}[1]{\begin{equation}\begin{split}#1\end{split}\end{equation}}
\newcommand{\eqnref}[1]{Eq.\,\eqref{#1}}
\newcommand{\figref}[1]{Fig.\,\ref{#1}}
\newcommand{\tabref}[1]{Tab.\,\ref{#1}}
\newcommand{\secref}[1]{Sec.\,\ref{#1}}\newcommand{\appref}[1]{Appendix\,\ref{#1}}
\newcommand{\refcite}[1]{Ref.\,\cite{#1}}

\usepackage[mathlines]{lineno}

\begin{document}


\title{Markovian Entanglement Dynamics under Locally Scrambled Quantum Evolution}

\author{Wei-Ting Kuo}
\author{A. A. Akhtar}
\author{Daniel P. Arovas}
\author{Yi-Zhuang You}
\affiliation{Department of Physics, University of California San Diego, La Jolla, CA 92093, USA}

\date{\today}

\begin{abstract}
We study the time evolution of quantum entanglement for a specific class of quantum dynamics, namely the locally scrambled quantum dynamics, where each step of the unitary evolution is drawn from a random ensemble that is invariant under local (on-site) basis transformations. In this case, the average entanglement entropy follows a Markovian dynamics, such that the entanglement property of the future state can be inferred from the entanglement property of the unitary operator of the underlying quantum dynamics. We introduce the entanglement feature formulation to concisely organize the entanglement entropies over all subsystems into a many-body  wave function, which allows us to describe the entanglement dynamics using an imaginary-time Schr\"odinger equation, such that various tools developed in quantum many-body physics can be applied. The framework enables us to investigate a variety of random quantum dynamics beyond  Haar random circuits and Brownian circuits. We perform numerical simulations for these models and demonstrate the validity and prediction power of the entanglement feature approach.
\end{abstract}

\pacs{Valid PACS appear here}
\maketitle


\section{\label{sec:2} Introduction}

Quantum entanglement dynamics\cite{Calabrese2005EEEOS,Kim2013BSEDNS,Liu2014ETUSHT,Kaufman2016QTTEIMS,Ho2017Entanglement} is an emerging field that ties several interesting topics together, including non-equilibrium and driven quantum systems\cite{Eisert2015QMSE,Ponte2015PDEMLQS,Moessner2017EOQFM}, many-body localization and thermalization\cite{Bardarson2012UGEMML,Kjall2014Many-Body,Luitz2015Many-body,Nandkishore2015MLTQSM,Vasseur2016Nonequilibrium,Abanin2019CMLTE}, quantum chaos and holography\cite{Wang2004ESQC,Hosur2016Chaos,Mezei2017On,Bertini2019ESMMMMQC,Gharibyan2019Quantum}. The central theme is to understand the production and propagation of quantum entanglement in quantum many-body systems. For pure states, the amount of quantum entanglement between a subsystem $A$ and its environment $\bar{A}$ can be quantified by the (R\'enyi) \emph{entanglement entropy} (EE) $S^{(n)}(A)=\frac{1}{1-n}\log\Tr_A\rho_A^n$ where $\rho_A=\Tr_{\bar{A}}\ket{\Psi}\bra{\Psi}$ is the reduced density matrix of subsystem $A$. Various quantum information measures (such as mutual and tripartite information) can be constructed from the EE over different regions. Here, we would like to focus on the 2nd R\'enyi entropies $S^{(2)}(A)$ and establish their dynamic equations under quantum evolution.

As a quantum state $\ket{\Psi}$ evolves in time, its EE's $S^{(2)}(A)$ over different regions $A$ will also change with respect to time in general. It is desired to understand how the unitary evolution of the quantum state induces the dynamics of quantum entanglement. There have been several works on the entanglement growth in quantum many-body systems\cite{Chandran2015Finite,Nahum2017Quantum,Zhou2017Operator,Ho2017Entanglement,Mezei2017On,Jonay2018CDOSE,Keyserlingk2018Operator,Nahum2018Operator,Nahum2018Dynamics,Rakovszky2018Diffusive,Khemani2018Operator,Gopalakrishnan2018Hydrodynamics,Rakovszky2019Entanglement}. The main focus has been on the half-system (or a single region) EE. To gain more resolution of the many-body entanglement structure, we extend our scope to all possible bipartitions of the system (including multiple disconnected entanglement regions). The question we would like to address is that given $S^{(2)}(A)$ at initial time over all possible subsystems $A$, what will be the equation of motion governing the evolution for all of them jointly in later time?

However, EE's over all regions contain a large amount of data, because the number of possible bipartitions $2^L$ grows exponentially in system size $L$. We need a conceptually concise way to organize these entropy data, in order to make progress in describing their dynamics. In \refcite{You2018Entanglement}, it was proposed that all these EE's can be organized into ``entanglement features'', which admit compact representations in terms of Boltzmann weights of Ising models. The key idea is to label each entanglement region $A$ by a set of Ising variables $\vect{\sigma}=(\sigma_1,\sigma_2,\cdots)$, such that $\sigma_i=\downarrow$ (or $\uparrow$) corresponds to $i\in A$ (or $i\in \bar{A}$) for each site $i$. Then the EE $S^{(2)}(A)\equiv S^{(2)}[\vect{\sigma}]$ can be treated as a free energy associated to the Ising configuration $\vect{\sigma}$, and the \emph{entanglement feature} (EF) refers to the corresponding Boltzmann weight $W[\vect{\sigma}]=e^{-S^{(2)}[\vect{\sigma}]}=\Tr\rho_A^2$, which is simply the purity for the 2nd R\'enyi case. Its time evolution can be related to the Loschmidt echo on the duplicated system,\cite{Ho2017Entanglement} which could be of experimental relevance. In this work, we further develop the Ising formulation by encoding the EF as a fictitious spin state $\ket{W}=\sum_{\vect{\sigma}}W[\vect{\sigma}]\ket{\vect{\sigma}}$, which we called the EF state. This rewriting packs the exponentially many entanglement data into a single EF state (as a many-body wave function). This conceptual simplification enables us to formulate the entanglement dynamics in a concise form of imaginary-time Hamiltonian evolution of the EF state 
\eq{\label{eq:Sch Eq intro}\partial_t\ket{W}=-\hat{H}_\text{EF}\ket{W},}
which can be further analyzed using powerful tools that have been developed in quantum many-body physics. Our development is along the line of mapping entanglement dynamics to statistical mechanical problems, as discussed in a few recent works
\cite{Hayden2016Holographic,Nahum2017Quantum,Jonay2018CDOSE,Zhou2018Emergent,Nahum2018Operator,Nahum2018Dynamics,Mezei2018Membrane,Vasseur2018Entanglement}. Given the equivalence between statistical mechanics and imaginary-time quantum mechanics, it is not surprising that the entanglement dynamics could admit a quantum mechanical formulation as \eqnref{eq:Sch Eq intro}.

Treating the EF $W[\vect{\sigma}]$ as an (unnormalized) probability distribution of entanglement regions $\vect{\sigma}$, the proposed dynamic equation in \eqnref{eq:Sch Eq intro} could be interpreted as a Markov equation. The assumption behind this equation is that the future EF of a many-body state can be entirely determined based on the current EF without the need to know about the past EF or about other information beyond the EF. Unfortunately, this assumption does \emph{not} hold in general! In fact, the entanglement dynamics is generally non-Markovian, meaning that knowing the present EE's even for all possible regions is still insufficient to determine their evolution in the future,\footnote{One can easily construct examples like $\ket{\uparrow\uparrow}$ and $\ket{\uparrow\downarrow}$, which are both product states, but their evolution under the same Hamiltonian $H=\vect{S}_1\cdot\vect{S}_2$ will result in states of different entanglements.} so we should not expect \eqnref{eq:Sch Eq intro} to work in general. In this work, we point out a specific yet rich enough class of quantum dynamics, called the \emph{locally scrambled} quantum dynamics, whose entanglement dynamics can be described by \eqnref{eq:Sch Eq intro} (or some discrete version of it). Quantum dynamics can always be formulated as a unitary evolution $U=\prod_t U_t$ that can be chopped up into products of simpler unitaries $U_t$ at each time slice $t$ following a time ordering. A quantum dynamics is said to be locally scrambled, if for every time step, the unitary $U_t$ is drawn from a random unitary ensemble that is invariant under local (on-site) basis transformations, and $U_t$ at different time $t$ are sampled independently. Such dynamics can be constructed by inserting local scramblers (product of on-site Haar random unitaries) between every time step, as if the system constantly forget about the choice of local basis from one time step to another. It can be used to model those quantum many-body systems with fast and random dynamics on each site, such that the quantum information is scrambled on each site quickly and sufficiently during each step of the time evolution. One famous example in this class is the Haar random unitary circuit\cite{Nahum2017Quantum,Zhou2018Emergent,Keyserlingk2018Operator,Nahum2018Operator}. We will provide more examples of locally scrambled quantum dynamics in this work. 

The reason that the future EE can be uniquely determined by the present EE under the locally scrambled quantum dynamics is related to the fact that the EE is a local-basis-independent quantity. As the local scramblers constantly remove the local-basis-dependent information in the quantum many-body state, only the local-basis-independent information can survive in time to govern the future evolution. Such local-basis-independent information can be captured by EE's over all possible entanglement regions, which are summarized as the EF of the quantum many-body state. In this work, we develop the theoretical framework to derive the dynamic equation governing the evolution of the EF under locally scrambled quantum dynamics. We establish a systematic approach to construct the EF Hamiltonian $\hat{H}_\text{EF}$ based on the entanglement properties of the physical Hamiltonian or unitary operators that describe the quantum dynamics.  We also carry out numerical simulations to justify the assumptions made in the theoretical development, and demonstrate the prediction power of the EF approach.

The paper is organized as follows. In \secref{sec:theory}, we first develop the theoretical framework for the EF and its application to the locally scrambled quantum dynamics. We start with the definition of EF for both quantum many-body state and quantum unitary circuits in \secref{sec:definition}. We then promote these notions to their quantum mechanical versions, introducing the EF states and EF operators in \secref{sec:operator}. With this setup, in \secref{sec:relation}, we prove an important relation between the the state and the unitary EF's, thereby mapping the unitary evolution of the quantum state to the dissipative evolution of the EF state under the corresponding entanglement dynamics in \secref{sec:dynamics}. Taking the continuum limit, we obtain the Schr\"odinger equation for EF state and derived the most generic form of the EF Hamiltonian in \secref{sec:Hamiltonian}. We analyze the spectral properties of the EF Hamiltonian and their consequences on the universal behavior of entanglement dynamics in \secref{sec:universal}. We investigate the excitation spectrum of the EF Hamiltonian and obtain the quasiparticle dispersion in \secref{sec:excitation}, which allows us to predict the long-time saturation behavior of the entanglement. We will provide numerical evidences in \secref{sec:numerics} to demonstrate the validity of the EF approach. We first introduce two models of locally scrambled quantum dynamics in \secref{sec:models}, namely the locally scrambled quantum circuit and the locally scrambled Hamiltonian dynamics, which are further discussed in details in \secref{sec:RUC} and \secref{sec:LSHD} separately. We sum up in \secref{sec:summary} making connections to related topics and potential future development.

\section{\label{sec:theory}Theoretical Framework}

\subsection{Definition of Entanglement Features}\label{sec:definition}

Let us consider a quantum many-body system consisting of $L$ qudits, where each qudit ($d$-dimensional quantum system) has a $d$-dimensional physical Hilbert space, such that the total Hilbert space dimension is $d^L$. To define the 2nd R\'enyi entropy, we will need to duplicate the system and evaluate the expectation value of swap operators within a subsystem $A$ of interest. There are altogether $2^L$ possible choices of a subsystem $A$, as each qudit can independent decide to be included in $A$ or not. To label the $2^L$ different bipartitions of the system, we introduce a set of classical Ising variables $\vect{\sigma}=(\sigma_1,\sigma_2,\cdots,\sigma_{L})$, such that the Ising variable $\sigma_i$ determines if the $i$th qudit belongs to region $A$ or its complement $\bar{A}$, following
\eq{\label{eq:def sigma}\sigma_i=\left\{\begin{array}{ll}
\uparrow & i\in\bar{A},\\
\downarrow & i\in{A}.
\end{array}\right.}
These Ising variables do not correspond to any degrees of freedom of the underlying quantum many-body system. Instead, they represent the identity or swap operator supported on the duplicated system, which are used to define the 2nd R\'enyi entropy. To be more specific, we define a permutation operator $\scX_{\sigma_i}$ acting on the duplicated Hilbert space of the $i$th qudit,
\eq{\scX_{\sigma_i}=\left\{\begin{array}{ll}
\dia{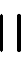}{12}{-3}_i\equiv\sum_{\alpha,\beta=1}^d\ket{\alpha\beta}_i\bra{\alpha\beta}_i & \text{if }\sigma_i=\uparrow,\\
\dia{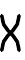}{12}{-3}_i\equiv\sum_{\alpha,\beta=1}^d\ket{\alpha\beta}_i\bra{\beta\alpha}_i & \text{if }\sigma_i=\downarrow,
\end{array}\right.}
which is assigned to the identity operator $\dia{II}{11}{-2}_i$ or the swap operator $\dia{X}{11}{-2}_i$ depending on the Ising variable $\sigma_i$. Assembling these permutation operators together, we define $\scX_\vect{\sigma}=\bigotimes_{i=1}^L\scX_{\sigma_i}$ for the  duplicated $L$-qudit system, which implements swap operations in the region $A$ specified by the Ising configuration $\vect{\sigma}$ .

\begin{figure}[htbp]
\begin{center}
\includegraphics[width=0.8\columnwidth]{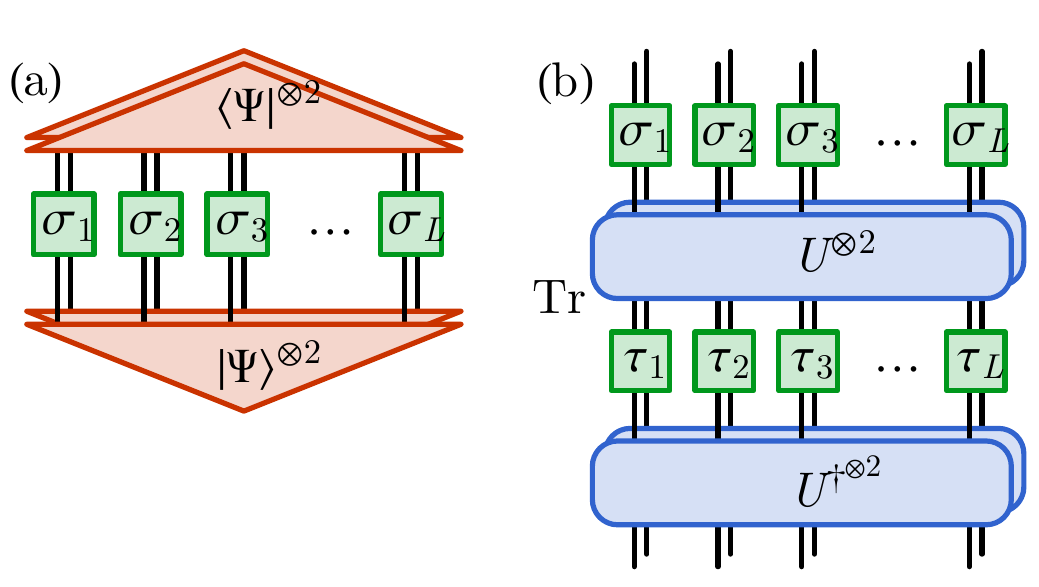}
\caption{Diagrammatic representation of (a) the state EF $W_\ket{\Psi}[\vect{\sigma}]$ and (b) the unitary EF $W_U[\vect{\sigma},\vect{\tau}]$. The Tr operator contracts the dangling bottom legs with the corresponding dangling top legs.}
\label{fig:EF}
\end{center}
\end{figure}

With these notation setup, we can define the \emph{entanglement feature} (EF) of quantum many-body states and time-evolution unitary circuits\cite{You2018Machine,You2018Entanglement}. The EF of a many-body pure state $\ket{\Psi}$ is defined as
\eq{\label{eq:WPsi}W_\ket{\Psi}[\vect{\sigma}]\equiv e^{-S^{(2)}[\vect{\sigma}]}=\Tr\big(\scX_{\vect{\sigma}}(\ket{\Psi}\bra{\Psi})^{\otimes2}\big),}
which resembles Boltzmann weights for Ising configurations $\vect{\sigma}$ labeling different entanglement regions. In terms of the tensor network representation, the state EF can be depicted as \figref{fig:EF}(a). Not only for quantum states, the EF can also be defined for unitary circuits under the state-operator correspondence.\cite{Hosur2016Chaos,Nie2018Signature,Kudler-Flam2019Quantum} The EF of a unitary circuit $U$ is defined as
\eq{\label{eq:WU}W_{U}[\vect{\sigma},\vect{\tau}]=\Tr\big(\scX_\vect{\sigma}U^{\otimes2}\scX_\vect{\tau}U^{\dagger\otimes2}\big),}
which depends on two sets of Ising configurations $\vect{\sigma}$ and $\vect{\tau}$ that separately specifies the entanglement regions on the past (input) and the future (output) sides of the unitary circuit, as illustrated in \figref{fig:EF}(b). The state EF $W_\ket{\Psi}[\vect{\sigma}]$ provides a comprehensive description of the entanglement properties of the pure state $\ket{\Psi}$, which contains the information about EE, mutual information and multipartite information among different subsystems. Similarly, the unitary EF $W_U[\vect{\sigma},\vect{\tau}]$ characterizes the entanglement properties of the unitary circuit $U$, including the EE and mutual information between past and future degrees of freedoms, which are also closely related to the operator-averaged out-of-time ordered correlator (OTOC)\cite{Hosur2016Chaos,Fan2017OTOC,Lensky2018Chaos} under the quantum dynamics $U$. 

It worth mention that entanglement features are invariant under \emph{local basis transformations}. A generic local basis transformation takes the form of  $V=\bigotimes_{i=1}^{L}V_i$ with $V_i$ being a unitary operator acting on the $i$th qudit. It is easy to see that both the state EF and the unitary EF are independent of the choice of local basis, i.e.
\eq{\label{eq:invariance}W_{V\ket{\Psi}}=W_{\ket{\Psi}},\quad W_{V^\dagger U V}=W_{U}.}
In this way, the EF forgets about the local basis dependent information in quantum states or unitary circuits, and only captures the entanglement properties that are universal to local basis choices. 

\subsection{Operator Formalism of Entanglement Features}\label{sec:operator}

To make our notation more concise, let us introduce a set of Ising basis $\ket{\vect{\sigma}}$, then we can pack $W_\ket{\Psi}$ to an \emph{entanglement feature state} (EF state) $\ket{W_\Psi}$ as
\eq{\ket{W_\Psi}=\sum_{\vect{\sigma}}W_\ket{\Psi}[\vect{\sigma}]\ket{\vect{\sigma}},}
and $W_U$ to an \emph{entanglement feature operator} (EF operator) $\hat{W}_U$ as
\eq{\hat{W}_U=\sum_{\vect{\sigma},\vect{\tau}}\ket{\vect{\sigma}}W_U[\vect{\sigma},\vect{\tau}]\bra{\vect{\tau}}.}
The Ising basis $\ket{\vect{\sigma}}$ span a $2^L$-dimensional Hilbert space of $L$ qubits, called the \emph{entanglement feature Hilbert space} (EF Hilbert space). It should not be confused with the $d^L$-dimensional physical Hilbert space of the underlying quantum many-body system. Each Ising basis state $\ket{\vect{\sigma}}$ in the EF Hilbert space simply corresponds to a bipartition of the $L$ physical qudits following \eqnref{eq:def sigma}.

Given the EF state $\ket{W_\Psi}$, the EE $S^{(2)}[\vect{\sigma}]$ over all regions can be retrieved from the inner product of $\ket{W_\Psi}$ with the corresponding Ising basis state
\eq{\label{eq:EE and EF}e^{-S^{(2)}[\vect{\sigma}]}=W_{\ket{\Psi}}[\vect{\sigma}]=\braket{\vect{\sigma}}{W_\Psi}.}
In particular, a \emph{product state} $\ket{\Psi_\text{prod}}=\bigotimes_{i}\ket{\psi_i}$ has zero EE in any region ($\forall\vect{\sigma}:S^{(2)}[\vect{\sigma}]=0$), so its EF state is therefore a equal weight superposition of all Ising configurations, 
\eq{\label{eq:Wprod}\ket{W_\text{prod}}=\sum_{\vect{\sigma}}\ket{\vect{\sigma}}\quad\text{(product state)},}
which corresponds to the (ideal) paramagnetic state of Ising spins. On the other hand, a \emph{Page state}\cite{Page:1993fv} $\ket{\Psi_\text{Page}}$  exhibits the maximal volume-law EE, whose EF state is given by
\eq{\label{eq:WPage}\ket{W_\text{Page}}=\sum_{\vect{\sigma}}\frac{\cosh(\eta\sum_{i=1}^L\sigma_i)}{\cosh(\eta L)}\ket{\vect{\sigma}}\quad\text{(Page state)},}
where $\eta=\frac{1}{2}\log d$ and we have adopted $\sigma_i=\pm1$ in the formula to represent $\uparrow,\downarrow$ spins. This result follows from the definition. Its detailed derivation can be found in \appref{app:Page}. The state $\ket{W_\text{Page}}$ contains extensive ferromagnetic correlations among Ising spins. In this picture, the process of quantum state thermalization corresponds to the process of building up ferromagnetic correlations in the EF state (until saturation to the Page state).

Let us also provide some examples for the EF of unitary gates which will be useful later. The EF of a single-qudit identity operator is straight forward to calculate based on the definition in \eqnref{eq:WU},
\eqs{\label{eq:W1 single}\hat{W}_\id&=d^2(\ket{\uparrow}\bra{\uparrow}+\ket{\downarrow}\bra{\downarrow})+d(\ket{\uparrow}\bra{\downarrow}+\ket{\downarrow}\bra{\uparrow}),\\
&=d(d+X),}
where $X$ denotes the Pauli-$X$ operator acting on the qudit site (acting in the EF Hilbert space, not in the qudit Hilbert space), and $d$ is the qudit dimension. A more non-trivial example is the EF of a two-qudit Haar random unitary gate $U_{ij}$ (averaged over Haar ensemble) that acts on qudits $i$ and $j$,
\eqs{\label{eq:WHaar}\hat{W}_\text{Haar}=&d^2(d+X_i)(d+X_j)\\
&-\frac{d^2(d^2-1)}{d^2+1}\frac{1-Z_iZ_j}{2}(d^2-X_iX_j),}
where $X_i$ and $Z_i$ are Pauli-$X$ and $Z$ operators acting on site $i$. The derivation can be found in \appref{app:EF Haar}.

\begin{figure}[htbp]
\begin{center}
\includegraphics[width=0.9\columnwidth]{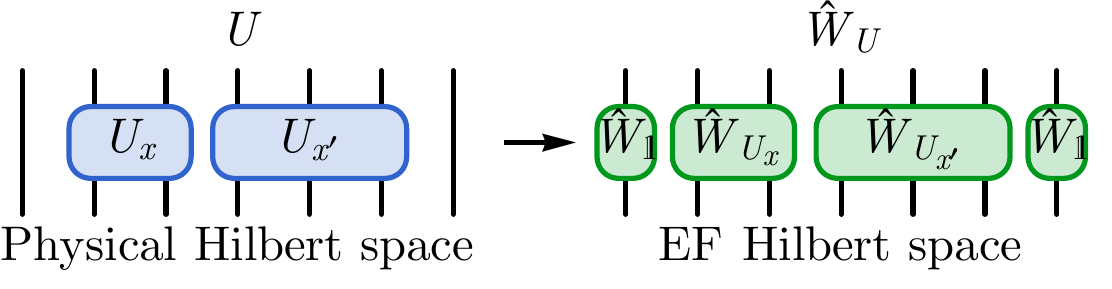}
\caption{The mapping from the unitary operator in the physical Hilbert space to the corresponding EF operator in the EF Hilbert space. Locality is preserved under the mapping, enabling us to factorize the operators in the same manner on both sides.}
\label{fig:factorization}
\end{center}
\end{figure}

Unitary gates are the building blocks to construct more complicated unitary circuits. One nice property of the EF operator is that it preserves the locality in space, meaning that if a unitary $U$ operator can be factorized to smaller unitaries $U_{x}$ over the space $x$, its corresponding EF operator $\hat{W}_{U}$ is also factorized in the same manner
\eq{\label{eq:factorization}U=\bigotimes_{x}U_{x}\quad\Rightarrow\quad \hat{W}_U=\bigotimes_{x}\hat{W}_{U_{x}},}
as examplified in \figref{fig:factorization}. This property allows us to assemble the local EF operators together. For example, the EF operator $\hat{W}_{\id}$ of the identity operator for a $L$ qudit system be obtained by assembling the single-qudit result in \eqnref{eq:W1 single} together
\eq{\label{eq:W1}\hat{W}_{\id}=\prod_{i=1}^Ld(d+X_i)=(\coth\delta\mathop{\mathrm{csch}}\delta)^L\prod_{i=1}^Le^{\delta X_i},}
where we have introduced the constant $\delta$
\eq{\delta\equiv\mathop{\mathrm{arccoth}} d=\frac{1}{2}\log \frac{d+1}{d-1}}
to exponentiate the operator. The exponential form allows us to take the operator inverse easily, such that
\eq{\label{eq:invW1}\hat{W}_\id^{-1}=\prod_{i=1}^{L}\frac{1-d^{-1}X_i}{d^2-1}=(\tanh\delta\sinh\delta)^L\prod_{i=1}^Le^{-\delta X_i}.}
These results will be useful in later discussions. In the following, we will show how the evolution of the EF state can be inferred from the EF operator of the unitary circuit.

\subsection{Relation between State and Unitary Entanglement Features}\label{sec:relation}

Suppose $U$ describes a unitary circuit that evolves an initial quantum many-body state $\ket{\Psi}$ to the final state $U\ket{\Psi}$. This quantum dynamics will induce a corresponding entanglement dynamics, under which the EF of the initial state $W_\ket{\Psi}$ evolves to that of the final state $W_{U\ket{\Psi}}$. Can we predict the final state EF $W_{U\ket{\Psi}}$ based on our knowledge about the initial state EF $W_\ket{\Psi}$ and the EF $W_U$ of the unitary evolution? 

In general, this problem is not tractable. Because $U$ and $\ket{\Psi}$ contain many ``non-universal'' features that are specific to the choice of local basis, such features may affect the final state entanglement, but they are not captured by the EF, as the EF is invariant under local basis transformations. Therefore, the finial state EF can not be inferred from the initial state EF and the unitary EF in general. However, instead of dealing with a specific unitary circuit $U$, we consider an ensemble of unitary circuits $U'=V^\dagger U V$ related to $U$ by local basis transformations $V$, denoted by
\eq{\label{eq:LSUE}\scE_U=\Big\{V^\dagger U V\Big|V=\bigotimes_{i=1}^{L}V_i, V_i\in\text{Haar}\Big\},}
where each $V_i$ is independently drawn from the Haar random unitary ensemble defined on the $i$th qudit. We will call $\scE_U$ the \emph{locally scrambled} unitary ensemble associated with $U$. According to \eqnref{eq:invariance}, one immediately see that all unitary operators $U'\in\scE_U$ in the ensemble share the \emph{same} entanglement feature as that of $U$, i.e. $W_{U'}=W_{U}$. Rather than asking about the EF of a specific final state $U\ket{\Psi}$, if we are allowed to consider the ensemble average of the EF over all  final states $U'\ket{\Psi}$ with $U'\in\scE_U$, the final state EF $W_{U'\ket{\Psi}}$ will indeed be constructable from the initial state EF $W_\ket{\Psi}$ and the unitary EF $W_{U'}=W_{U}$ on the average level. Using the operator formalism, the relation can be written in a concise form as
\eq{\label{eq:EF relation}\mathop{\dsE}_{U'\in\scE_U}\ket{W_{U'\Psi}}=\hat{W}_U\hat{W}_\id^{-1}\ket{W_\Psi},}
where $\hat{W}_\id$ is the EF operator for the identity evolution $\id$ and $\hat{W}_\id^{-1}$ is its inverse, which was given in \eqnref{eq:invW1} explicitly. One can derive \eqnref{eq:EF relation} using tensor network diagrams, see \appref{app:diagram} for details. To simplify the notation, we may suppress spelling out the ensemble average $\mathop{\dsE}_{U'\in\scE_U}$ explicitly in later discussions, with the understanding that in this work any unitary operator appeared in the subscript of the EF operator will be implicitly averaged over local basis transformations. \eqnref{eq:EF relation} establishes an important relation between the state and the unitary EF's, which enables us to compute the evolution of the state EF induced by the underlying quantum dynamics, given the EF of the corresponding unitary evolution $U$. A special case of \eqnref{eq:EF relation} has been discussed in \refcite{You2018Entanglement,Lensky2018Chaos}, where the initial state is restricted to product states.

As a side remark, we would like to provide some justifications for the use of locally scrambled unitary ensembles $\scE_U$. Technically speaking, working with these ensembles enables us to predict the future evolution of EE's purely based on their current data, because the local-basis-dependent features of a quantum state are removed by local scrambling and the remaining local-basis-independent features are captured by the EF\footnote{Strictly speaking, all the local-basis-independent features are capture by the full set of R\'enyi entropy to all R\'enyi orders. But here we only focus on a subset described by the 2nd R\'enyi entropy.}. This setup allows us to make progress in understanding the entanglement dynamics with a tractable theoretical limit. Physically speaking, we can imagine systems with separating time scales between the on-site and the inter-site quantum dynamics. Suppose the on-site dynamics is fast and random, then the quantum information would be sufficiently scrambled on every site, before it can spread out to other sites at a longer time scale. So the overall unitary evolution will constantly be interrupted by the insertion of local scramblers $V_i\in\text{Haar}$, making the evolution effectively local-basis-independent. In fact, many well explored random unitary ensembles in the field of entanglement dynamics are local-basis-independent (or ``locally scrambled'' in our language), including random unitary dynamics\cite{Oliveira2007Generic,Znidaric2007Optimal,Nahum2017Quantum}, random Hamiltonian dynamics\cite{You2018Entanglement,Vijay2018Finite-Temperature,Liu2018Entanglement,Rowlands2018Noisy} and random Floquet dynamics\cite{Chan2018Spectral,Chan2018Solution}. This strategy has also been adopted in the discussion of operator dynamics\cite{Jonay2018CDOSE,Keyserlingk2018Operator,Nahum2018Operator,Khemani2018Operator,Gopalakrishnan2018Hydrodynamics} and random tensor networks\cite{Hayden2016Holographic,You2018Machine,Qi2017Holographic,Vasseur2018Entanglement}. Historically, the study of these models has advanced our understanding about  the universal behavior of entanglement dynamics, so we would like to carry on this line of research. 

\subsection{Markovian Entanglement Dynamics}\label{sec:dynamics}

As long as we know how to construct the EF operator $\hat{W}_U$ for any unitary evolution $U$ of interest, we can apply the operator formalism in \eqnref{eq:EF relation} to compute the entanglement dynamics. However, calculating the EF for a large and deep unitary circuit is a difficult many-body problem, hence the relation \eqnref{eq:EF relation} is still hard to apply. But if all unitary gates in the unitary circuit are independently drawn from locally scrambled unitary ensembles, they will be decoupled in time, such that we can apply the EF operator iteratively to drive the evolution of the EF state.

\begin{figure}[htbp]
\begin{center}
\includegraphics[width=0.68\columnwidth]{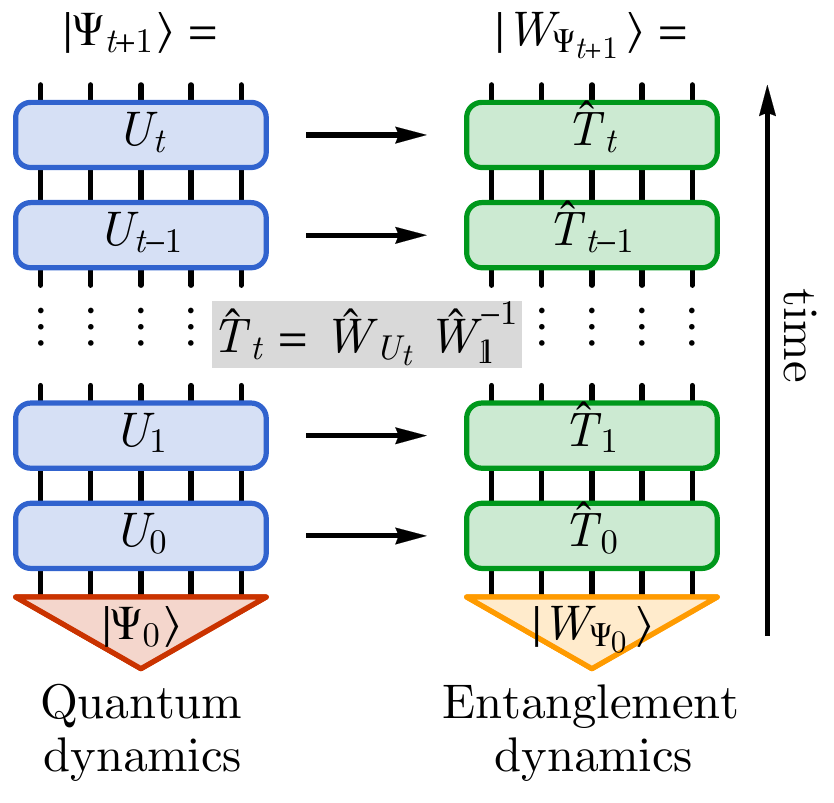}
\caption{Quantum dynamics induces entanglement dynamics, assuming each unitary $U_{t}$ is drawn from local basis invariant ensemble independently. The operator entanglement property of $U_{t}$ determines the transfer matrix $\hat{T}_{t}$ that evolves the EF state via \eqnref{eq:T def}, and the EF state $\ket{W_{\Psi_{t}}}$ encodes the entanglement properties of the quantum state $\ket{\Psi_{t}}$.}
\label{fig:dynamics}
\end{center}
\end{figure}

To be more concrete, let us consider the case where the full unitary evolution can be broken up into discrete time steps (or layers), and each single-step unitary evolution at time $t$ is described by $U_{t}$, as illustrated on the left of \figref{fig:dynamics}. Then the quantum many-body state $\ket{\Psi_{t}}$ evolves from step to step following
\eq{\ket{\Psi_{t+1}}=U_{t}\ket{\Psi_{t}}.}
Suppose $U_{t}$ at different time $t$ are independently drawn from random unitary ensembles (not necessary Haar random) which are invariant under local basis transformation, the full unitary evolution 
\eq{U=\prod_tU_t=U_{t}U_{t-1}\cdots U_{1}U_{0}}
will form a random unitary circuit that defines a \emph{locally scrambled quantum dynamics}. If we spell out the local basis transformations $V_t$ that has been made at each time step, i.e.~$U_t=V_t^\dagger U'_t V_t$,
\eq{\label{eq:U seq}U=V_{t}^\dagger\;U'_{t}\;V_{t}V_{t-1}^\dagger\;U'_{t-1}\;V_{t-1}\cdots,}
we can see that the neighboring transformations $V_{t}V_{t-1}^\dagger$ can merge into a single layer of local scramblers. Therefore a locally scrambled quantum dynamics can also be viewed as repeatedly applying the on-site scrambling $V_{t}V_{t-1}^\dagger$ followed by the inter-site unitary $U'_{t}$. In this way, the quantum many-body state is always sufficiently scrambled on each qudit and the scrambling is uncorrelated in time, such that the information about local basis choice does not pass on from step to step. Separating each step of the unitary evolution by local scramblers is our key assumption about the quantum dynamics, which enables us to proceed. 

The entanglement dynamics induced by the locally scrambled quantum dynamics is Markovian, and admits a simple transfer matrix description. To see this, we evaluate the final state EF averaging over all locally scrambled unitary ensembles at different steps
\eq{\label{eq:average}\ket{W_{\Psi_{t+1}}}=\mathop{\dsE}_{U_{t}}\mathop{\dsE}_{U_{t-1}}\cdots\ket{W_{U_{t}U_{t-1}\cdots\Psi_{0}}}.}
Applying \eqnref{eq:EF relation}, we arrive at the recurrent equation for the ensemble averaged EF state
\eq{\label{eq:transfer}\ket{W_{\Psi_{t+1}}}=\hat{T}_{t}\ket{W_{\Psi_{t}}},}
where we have introduced the transfer matrix
\eq{\label{eq:T def}\hat{T}_{t}=\hat{W}_{U_{t}}\hat{W}_{\id}^{-1}}
to evolve the EF state $\ket{W_{\Psi_{t}}}$ according to the EF of the single-step unitary $U_{t}$. As summarized in \figref{fig:dynamics}, \eqnref{eq:T def} is the key equation that bridges the quantum dynamics and entanglement dynamics, allowing us to predict the evolution of entanglement properties of a quantum state based on the entanglement properties of the unitary operator applied at each time step. If we further assume \emph{locality} of the quantum dynamics such that $U_{t}=\bigotimes_{x}U_{t,x}$ can be decomposed into products of non-overlapping local unitary gates $U_{t,x}$ (each gate only acts on a few qudits and its spatial position is labeled by $x$), the EF operator $\hat{W}_U$ can be factorized in the same manner following \eqnref{eq:factorization}
\eq{\label{eq:W locality}\hat{W}_{U_{t}}=\bigotimes_{x}\hat{W}_{U_{t,x}},}
where $\hat{W}_{U_{t,x}}$ is the EF operator for each local unitary gate, which can be easily computed (as it only involves a few qudits). Along this line, the transfer matrix $\hat{T}_t$ can be constructed purely based on our knowledge about the EF of each unitary gate involved in the quantum dynamics.

Using \eqnref{eq:transfer}, we can evolve the EF of any initial quantum state in time, given the locally scrambled quantum dynamics. The time evolution of the (2nd R\'enyi) EE can be read out from the EF by
\eq{\label{eq:S def}S^{(2)}[\vect{\sigma}](t)=-\log\braket{\vect{\sigma}}{W_{\Psi_{t}}},}
following \eqnref{eq:EE and EF}. Strictly speaking, there is a subtle issue about exchanging the order of the logarithm with all the ensemble average in \eqnref{eq:average}. We are typically more interested in the ensemble average of the EE other than the EF. So the correct average for the EF should be the geometric mean $\exp(\dsE \log W[\vect{\sigma}])$, but we are replacing it by the algebraic mean $\dsE W[\vect{\sigma}]$ in \eqnref{eq:average}, which always over estimates the EF and hence underestimates the EE. So the EE obtained in \eqnref{eq:S def} only serves as a lower bound of the ensemble averaged EE. We may treat this lower bound as an approximation, but we can not claim that it is always a good approximation, because there are known scenarios where this approximation is problematic. For example, near the entanglement transition\cite{Vasseur2018Entanglement,Jian2019MCRQC,Bao2019TPTRUCWM} where critical fluctuation is important, this approximate treatment gives wrong answers about the universality class and critical exponents. There have been more rigorous treatments developed in \refcite{Vasseur2018Entanglement,Skinner2019MPTDE} using replica tricks, but we will not pursuit that direction in this paper. For thermalizing dynamics and volume-law states, we believe that the lower bound estimation in \eqnref{eq:S def} will provide a decent approximation, because the EF of thermalizing state contains strong ferromagnetic correlation to suppress the spin fluctuation, which allows us to replace the geometric mean by the algebraic mean as the fluctuation is small. We will rely on numerical simulations in \secref{sec:numerics} to justify this assumption.

To conclude, the EF formalism provides a concise description for the entanglement dynamics, when the underlying quantum dynamics is locally scrambled. However, there are also several limitations of locally scrambled quantum dynamics. First of all, the dynamics is not translation invariant in time, because the local scrambles at each step must be sampled independently. As a result, energy is not conserved under such dynamics. Secondly, global symmetry\cite{Khemani2018Operator,Pai2019LFRC} can not be implemented in the current scheme, because symmetry representations on each site will all be scrambled together, such that the symmetry can not be preserved. Finally, in lack of the local-basis-specific information, we can not discuss the operator dynamics for specific local operators\cite{Parker2018UOGH} (but we can discuss operator averaged behaviors). To go beyond the local scrambling assumption, one idea could be to gradually introduce the correlation of unitary gates in time. But we will leave that for future study. We believe that our discussion of the locally scrambled quantum dynamics will set a cornerstone for future developments.

\subsection{Entanglement Feature Hamiltonian}\label{sec:Hamiltonian}

In the previous section, we have derived the dynamic equation \eqnref{eq:transfer} for EF states under discrete time dynamics. We can also consider the continuum limit of the dynamics, where we refine the time step and take $U_{t}$  to be close to identity (up to local basis transformation).

For example, we can consider generating $U_{t}$ by a local Hamiltonian for a short amount of ``time'' $\epsilon\ll 1$ with the local basis scrambled
\eq{\label{eq:Ut of Hdyn}U_{t}=V_t^\dagger e^{-\ii \epsilon H} V_t,}
where $V_t=\bigotimes_{i=1}^LV_{t,i}$ is a layer of local scramblers and each scrambler $V_{t,i}$ is an on-site unitary operator independently drawn from Haar random ensemble. The full unitary evolution $U=\prod_t U_t$ is given by the time-order product. The onsite scrambling does not generate entanglement (among different sites). The entanglement generation and propagation all depend on the inter-site couplings in the Hamiltonian $H$. As $\epsilon$ is small, the entanglement dynamics will be slow (smooth) enough that admits a continuum time description. We will study this model in more details later, but the goal here is to first establish a Hamiltonian formulation for the evolution of EF state in the continuum limit.

When $U_{t}$ is close to an identity operator (up to local basis transformations), its EF operator $\hat{W}_{U_{t}}$ will approach $\hat{W}_{\id}$, hence the transfer matrix $\hat{T}_{t}=\hat{W}_{U_{t}}\hat{W}_{\id}^{-1}$ will also be close to the identity operator $\hat{\id}$ (in the EF Hilbert space). It turns out that the difference between $\hat{T}_{t}$ and $\hat{\id}$ is of the order $\epsilon^2$ (not $\epsilon$ as one may expect). A general argument for this property is as follows. Given $U_t$ in \eqnref{eq:Ut of Hdyn}, its EF is described by
\eq{\label{eq:WUt}W_{U_t}[\vect{\sigma},\vect{\tau}]=\Tr(\scX_\vect{\sigma}e^{-\ii\epsilon \dsH}\scX_\vect{\tau}e^{\ii\epsilon \dsH}),} 
with $\dsH=H\otimes\id+\id\otimes H$. It can be shown that $W_{U_t}[\vect{\sigma},\vect{\tau}]$ must be even in $\epsilon$, because it is real by definition but $\epsilon$ comes with the imaginary unit in \eqnref{eq:WUt}, thus the odd-power expansions of $W_{U_t}[\vect{\sigma},\vect{\tau}]$ in $\epsilon$ could only be imaginary, and must therefore vanish altogether. So the operators $\hat{W}_{U_t}$ and $\hat{T}_t$ are even in $\epsilon$, hence the leading order deviation of $\hat{T}_t$ from $\hat{\id}$ is of the order $\epsilon^2$.

Given this, we expand $\hat{T}_{t}$ around the identity operator $\hat{\id}$ and define the \emph{entanglement feature Hamiltonian} (EF Hamiltonian) 
\eq{\label{eq:HEF def}\hat{H}_\text{EF}=\frac{1}{\epsilon^2}(\hat{\id}-\hat{T}_{t})=\frac{1}{\epsilon^2}(\hat{\id}-\hat{W}_{U_t}\hat{W}_{\id}^{-1}),}
such that the recurrent equation \eqnref{eq:transfer} transforms to an imaginary-time Schr\"odinger equation in the continuum limit of $\epsilon\ll 1$,
\eq{\label{eq:Sch Eq}\partial_{t}\ket{W_{\Psi_{t}}} = -\hat{H}_\text{EF}\ket{W_{\Psi_{t}}}.}
The differentiation $\partial_t\ket{W_{\Psi_{t}}}$ should be considered as the limit of $(\ket{W_{\Psi_{t+\epsilon^2}}}-\ket{W_{\Psi_{t}}})/\epsilon^2$, where $\epsilon^2$ serves as the infinitesimal time step. In general, $\hat{H}_\text{EF}$ can be time-dependent, but let us omit the explicit time dependence for simplicity. The locality of the EF operator $\hat{W}_{U_t}$ as discussed in \eqnref{eq:W locality} translates to the locality of the EF Hamiltonian $\hat{H}_\text{EF}$, which allow us to write $\hat{H}_\text{EF}=\sum_{x}\hat{H}_{x}$ as sum of local terms. In principle, the specific form of these local terms $\hat{H}_x$ can be derived from the terms in the quantum many-body Hamiltonian $H$ that drives the quantum dynamics, which we will demonstrate later in \secref{sec:LSHD}. However, even if we have no specific knowledge about $H$, we can already learn a lot about $\hat{H}_\text{EF}$ based on the general properties of entanglement dynamics. In the following, we will show how the physical constraint of entanglement dynamics can pin down the general form of the EF Hamiltonian.

Let us consider the two-local EF Hamiltonian, meaning that the local terms $\hat{H}_x$ span over two sites at most. We find that the most general two-local EF Hamiltonian should take the following form
\eq{\label{eq:HEF form}
\hat{H}_\text{EF} = \sum_{i,j}g_{ij}\frac{1-{Z}_{i}{Z}_{j}}{2}e^{-\beta_{ij}X_iX_j-\delta(X_i+X_j)},} 
where $g_{ij}\geq 0$ and $\beta_{ij}\in\dsR$ are model parameters and the constant $\delta$ is fixed by the qudit dimension $d$ via $\coth\delta=d$. Here $X_{i},Z_{i}$ are Pauli operators acting on the $i$th Ising spin (that labels the entanglement region). Each local term in the Hamiltonian consists of a term $e^{-\beta_{ij}X_iX_j-\delta(X_i+X_j)}$ that fluctuates Ising spins, followed by a ferromagnetic projection operator $(1-Z_iZ_j)/2$. Although we call $\hat{H}_\text{EF}$ a Hamiltonian, it is not a Hermitian operator as expected in conventional quantum mechanics, because fluctuation term and the projection term do not commute. As a result, the left- and the right-eigenstates of $\hat{H}_\text{EF}$ could be different. The coupling strength $g_{ij}$ describes the entangling power of the quantum dynamics, i.e.~the velocity that the entanglement builds up between sites $i$ and $j$ if initialized from a product state.

The postulated form of $\hat{H}_\text{EF}$ in \eqnref{eq:HEF form} is constrained by the following physical requirements (or assumptions).
\begin{itemize}[]
\item Pure state remains pure under quantum dynamics (i.e.~a $\dsZ_2$ Ising symmetry),
\eq{\label{eq:Ising symm}[\hat{H}_\text{EF},\prod_{i}{X}_{i}]=0.}
An important entanglement property of pure states is that the EE of a region $A$ should be the same as that of its complement $\bar{A}$, therefore the pure state EF must be invariant under Ising symmetry, i.e.~$W_{\ket{\Psi}}[\vect{\sigma}]=W_{\ket{\Psi}}[-\vect{\sigma}]$, which can be equivalently written as $\ket{W_{\Psi}}=\prod_{i}X_i\ket{W_{\Psi}}$. Since any quantum dynamics (described by a unitary evolution) will preserve the purity of the quantum state, the entanglement dynamics should also respect this Ising symmetry, such that the EF Hamiltonian $\hat{H}_\text{EF}$ must commute with the symmetry operator $\prod_{i}X_i$ as asserted in \eqnref{eq:Ising symm}.

\item EE must vanish for empty entanglement regions,
\eq{\label{eq:left null}\bra{\uparrow\uparrow\uparrow\cdots}\hat{H}_\text{EF}=0.}
By empty entanglement region, we mean $A=\emptyset$ is an empty set, which correspond to the Ising configuration $\vect{\sigma}=\;\uparrow\uparrow\uparrow\cdots\equiv\;\Uparrow$ (i.e. $\forall i:\sigma_i=+1$). Hereinafter we use the symbol $\Uparrow$ to denote the all-up configuration to simplify the notation. When the entanglement region is empty, the EE must be zero, i.e.~$S^{(2)}[\Uparrow]=0$. This requires $\braket{\Uparrow}{W_\Psi}=W_{\ket{\Psi}}[\Uparrow]=e^{-S^{(2)}[\Uparrow]}=1$ to remain at constant under any entanglement dynamics. Now suppose $\ket{W_{\Psi_{t}}}$ is time dependent under the entanglement dynamics. Taking the time derivative on both sides of $\braket{\Uparrow}{W_{\Psi_{t}}}=1$ and apply the dynamic equation \eqnref{eq:Sch Eq}, we can see that $\bra{\Uparrow}\partial_{t}\ket{W_{\Psi_{t}}}=-\bra{\Uparrow}\hat{H}_\text{EF}\ket{W_{\Psi_{t}}}=0$ must hold for any EF state $\ket{W_{\Psi_{t}}}$, therefore we must require $\bra{\Uparrow}\hat{H}_\text{EF}=0$ as claimed in \eqnref{eq:left null}.

\item Statistical time-reversal symmetry of random unitary ensembles,
\eq{\label{eq:WH=HW}\hat{W}_{\id}\hat{H}_\text{EF}^\intercal=\hat{H}_\text{EF}\hat{W}_{\id}.}
We assume that the random unitary gates in the circuit are statistically invariant under time-reversal, meaning that $U_{t}$ and $U_{t}^\dagger$ will appear with equal probability in the unitary ensemble. Then according to the definition of unitary EF in \eqnref{eq:WU}, the time-reversal symmetry implies to $W_{U}[\vect{\sigma},\vect{\tau}]=W_{U}[\vect{\tau},\vect{\sigma}]$, i.e.~$\hat{W}_{U}^\intercal=\hat{W}_{U}$. As a special case, we also have $\hat{W}_{\id}^\intercal=\hat{W}_{\id}$ by definition. Transposing both sides of $\hat{W}_U\hat{W}_{\id}^{-1}=\hat{\id}-\epsilon^2\hat{H}_\text{EF}$, we obtain $\hat{W}_{\id}^{-1}\hat{W}_U=\hat{\id}-\epsilon^2\hat{H}_\text{EF}^\intercal$. Therefore $\hat{H}_\text{EF}^\intercal$ and $\hat{H}_\text{EF}$ must be related by $\hat{W}_{\id}\hat{H}_\text{EF}^\intercal=\hat{H}_\text{EF}\hat{W}_{\id}$ as stated in \eqnref{eq:WH=HW}. One known scenario that the statistical time-reversal symmetry is broken is that the unitary operators cyclically permute the qudit along one direction, which describes a quantum dynamics that has dynamic anomaly.\cite{Po2016Chiral,Lee2019Topological} We conjecture that the statistical time-reversal symmetry effectively restricts the quantum dynamics to be anomaly free.
\end{itemize}

With these conditions, we can start from a generic two-local Hamiltonian $\hat{H}_\text{EF}=\sum_{i,j}\hat{H}_{ij}$ and derive the generic form of \eqnref{eq:HEF form}. First of all, the Ising symmetry in \eqnref{eq:Ising symm} restricts $\hat{H}_{ij}$ to be a linear combination of the following operators $\hat{H}_{ij}=x_1 + x_2 X_j +x_3 X_i+ x_4 X_iX_j+x_5 Y_iY_j+x_6 Y_iZ_j+x_7 Z_iY_j+x_8 Z_iZ_j$, which contains all the two-local operators that commute with $X_iX_j$. Then the left-null-state requirement in \eqnref{eq:left null} further requires $x_1=-x_8, x_2=\ii x_7, x_3=\ii x_6, x_4=x_5$, which reduce $\hat{H}_{ij}$ to $(1-Z_iZ_j)(x_1+x_2 X_j+x_3 X_i+x_4 X_iX_j)$. Finally, the statistical time-reversal symmetry in \eqnref{eq:WH=HW} requires 
\eq{\label{eq:x relation}x_2=x_3=-\frac{d(x_1+x_4)}{d^2+1},}
leaving only two independent parameters $x_1$ and $x_4$. This relation can be resolved by introducing another two parameters $g$ and $\beta$ to parametrize $x_1+x_2 X_j+x_3 X_i+x_4 X_iX_j=\frac{g}{2} e^{-\beta X_iX_j-\delta(X_i+X_j)}$ with $\coth\delta=d$ fixed, such that 
\eqs{x_1&=\frac{g(d^2\cosh\beta-\sinh\beta)}{2(d^2-1)},\\
x_2=x_3&=-\frac{gde^{-\beta}}{2(d^2-1)},\\
x_4&=\frac{g(\cosh\beta-d^2\sinh\beta)}{2(d^2-1)},}
automatically satisfies \eqnref{eq:x relation}. The resulting local term reads $\hat{H}_{ij}=g \frac{1-Z_iZ_j}{2}e^{-\beta X_iX_j-\delta(X_i+X_j)}$, which matches the form of \eqnref{eq:HEF form}. 

\subsection{Universal Behaviors of Entanglement Dynamics}\label{sec:universal}

The generic form of the EF Hamiltonian $\hat{H}_\text{EF}$ in \eqnref{eq:HEF form} is already useful to illustrate several universal behaviors about the entanglement dynamics. Suppose the EF Hamiltonian admits the following spectral decomposition
\eq{\hat{H}_\text{EF}=\sum_{a}\ket{R_a}\lambda_a\bra{L_a},}
where $\ket{R_a}$ and $\bra{L_a}$ are respectively the right- and left-eigenstate of the eigenvalue $\lambda_a$. The right-eigenstate is related to the corresponding left-eigenstate by $\ket{R_a}\propto(\bra{L_a} \hat{W}_{\id})^\intercal$, which follows from \eqnref{eq:WH=HW}. Then the Schr\"odinger equation for EF state \eqnref{eq:Sch Eq} can be formally solved as
\eq{\label{eq:solution}\ket{W_{\Psi_{t}}}=\sum_{a}e^{-\lambda_a t}\ket{R_a}\braket{L_a}{W_{\Psi_{0}}}.}
The dynamics of the EE can be inferred from \eqnref{eq:S def} as 
\eqs{\label{eq:S sol}S^{(2)}[\vect{\sigma}](t)&=-\log\braket{\vect{\sigma}}{W_{\Psi_{t}}}\\
&=-\log\sum_{a}e^{-\lambda_a t}\braket{\vect{\sigma}}{R_a}\braket{L_a}{W_{\Psi_{0}}}.}

Independent of the choice of model parameters $g_{ij},\beta_{ij}$, the EF Hamiltonian $\hat{H}_\text{EF}$ has the following spectral properties:
\begin{itemize}[]
\item $\hat{H}_\text{EF}$ is positive semi-definite (all its eigenvalues $\lambda_a\geq 0$ are real and non-negative),

\item $\hat{H}_\text{EF}$ always has (at least) a zero eigenvalue $\lambda_0=0$ in the $\dsZ_2$ (Ising parity) even sector, whose left- and right-eigenstates are
\eqs{\label{eq:L0R0}
\bra{L_0}&=\frac{\bra{\Uparrow}+\bra{\Downarrow}}{2},\\
\ket{R_0}&=\ket{W_\text{Page}}.}
The left zero mode $\bra{L_0}$ is the Ising symmetric superposition of the all-up and the all-down states. The right zero mode $\ket{R_0}$ is the Page EF state given in \eqnref{eq:WPage}.

\end{itemize}
The proof can be found in \appref{app:HEF}. With these results, we can obtain several universal behaviors of  entanglement dynamics with local scrambling in the short-time and long-time limit.

In the short-time limit ($t\to0$), expanding the solution of EF state in \eqnref{eq:S sol} to first order in $t$, we can show that the EE grows \emph{linearly} in time,
\eq{S^{(2)}[\vect{\sigma}](t)=S^{(2)}[\vect{\sigma}](0)+v_\text{E}^{(2)}[\vect{\sigma}]\times t+\scO(t^2),}
where the linear-time coefficient $v_\text{E}^{(2)}[\vect{\sigma}]$ is the entanglement growth rate, which is related to the entanglement velocity introduced in Ref.\onlinecite{Hartman2013Time,Liu2014Entanglement}
\eq{\label{eq:vE=HEF}
v_\text{E}^{(2)}[\vect{\sigma}]=\partial_tS^{(2)}[\vect{\sigma}](0)=\frac{\bra{\vect{\sigma}}\hat{H}_\text{EF}\ket{W_{\Psi_{0}}}}{\braket{\vect{\sigma}}{W_{\Psi_{0}}}}.}
The entanglement velocity $v_\text{E}^{(2)}[\vect{\sigma}]$ characterizes how fast the EE grows in a given entanglement region specified by $\vect{\sigma}$. It is proportional to the matrix element of the EF Hamiltonian $\hat{H}_\text{EF}$, as can be seen in \eqnref{eq:vE=HEF}, because $\hat{H}_\text{EF}$ is the time-evolution generator that drives the entanglement dynamics. In particular, if the initial state is a generic product state, i.e.~$\ket{W_{\Psi_{0}}}=\ket{W_\text{prod}}=\sum_{\vect{\sigma}}\ket{\vect{\sigma}}$ as given in \eqnref{eq:Wprod}, the entanglement velocity $v_\text{E}^{(2)}[\vect{\sigma}]$ admits an explicit formula
\eq{\label{eq:vE}v_\text{E}^{(2)}[\vect{\sigma}]=\sum_{\langle ij\rangle}\tilde{g}_{ij}\frac{1-\sigma_i\sigma_j}{2},}
where $\tilde{g}_{ij}=g_{ij}e^{-\beta_{ij}-2\delta}\geq 0$ is the effective coupling. \eqnref{eq:vE} describes how the entanglement velocity $v_\text{E}^{(2)}$ depends on the choice of the entanglement region $\vect{\sigma}$. It is obvious that the entanglement velocity $v_\text{E}^{(2)}[\vect{\sigma}]\geq 0$ is non-negative for all choices of entanglement regions, because the EE can only \emph{grow} from an unentangled product state. If $\tilde{g}_{ij}=\tilde{g}$ is uniform through out the system, $v_\text{E}^{(2)}[\vect{\sigma}]$ will simply be proportional to the number of domain walls in the Ising configuration $\vect{\sigma}$, which is also the area $|\partial A|$ of the entanglement region $A$. Therefore the entanglement velocity follows the \emph{area-law} scaling,
\eq{v_\text{E}^{(2)}=\tilde{g}|\partial A|,}
which can be expected from the locality of the entanglement dynamics in our setup.

In the long-time limit ($t\to\infty$), the EF state is dominated by the zero mode (assuming the zero mode is unique) and all the other modes decays exponentially with time. The positive semi-definite property of the EF Hamiltonian, i.e. $\lambda_a\geq 0$, ensures that all modes (except the zero mode) will decay exponentially in time. As $t\to\infty$, \eqnref{eq:solution} reduces to
\eq{\label{eq:W project}\ket{W_{\Psi_{\infty}}}=\ket{R_0}\braket{L_0}{W_{\Psi_{0}}},}
with the left and right zero modes given by \eqnref{eq:L0R0}. Given that the EE vanishes in trivial regions, $\braket{\Uparrow}{W_{\Psi}}=\braket{\Downarrow}{W_{\Psi}}=1$, so $\braket{L_0}{W_{\Psi}}=1$ for any EF state $\ket{W_\Psi}$. Then \eqnref{eq:W project} results in
\eq{\ket{W_{\Psi_{\infty}}}=\ket{R_0}=\ket{W_\text{Page}},}
meaning that the EF always converge to that of the Page state in the long-time limit regardless what the initial state is. All states are doomed to thermalize under the quantum dynamics with local scrambling. The Page state will be their final destiny, whose EE reads
\eq{S^{(2)}[\vect{\sigma}]=-\log\frac{\cosh(\eta\sum_{i=1}^L\sigma_i)}{\cosh(\eta L)},}
which follows from \eqnref{eq:WPage}. For $|A|\ll L$, the EE exhibits the \emph{volume-law} scaling
\eq{S^{(2)}(A)=2\eta|A|,}
with the volume law coefficient given by $2\eta=\log d$. It worth mention that the above conclusion is based on the assumption that the zero mode is unique. If there are other degenerated zero modes (other than $\ket{W_\text{Page}}$), the final state may not converge to the Page state and the system can evade thermalization. We will discuss such a possibility later with a more concrete model in \secref{sec:RUC}.

\subsection{Excitation Spectrum of the Entanglement Feature Hamiltonian}\label{sec:excitation}

Having discussed the ground state property of the EF Hamiltonian $\hat{H}_\text{EF}$, let us turn to the low-lying excited states of $\hat{H}_\text{EF}$. According to \eqnref{eq:solution}, every eigenmode with finite eigenenergy $\lambda_a$ will decay exponentially in time as $e^{-\lambda_a t}$. Eventually, only the ground state with zero eigenenergy ($\lambda_0=0$) would survive, and the system thermalizes to Page states. Hence the low-energy excitation spectrum determines how the EE approaches to its thermal limit in the late-time regime. Here we will focus on the spectrum of two kinds of excitations, namely the two-domain-wall excitation and the single spin-flip excitation, which dominate the low-energy excitations. We obtain the analytical expression of their dispersion relations, from which we can estimate the excitation gap and determine the relaxation time. In \secref{sec:LSHD}, we further compare the analytically estimated relaxation time with the numerical fitted one, and find good consistency.

\begin{figure}[htbp]
	\begin{center}
		\includegraphics[width=0.70\columnwidth]{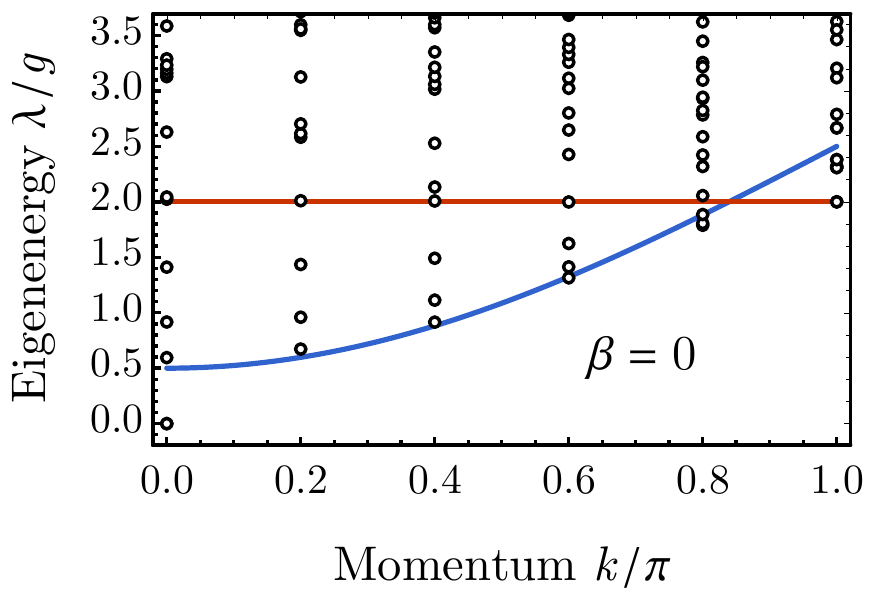}
		\caption{We perform exact diagonalization for the EF Hamiltonian $\hat{H}_\text{EF}$ with $\beta=0, L =10$. Each small circle represents an eigenstate label by its eigenenergy $\lambda$ and its crystal momentum $k$. The blue curve is the analytical result of two-domain-wall ansatz  \eqnref{eq:dispersion}. The red curve is the analytical result of single spin-flip ansatz \eqnref{eq:SMAdispersion}.   }
		\label{fig:betadispersion}
	\end{center}
\end{figure}

For simplicity, we assume the parameters $g_{ij},\beta_{ij}$ in the EF Hamiltonian $\hat{H}_\text{EF}$ are spatially homogeneous (i.e $g_{ij}=g,\beta_{ij}=\beta$). For the parameter $\beta$, any unitary evolution generated from Hamiltonian $e^{-\ii\epsilon H}$ would have nonzero $\beta$ only at the order of $\scO(\epsilon^{2})$ in small $\epsilon$ limit (see \appref{app:epsilon expansion} for details). Hence, we will take $\beta=0$ in the following. More general results for $\beta\neq0$ can be found in \appref{app:dispersiontwodomain} and \appref{app:dispersionSMA}. To first gain some intuitions about the excitation spectrum, we start with the exact diagonalization (ED) of EF Hamiltonian. The result is shown in \figref{fig:betadispersion}. Apart from the eigenenergy $\lambda_a$, every state $\ket{R_a}$ is also label by its crystal momentum $k_a$, which is defined through its translation eigenvalue as $\mathsf{T}\ket{R_a}=e^{\ii k_a}\ket{R_a}$, where the translation operator $\mathsf{T}$ is defined by its action on the Ising basis $\mathsf{T}\ket{\sigma_1\sigma_2\cdots\sigma_L}=\ket{\sigma_L\sigma_1\cdots\sigma_{L-1}}$. One can see that above the ground state at $\lambda_0=0$ and $k_0=0$, there is a continuum of excited states.

\begin{figure}[htbp]
	\begin{center}
		\includegraphics[width=0.72\columnwidth]{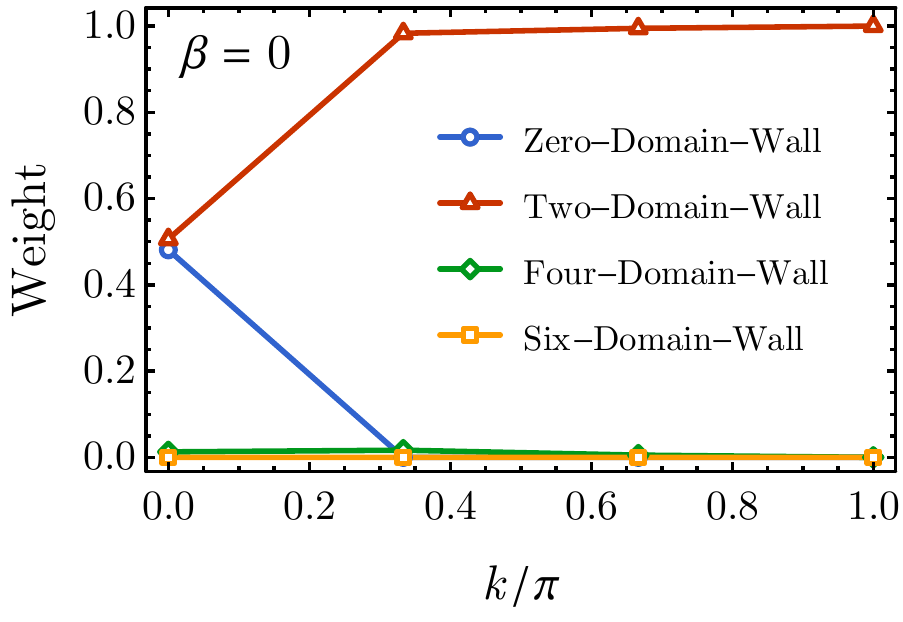}
		\caption{We perform exact diagonalization for the EF Hamiltonian $\hat{H}_\text{EF}$ with $g=1,\beta=0, L =6$. The weight is defined as follows: the left excited state $\bra{L}$ can be expressed as the linear combination of two-, four- and six-domain-wall states with the coefficient $(c_{2},c_{4},c_{6})$. The weight of individual type is equal to $|c_{n}|^{2}$. For $k=0$, zero-domain wall states take half of the weight in the lowest-energy excited state. However, they have no contribution in dispersion relation since their eigenenergy is zero.}
		\label{fig:weightdis}
	\end{center}
\end{figure}

To better understand these excited states, we look into their wave function. We realize that the excitation can be classified based on the number of domain walls in the \emph{left}-eigenstate. For instance, $\bra{\uparrow\cdots\uparrow \downarrow\cdots\downarrow\uparrow\cdots\uparrow }$ is an example of two-domain-wall states. As mentioned in \eqnref{eq:L0R0}, the left ground state 
$\bra{L_{0}}=(\bra{\Uparrow}+\bra{\Downarrow})/2$ contains no domain wall and hence no excitation. Other excited left-eigenstate will be a superposition of states of different domain-wall number. Note that the corresponding right eigenstate can be obtained from $\ket{R} = (\bra{L}\hat{W}_\id)^\intercal$. \figref{fig:weightdis} shows the weights of different domain-wall states in the lowest-energy excited state of various momenta. The ED result indicates that the lowest-energy excited state mainly consists of two-domain-wall states, so we will focus on them in the following.

Based on the numerical observation, we approximate low energy excitation by the two-domain-wall (2DW) ansatz state as follows,
\eq{\bra{k}\propto\sum_{i_1,i_2}e^{\ii k\frac{i_1+i_2}{2}}\phi_{i_2-i_1}^*\bra{i_1,i_2},}
where $\bra{i_1,i_2}=\bra{\Uparrow}\prod_{i=i_1}^{i_2-1}X_{i}$ is a two-domain-wall state with domain walls located at $i_1$ and $i_2$. $k$ labels the center of mass momentum of the pair of domain walls. $\phi_{\Delta i}$ is a variational wave function that describes the relative motion between the domain walls. We can then evaluate the energy expectation value $\lambda(k)$ on the ansatz state $\bra{k}$, 
\eq{\lambda_\text{2DW}(k)=\frac{\bra{k}\hat{H}_\text{EF}\hat{W}_{\id}\ket{k}}{\bra{k}\hat{W}_{\id}\ket{k}},}
where $\hat{W}_{\id}\ket{k}$ is understood as the corresponding right-state of the ansatz left-state $\bra{k}$. Two assumptions are made to derive the analytical expression of the dispersion relation. The first assumption is that these domain walls have no interaction with each other and thus $\phi_{\Delta i}$ can be approximated by plane waves. The second assumption is the thermodynamic limit $L\rightarrow \infty$, which would simplify the calculation but suppress the contribution from short two-domain-wall states (see \appref{app:dispersiontwodomain} for details). Based on these assumptions, the dispersion relation for $\beta=0$ can be derived as,
\eq{\label{eq:dispersion}
\lambda_\text{2DW}(k)=2g\Big(1+\frac{1}{d^2}\Big)-\frac{4g}{d}\cos\frac{k}{2}+\scO(d^{-3}).}
The band minimum is at $k=0$, which defines the excitation gap
\eq{\label{eq:gap}\Delta =\min_k\lambda(k)= 2g\Big(1-\frac{1}{d}\Big)^{2}+\scO(d^{-3}).}
It turns out that the gap remains open (i.e. $\Delta>0$) for any finite $g>0$. 

\begin{figure}[htbp]
	\begin{center}
		\includegraphics[width=0.65\columnwidth]{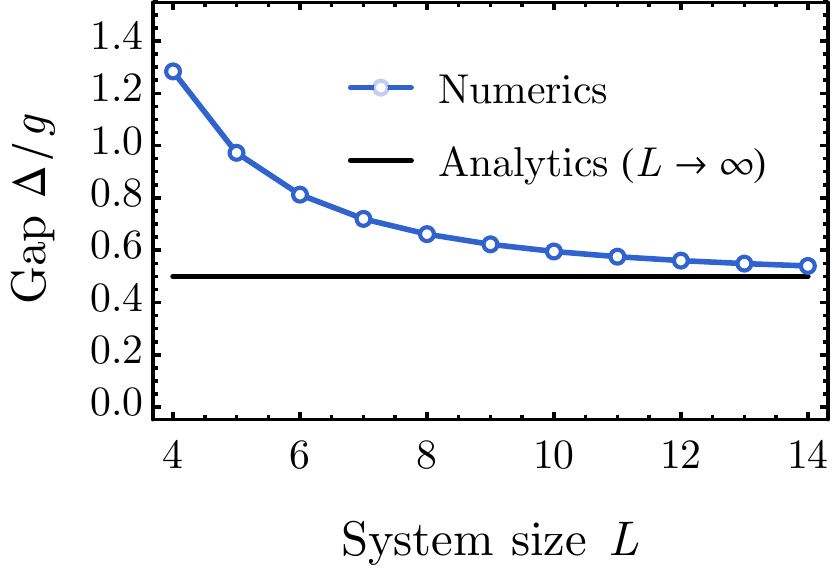}
		\caption{Comparison of the excitation gap between the finite-size ED result and the analytical result of two-domain-wall ansatz in the thermodynamic limit for the qudit dimension $d=2$. The analytical result $\Delta=g/2$ is given by \eqnref{eq:gap}.}
		\label{fig:gapED}
	\end{center}
\end{figure}

The comparison between ED result (black circles) and our analytical expression (blue curve) is shown in \figref{fig:betadispersion}. The lower-edge of the excitation spectrum is pretty well captured by the two-domain-wall ansatz. The comparison also reveals a finite-size-effect in the spectrum. In \figref{fig:gapED}, we show how the gap at $k=0$ (from ED) approaches to the analytic result of \eqnref{eq:gap} with increasing system size $L$. We also observe a systematic deviation of our analytical result from the excitation edge near $k=\pi$. The reason is that the eigenstate around $k=\pi$ is dominated by single-site excitations, where the domain-walls are next to each other such that their interaction can not be ignored. To capture the interaction effect, we switch to another ansatz state, which describes the motion of a tightly-bound domain-wall pair, or equivalently a single spin-flip (SSF) excitation (see \appref{app:dispersionSMA} for details). The dispersion of the SSF excitation reads 
\eq{\label{eq:SMAdispersion}
	\lambda_\text{SSF}(k)=2g,}
which turns out to be independent of the qudit dimension $d$ and the momentum $k$. This dispersion relation basically passes a series of points in \figref{fig:betadispersion} and only becomes the lowest excited state around $k=\pi$.

\section{Applications and Numerics}
\label{sec:numerics}

\subsection{Models of Locally Scrambled Quantum Dynamics}
\label{sec:models}

\begin{figure}[htbp]
\begin{center}
\includegraphics[width=0.8\columnwidth]{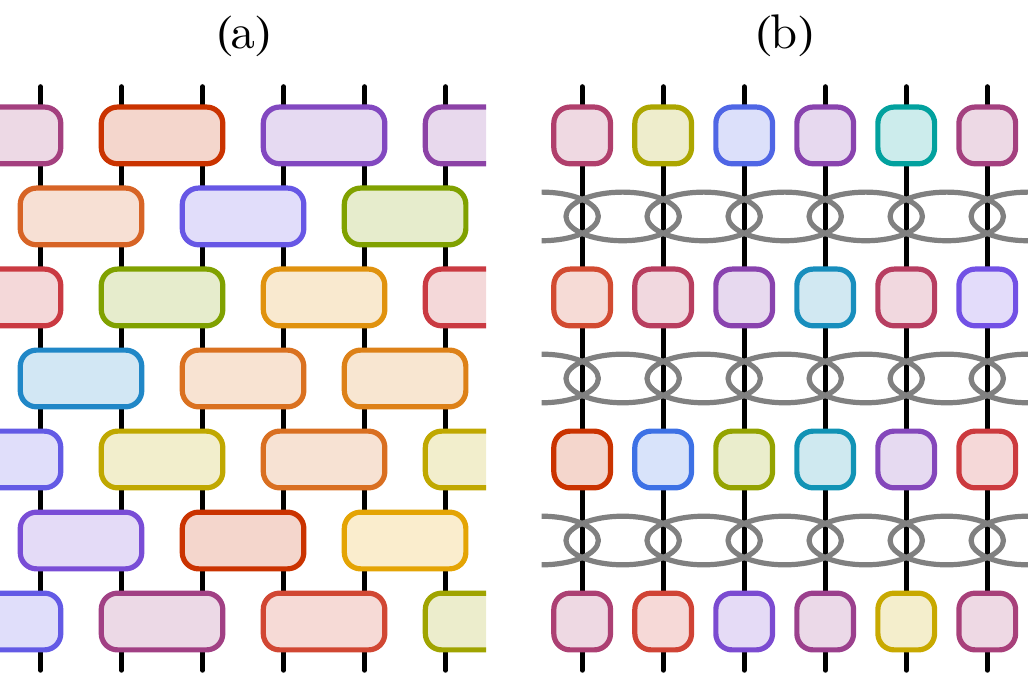}
\caption{(a) Locally scrambled random circuit. The gates are drawn independently in both space and time (as indicated by different colors). (b) Locally scrambled Hamiltonian dynamics. The unitary operators generated by the local Hamiltonian are overlapping gray ovals in each layer. The on-site scramblers are uncorrelated in both space and time (as indicated by different colors).}
\label{fig:models}
\end{center}
\end{figure}

In the following, we will apply the entanglement feature formalism to several scenarios of locally scrambled quantum dynamics. We will consider two types of models: random circuit models with discrete time as in \figref{fig:models}(a), and Hamiltonian generated evolutions with local scramblers in the limit of continuous time as in \figref{fig:models}(b). For the discrete time models, namely \emph{locally scrambled random circuits}, we will adopt the transfer matrix method to study the entanglement dynamics. For the continuous time models, namely \emph{locally scrambled Hamiltonian dynamics}, we will apply the EF Hamiltonian approach.

The random circuit we consider will be of the ``brick wall'' structure as shown in \figref{fig:models}(a). The entire unitary circuit $U=\prod_t U_t$ is constructed by stacking layers of unitary gates. Each layer $U_t$ is described by
\eq{U_t=\left\{\begin{array}{ll}\bigotimes_{x} U_{t;2x-1,2x} & t\in\text{odd},\\
\bigotimes_{x} U_{t;2x,2x+1} & t\in\text{even},
\end{array}\right.}
where $U_{t;ij}$ denotes the two-qudit unitary gate acting on sites $i$ and $j$ at time $t$. Each gate $U_{t;ij}$ is independently sampled from a locally scrambled unitary ensemble, so the quantum circuit $U$ will be dubbed as a locally scrambled random circuit. In fact, any gate can be made locally scrambled by symmetrizing over local basis transformations as constructed in \eqnref{eq:LSUE}. The construction here is more general than the Haar random unitary circuit\cite{Nahum2017Quantum}, as the unitary gate here does not need to be Haar random. As the quantum state evolves by $\ket{\Psi_{t+1}}=U_t\ket{\Psi_t}$, the corresponding EF state evolves by $\ket{W_{\Psi_{t+1}}}=\hat{T}_t\ket{W_{\Psi_t}}$. The transfer matrix $\hat{T}_t$ follows the same structure as $U_t$,
\eq{\label{eq:T RUC}\hat{T}_t=\left\{\begin{array}{ll}\bigotimes_{x} \hat{T}_{2x-1,2x} & t\in\text{odd},\\
\bigotimes_{x} \hat{T}_{2x,2x+1} & t\in\text{even}.
\end{array}\right.}
According to \eqnref{eq:T def}, $\hat{T}_{ij}$ is fully determined by the EF of $U_{t;ij}$ via
\eq{\label{eq:Tij}\hat{T}_{ij}=\hat{W}_{U_{t;ij}}\hat{W}_{\id_{ij}}^{-1}.}
Here we have assumed that $U_{t;ij}$ are drawn from identical unitary ensembles, such that $\hat{T}_{ij}$ is time-independent (despite of the time-dependence in $U_{t;ij}$). In the following, we will provide examples of the locally scrambled two-qudit unitary ensemble. We will use the transfer matrix approach to calculate the entanglement dynamics. The result will be compared with exact numerics by explicitly constructing the random circuit and average the final state EE over random realizations.

Another type of locally scrambled quantum dynamics that we will consider is generated by a local Hamiltonian $H=\sum_{\langle ij\rangle} H_{ij}$, which is a sum of local terms $H_{ij}$ defined on nearest neighboring bonds $\langle ij\rangle$ along a 1D chain. Each step of the unitary evolution $U_t$ is independently drawn from the locally scrambled unitary ensemble $\scE_{e^{-\ii \epsilon H}}$ generated by the Hamiltonian $H$,
\eq{\label{eq:LSEH}\scE_{e^{-\ii \epsilon H}}=\{V^\dagger e^{-\ii \epsilon H}V|V=\bigotimes_{i=1}^{L}V_i, V_i\in\text{Haar}\},}
which may be simply denoted by $U_t=V_t^\dagger e^{-\ii\epsilon H}V_t$, as in \eqnref{eq:Ut of Hdyn}. Combining the adjacent local scramblers following \eqnref{eq:U seq}, the unitary evolution can be considered as repeatedly applying a short-time unitary evolution $e^{-\ii\epsilon H}$ followed by a layer of local scramblers, as illustrated in \figref{fig:models}(b). Such dynamics will be called the locally scrambled Hamiltonian dynamics. It is similar to the Brownian random circuit model\cite{Lashkari2013TFSC} in that each step of the evolution is driven by a different random Hamiltonian, but our construction is more general in that the random Hamiltonian ensemble only needs to be invariant under local basis transformations other than the full basis transformation of the many-body Hilbert space. For small $\epsilon$, we can take the continuous time approach to calculate the entanglement dynamics by solving the imaginary-time Schr\"odinger equation $\partial_t\ket{W_{\Psi_{t}}}=-\hat{H}_\text{EF}\ket{W_{\Psi_{t}}}$ in \eqnref{eq:Sch Eq}. It worth mentioning that the locally scrambled quantum dynamics we considered here should be distinguished from Trotterizing a Hamiltonian dynamics. Here, the short-time evolutions $e^{-\ii\epsilon H}$ are interrupted by local scramblers, such that they do not combine to a coherent long-time evolution generated by the same Hamiltonian $H$. The local scramblers destroy the original notion of time. In the quantum dynamics, $e^{-\ii\epsilon H}$ advances the quantum state by $\epsilon$ in time, but after the insertion of layers of local scramblers, the entanglement dynamics only progress by $\epsilon^2$, which is much slower. This phenomenon is analogous to the quantum Zeno effect due to the insertion of measurement. We conjecture that the local scramblers play a similar role as random local measurement in implementing random local basis transformations, such that the quantum dynamics is no longer coherent.

\subsection{Locally Scrambled Random Circuits}
\label{sec:RUC}

\begin{figure}[htbp]
\begin{center}
\includegraphics[width=0.88\columnwidth]{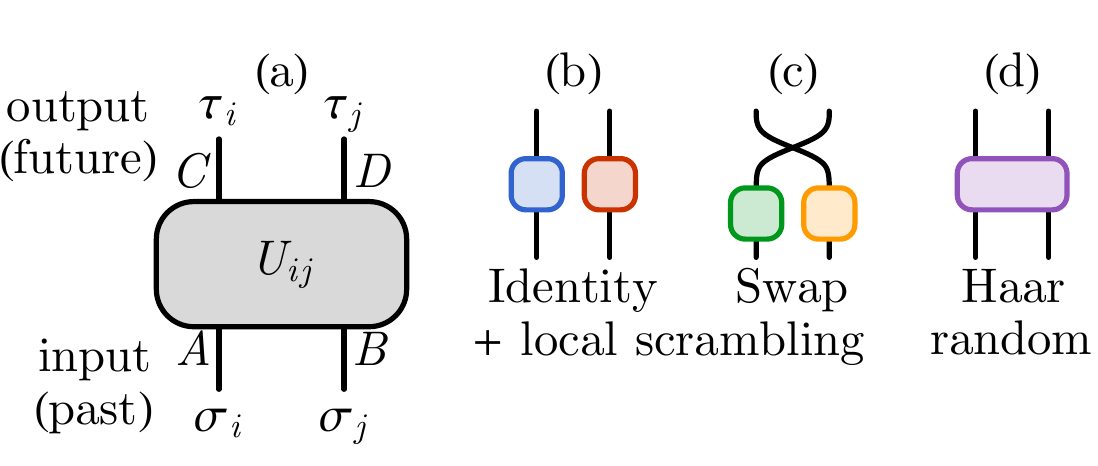}
\caption{(a) A generic two-qudit gate acting on qudits $i$ and $j$. The input channels are labeled by $A$ and $B$, and the output channels are labeled by $C$ and $D$. The EF of the gate will be labeled by the Ising configuration $\vect{\sigma}=(\sigma_i,\sigma_j)$ on the input side and $\vect{\tau}=(\tau_i,\tau_j)$ on the output side. (b-d) Examples of local basis independent ensembles of two-qudit gates: (b) identity gate with local scrambling, (c) swap gate with local scrambling, (d) Haar random unitary gate acting on both qudit (local basis automatically scrambled).}
\label{fig:gates}
\end{center}
\end{figure}

Let us first consider the locally scrambled random circuit as in \figref{fig:models}(a). The building blocks of the random circuit are two-qudit unitary gates. Each gate is independently drawn from local basis independent random ensembles. The EF of a two-qudit unitary operator $U_{ij}$ is completely characterized by two parameters: the cross channel mutual information $I_{ij}^\times$ and the tripartite information $I_{ij}^\triangledown$. Let us label the input and output channels of the two-qudit unitary by $A,B,C,D$ as shown in \figref{fig:gates}(a), then $I_{ij}^\times$ and $I_{ij}^\triangledown$ are defined as follows
\eqs{\label{eq:def I}
I_{ij}^\times&=I^{(2)}(A:D)=I^{(2)}(B:C),\\
I_{ij}^\triangledown&=I^{(2)}(A:C)+I^{(2)}(A:D)-I^{(2)}(A:CD).}
The mutual information, such as $I^{(2)}(A:D)=S^{(2)}_A+S^{(2)}_D-S^{(2)}_{AD}$, is understood by treating the unitary gate as a quantum state by bending the input and output legs to the same side, and calculating the operator EE following the definition in \refcite{Hosur2016Chaos,Nie2018Signature}.

In terms of these information measures $I_{ij}^\times$ and $I_{ij}^\triangledown$ of the unitary gate $U_{ij}$, the EF operator $\hat{W}_{U_{ij}}$ is given by
\eqs{\label{eq:WU general}
\hat{W}_{U_{ij}}&=d^2(d+X_i)(d+X_j)\\
&\phantom{=}-\frac{1-Z_iZ_j}{2}(A_{ij}-B_{ij}X_iX_j),\\
A_{ij}&=d^4(1-e^{I_{ij}^\triangledown-I_{ij}^\times}),\\
B_{ij}&=d^2(e^{I_{ij}^\times}-1).}
The cross channel mutual information $I_{ij}^\times\geq0$ is non-negative by the subadditivity\cite{Araki1970Entropy} of entropy. It describes the entanglement propagation, as it measures the amount of information transferred between site $i$ and $j$. The tripartite information $I_{ij}^\triangledown$ must be negative for unitary gates\cite{Hosur2016Chaos}, and therefore $I_{ij}^\times-I_{ij}^\triangledown\geq0$ holds. The negative tripartite information $(-I_{ij}^\triangledown)$ is proposed\cite{Hosur2016Chaos} to be a description of information scrambling, since it measures the amount of information about $A$ that is encoded in $C$ and $D$ jointly but can not be told by local measurements exclusively performed on $C$ or $D$. 

To gain more intuition about $I_{ij}^\times$ and $I_{ij}^\triangledown$, let us provide a few examples of local basis independent ensembles of two-qudit gates, as pictured in \figref{fig:gates}(b-d).

\begin{itemize}
\item Identity gate with local scrambling, i.e.~two on-site Haar random unitary gates direct product together, as \figref{fig:gates}(b). In this rather trivial case, we have
\eq{I_{ij}^\times=I_{ij}^\triangledown=0,}
such that the EF operator in \eqnref{eq:WU general} reduces to $\hat{W}_{\id}=d^2(d+X_i)(d+X_j)$, consistent with the previous result in \eqnref{eq:W1} by direct evaluation.

\item Swap gate with local scrambling, i.e.~two on-site Haar random unitary gates followed by an inter-site swap operator, as \figref{fig:gates}(c). In this case,
\eq{I_{ij}^\times=2\log d,\quad I_{ij}^\triangledown=0,}
such that the EF operator in \eqnref{eq:WU general} reduces to
\eqs{\label{eq:Wswap}\hat{W}_\text{swap}&=d^2(d+X_i)(d+X_j)\\
&\phantom{=}-d^2(d^2-1)\frac{1-Z_iZ_j}{2}(1-X_iX_j).}
The swap gate can generate and propagate quantum entanglement due to the non-vanishing cross channel information $I_{ij}^\times$. But there is no information scrambling happening \emph{between} the qudits (despite of the sufficient on-site scrambling), because the qubits are simply interchanged by the swap gate, such that local operators do not spread out other than being moved around in the space. The zero scrambling power of the swap gate is reflected in the zero tripartite information $I_{ij}^\triangledown$.

\item Haar random unitary gate acting on the two qudits, as \figref{fig:gates}(d). In this case,
\eq{I_{ij}^\times=\log\frac{2d^2}{d^2+1},\quad I_{ij}^\triangledown=\log\frac{4d^2}{(d^2+1)^2},}
such that the EF operator in \eqnref{eq:WU general} reduces to $\hat{W}_\text{Haar}$ given in \eqnref{eq:WHaar}, see \appref{app:EF Haar} for derivation. The Haar random unitary gate not only propagates quantum entanglement, but also scrambles the quantum information efficiently, as it has a negative tripartite information $I_{ij}^\triangledown$ (as long as $d>1$).
\end{itemize}

The above are examples of locally scrambled random unitary ensembles. Unitary gates drawn from such ensembles serve as the building block of locally scrambled random circuits. The entanglement dynamics of locally scrambled random circuits can be universally described by the transfer matrix approach as has been discussed in \secref{sec:models}. On the level of EF, the formulation is exact: the evolution of the average state EF can be precisely calculated from $\ket{W_{\Psi_{t+1}}}=\hat{W}_{U_t}\hat{W}_{\id}^{-1}\ket{W_{\Psi_t}}$ given the EF of the unitary. However, when applying the result to predict the EE, we rely on the assumption that the average EE can be approximated by the negative log of average EF following \eqnref{eq:S def}, where we effectively switch the order between the ensemble average and the logarithm. One major goal of the following is to provide numerical evidences to check this assumption in various different cases. It turns out that the negative log of EF generally provides a good estimate of the averaged EE, which makes our EF formulation useful in describing the entanglement dynamics for a broad class of random unitary circuits.

\begin{figure}[htbp]
\begin{center}
\includegraphics[width=0.6\columnwidth]{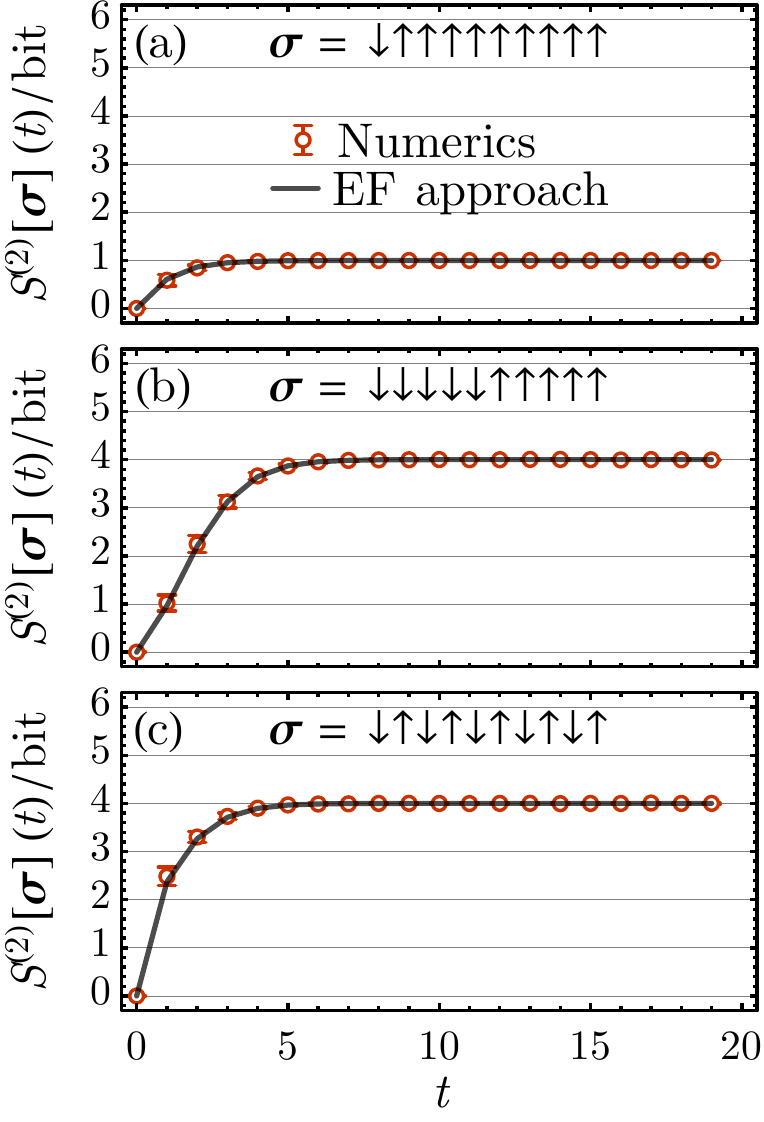}
\caption{The finial state EE of the Haar random circuit on a 10-site system for different choices of the entanglement regions: (a) single site, (b) half-system, (c) alternating\cite{Hsieh2014BESRQCWTS}. The qudit dimension is $d=2$ and the entropy is measured in unit of bit ($=\log 2$).}
\label{fig:RUC}
\end{center}
\end{figure}

Our first example is the standard Haar random unitary circuit, where each two-qudit gate is drawn from Haar random unitary ensemble independently. The model has be extensively studied in the literature,\cite{Nahum2017Quantum,Keyserlingk2018Operator,Khemani2018Operator,Nahum2018Operator} and the statistical mechanical model description has been developed by Zhou and Nahum in their pioneering work \refcite{Zhou2018Emergent}. We revisit this model to show that our formalism is equivalent to the Zhou-Nahum approach and can reproduce the known behaviors. Let us first calculate the transfer matrix $\hat{T}_{ij}$ of a single Haar random unitary gate $U_{ij}$ from its EF. Based on \eqnref{eq:WHaar} and \eqnref{eq:invW1}, we obtain
\eq{\label{eq:T Haar}\hat{T}_{ij}=\hat{W}_\text{Haar}\hat{W}_{\id}^{-1}=\Big(1+\frac{d(X_i+X_j)}{d^2+1}\Big)\frac{1+Z_iZ_j}{2}.}
Using the Ising basis $\ket{\sigma_i\sigma_j}$, \eqnref{eq:T Haar} can be expressed as
\eqs{\hat{T}_{ij}&=\ket{\uparrow\uparrow\,}\bra{\,\uparrow\uparrow}+\frac{d}{d^2+1}(\ket{\uparrow\downarrow\,}\bra{\,\uparrow\uparrow}+\ket{\downarrow\uparrow\,}\bra{\,\uparrow\uparrow})\\
&+\ket{\downarrow\downarrow\,}\bra{\,\downarrow\downarrow}+\frac{d}{d^2+1}(\ket{\downarrow\uparrow\,}\bra{\,\downarrow\downarrow}+\ket{\uparrow\downarrow\,}\bra{\,\downarrow\downarrow}),}
which is equivalent to the triangle weights $\dia{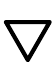}{11}{-2}=1$ and $\dia{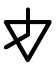}{11}{-2}=\dia{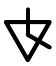}{11}{-2}=d/(d^2+1)$ that defines the Ising model in \refcite{Zhou2018Emergent}. An equivalent form of the transfer matrix \eqnref{eq:T Haar} was previously obtained in \refcite{Znidaric2008Exact}. Plugging \eqnref{eq:T Haar} to \eqnref{eq:T RUC}, we obtain the transfer matrix $\hat{T}_t$ that describes the EF state evolution under the quantum dynamics of the Haar random circuit. We assume the initial state is a product state, s.t. $\ket{W_0}=\ket{W_\text{prod}}$. We evolve the EF state by \eqnref{eq:transfer}. We can then compute the EE following \eqnref{eq:S def} and compare the result with the numerical simulation. In the simulation, we applied randomly sampled unitary gates to an initial product state and measure the final state EE, then perform the ensemble average of the EE. As shown in \figref{fig:RUC}, the EF approach provides pretty good prediction of the EE that matches the numerical result.

\begin{figure}[htbp]
\begin{center}
\includegraphics[width=0.85\columnwidth]{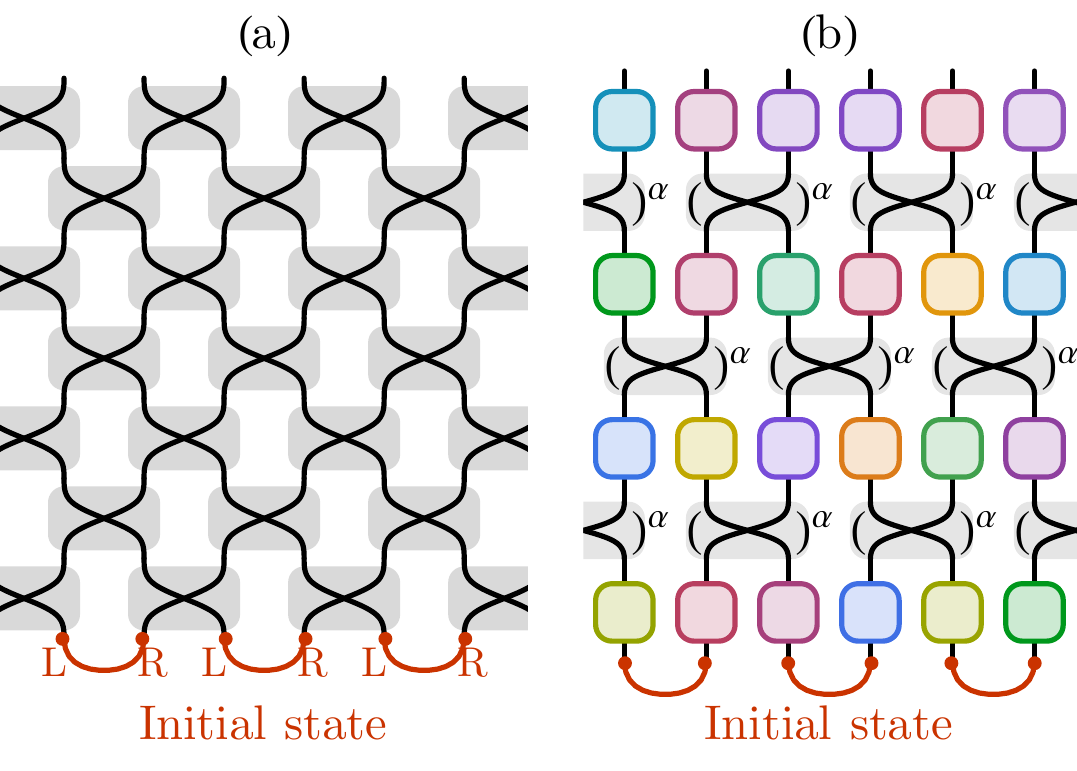}
\caption{(a) Swap gate circuit. Gray blocks mark out the swap gates. (b) Locally scrambled fractional swap gate circuit. Each swap gate is powered by the fraction $0<\alpha<1$.}
\label{fig:swap circuits}
\end{center}
\end{figure}

\begin{figure}[htbp]
\begin{center}
\includegraphics[width=0.9\columnwidth]{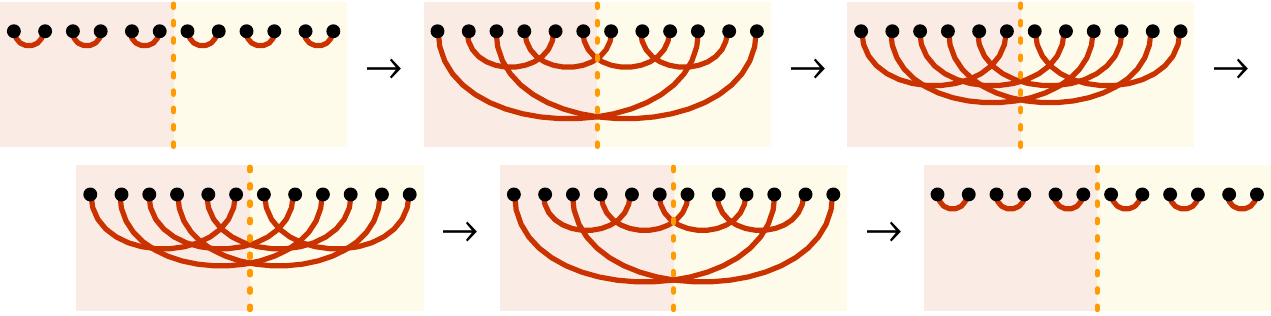}
\caption{Evolution of EPR pairs under the swap gate circuit on a 12-site chain with periodic boundary condition. The entanglement entropy between the left- and right-half system is proportional to the EPR pairs across the cut (indicated by dotted vertical line).}
\label{fig:EPR}
\end{center}
\end{figure}

Now let us turn to a new example of locally scrambled random circuits, namely the swap gate circuit, which is designed to mimic the entanglement dynamics in integrable conformal field theories (CFT) where entanglement spreads with the propagation of quasi-particles.\cite{Calabrese2005EEEOS,Nie2018Signature,Kudler-Flam2019Quantum} The circuit takes the architecture of the brick wall circuit in \figref{fig:models}(a) with gates drawn from the locally scrambled swap gate ensemble in \figref{fig:gates}(c), the resulting circuit is equivalent to an interweaving network as shown in \figref{fig:swap circuits}(a). The local scramblers in different layers can commute through the swap gates and combine to a single scrambling layer acting on the initial state, which can further be dropped as long as the initial state ensemble is already local basis invariant. For this model, we use a different initial state other than the product state. As illustrated in \figref{fig:swap circuits}(a), the initial state is chosen to be a product of Einstein-Podolsky-Rosen (EPR) pairs arranged along a one-dimensional chain, whose EF can be described by
\eq{\ket{W_0}=\prod_{x}\Big(1+\frac{1}{d}X_{2x-1}+\frac{1}{d}X_{2x}+X_{2x-1}X_{2x}\Big)\ket{\Uparrow}.}
For each EPR pair, the qudit labeled by $L$ (or $R$) will travel to the left (or right) in the swap gate circuit, which mimics the behaviors of left (or right) moving quasi-particles in an integrable CFT. In this way, entanglement spread out along the chain as EPR pairs  stretch out, following the steps depicted in \figref{fig:EPR}. On a finite-sized chain with periodic boundary condition, we expect to observe the half-system entanglement entropy to first grow and then decrease in time, and continue to oscillate like this. This recurrent behavior can be perfectly produced by the EF formulation, because, based on \eqnref{eq:Wswap}, the transfer matrix for a single swap gate turns out to be 
\eq{\hat{T}_{ij}=\hat{W}_\text{swap}\hat{W}_{\id}^{-1}=\frac{1}{2}(1+X_iX_j+Y_iY_j+Z_iZ_j),}
which is precisely the swap operator for Ising spins. In this way, the permutation of entangled qudits under the quantum dynamics is equivalently modeled by the permutation of correlated Ising spins in the EF formulation.

\begin{figure}[htbp]
\begin{center}
\includegraphics[width=0.6\columnwidth]{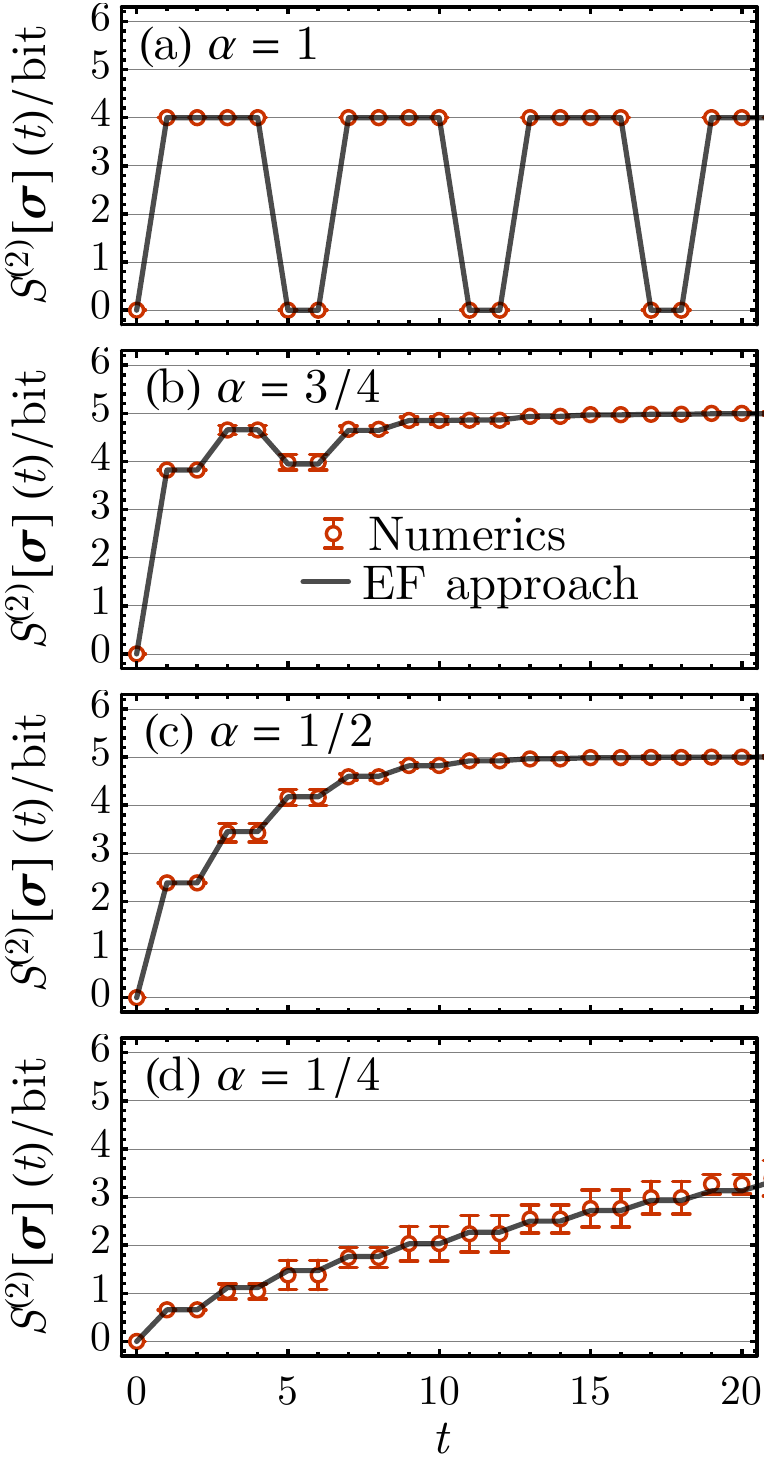}
\caption{Half-system entanglement entropy of the locally scrambled fractional swap gate circuit on a 12-site system with different fraction $\alpha$: (a) $\alpha=1$, (b) $\alpha=3/4$, (c) $\alpha=1/2$, (d) $\alpha=1/4$. The model is realized on a 12-site chain with periodic boundary condition. The entanglement region is chosen to be the first 6 sites. The qudit dimension is $d=2$ and the entropy is measured in unit of bit ($=\log2$).}
\label{fig:FSC}
\end{center}
\end{figure}

The recurrent (periodically oscillating) behavior of the half-system EE is demonstrated in \figref{fig:FSC}(a), where the EF approach matches the numerical simulation perfectly. The periodic recurrence of the low-entanglement state in the swap gate circuit seems to contradict with our previous conclusion in \secref{sec:universal} that locally scrambled quantum dynamics generally thermalize. The swap gate circuit evades thermalization because its corresponding EF transfer matrix admits more than one leading eigenstate. Let $\hat{T}=\bigotimes_x\hat{T}_{2x-1,2x}\bigotimes_x\hat{T}_{2x,2x+1}$ be two steps (one period) of the transfer matrix that translates the $L$ (or $R$) sublattice to the left (or right) by one unit-cell. On a chain of $2n$ sites, the operator $\hat{T}$ has $n^{-1}\sum_{d|n}\varphi(d)4^{n/d}$ fold degenerated eigenstates of eigenvalue $1$, with $\varphi(d)$ being the Euler totient function and $d$ running over all divisors of $n$. These eigenstates can be constructed by taking any Ising basis state and symmetrizing over the cyclic group generated by $\hat{T}$. Their degeneracy can be counted by mapping the problem to the number of $n$-bead necklaces with four colors,\cite{Gilbert1961STPS} where the four colors correspond to the four choices of $\uparrow\uparrow,\uparrow\downarrow,\downarrow\uparrow,\downarrow\downarrow$ configurations in each unit-cell. Therefore the Page state is not the unique state that can survive in the long-time limit, and thermalization is not the ultimate fate.

The swap gate circuit model can be generalized by introducing the \emph{fractional swap gate} that interpolates between the identity gate and the swap gate. The fractional swap gate can be written as a fractional power $\alpha$ of the swap gate with $0<\alpha<1$
\eq{\mathsf{SWAP}^\alpha=\frac{1+e^{\ii\alpha\pi}}{2}\;\dia{II}{14}{-4}+\frac{1-e^{\ii\alpha\pi}}{2}\;\dia{X}{14}{-4}.}
The fractional swap gate reduces to the identity gate (or the swap gate) at $\alpha=0$ (or $\alpha=1$). But unlike both identity and swap gates which do not scramble quantum information between the two qudits, the fractional swap gate does has finite scrambling power. We can construct a locally scrambled fractional swap gate circuit by starting from the architecture of the random circuit in \figref{fig:models}(a) and sampling every gate independently from local basis invariant fractional swap gate ensemble, as illustrated in \figref{fig:swap circuits}(b). The EF operator of the fractional swap gate follows the general form of \eqnref{eq:WU general} with parameters $A_{ij}$ and $B_{ij}$ given by
\eqs{A_{ij}&=d^2(d^2-1)\frac{3+\cos\alpha\pi}{2}\sin^2\frac{\alpha\pi}{2},\\
B_{ij}&=d^2(d^2-1)\sin^4\frac{\alpha\pi}{2}.}
Based on this result, the corresponding transfer matrix $\hat{T}_{ij}$ can be constructed by \eqnref{eq:Tij} and the evolution of the EF state can be calculated following the transfer matrix approach described in \eqnref{eq:T RUC}. In \figref{fig:FSC}(b-d), we compare the EE calculated based on the EF approach with the ensemble averaged EE from numerical simulation. They match perfectly for different values of $\alpha$. Because the fractional swap gate has finite scrambling power, the recurrence behavior no longer persist and the system can now thermalize. The entanglement dynamics is somewhat between that of the swap gate circuit and the Haar random circuit, in that the EE grows mostly linearly in time with small oscillations, until the EE eventually saturates to the thermal limit. As $\alpha$ becomes small, the system will take longer time (more steps) to thermalize. As shown in \figref{fig:FSC}(d), the oscillation of EE is suppressed and its growth curve is more smooth. In the $\alpha\to0$ limit, the entanglement dynamics approaches the continuum limit that can be described by the EF Hamiltonian, which is the topic of the following discussion.

\subsection{Locally Scrambled Hamiltonian Dynamics}
\label{sec:LSHD}

Now we turn to the locally scrambled Hamiltonian dynamics as illustrated in \figref{fig:models}(b). We consider the local Hamiltonian $H=\sum_{\langle ij\rangle} H_{ij}$ and assume that $H_{ij}$ on every bond is drawn from a local-basis-independent ensemble of two-qudit Hermitian operators. Equivalently, we can choose $H$ to be a fixed Hamiltonian and construct a locally scrambled unitary ensemble $\scE_{e^{-\ii\epsilon H}}$ by applying local basis transformations following \eqnref{eq:LSEH}. The quantum dynamics is described by the unitary 
\eq{\label{eq:U LSH}U=\prod_t \big(V_{t}e^{-\ii\epsilon H}\big),}
where $V_t$ describe the layer of local scramblers at time $t$, as illustrated in \figref{fig:models}(b). The corresponding entanglement dynamics is described by the imaginary-time Schr\"odinger equation \eqnref{eq:Sch Eq}, where the EF Hamiltonian takes the form of
\eq{\label{eq:HEF 1D}\hat{H}_\text{EF} = \sum_{\langle ij\rangle}g_{ij}\frac{1-{Z}_{i}{Z}_{j}}{2}e^{-\beta_{ij} X_iX_j-\delta(X_i+X_j)}.} 
It turns out that the parameters $\beta_{ij}\sim\scO(\epsilon^2)$ always vanish in the $\epsilon\to 0$ limit. The parameters $g_{ij}$ are the only non-trivial parameters to the leading order of $\epsilon$, which are determined by the local terms $H_{ij}$ in the Hamiltonian
\eqs{g_{ij}=&\frac{2}{d^2(d^2-1)}\big((\Tr H_{ij})^2+d^2\Tr(H_{ij}^2)\\&-d\Tr_j(\Tr_iH_{ij})^2-d\Tr_i(\Tr_jH_{ij})^2\big).}
The detailed derivation of these results can be found in \appref{app:epsilon expansion}.

One well-studied example of the locally scrambled Hamiltonian dynamics is the Brownian random circuit,\cite{Lashkari2013TFSC} where each step of the time evolution is generated by a random Hamiltonian drawn from the Gaussian unitary ensemble (GUE). The Hamiltonian can be written as a random $\U(d)$ spin model,
\eq{\label{eq:randH}H_t=\sum_{\langle ij\rangle}J_{t,ij}^{ab}T_i^a T_j^b,}
where $T_i^a$ (for $a=1,2,\cdots,d^2$) are $\U(d)$ generators on site $i$ with $\Tr T_i^{a\dagger}T_i^b=\delta^{ab}$. The coupling $J_{t,ij}^{ab}$ are independently drawn for each time $t$ and indices $i,j,a,b$ from the Gaussian distribution with zero mean and $d^{-2}$ variance. The quantum dynamics is described by $U=\prod_t e^{-\ii \epsilon H_t}$. The operator growth dynamics and the spectral form factor of the Brownian random circuit has been investigated in \refcite{Xu2018LQFS,Gharibyan2018ORMBSS,Zhou2019ODBQC,Chen2019QCDLPIS} recently, where differential equations governing the evolution of operator weight distribution were derived. Our approach also applies to the Brownian circuit model and results in similar differential equations for the evolution of EF state, whose EF Hamiltonian reads (see \appref{app:epsilon expansion} for derivation)
\eq{\label{eq:HEF Brownian}\hat{H}_\text{EF}=\frac{2(d^2-1)}{d^2}\sum_{\langle ij\rangle}\frac{1-Z_iZ_j}{2}e^{-\delta(X_i+X_j)}.}
We will not discuss this model in further details, given the extensive study of Brownian circuits in the literature. Instead, we will consider a new type of locally scrambled Hamiltonian dynamics.

\begin{figure}[htbp]
\begin{center}
\includegraphics[width=0.6\columnwidth]{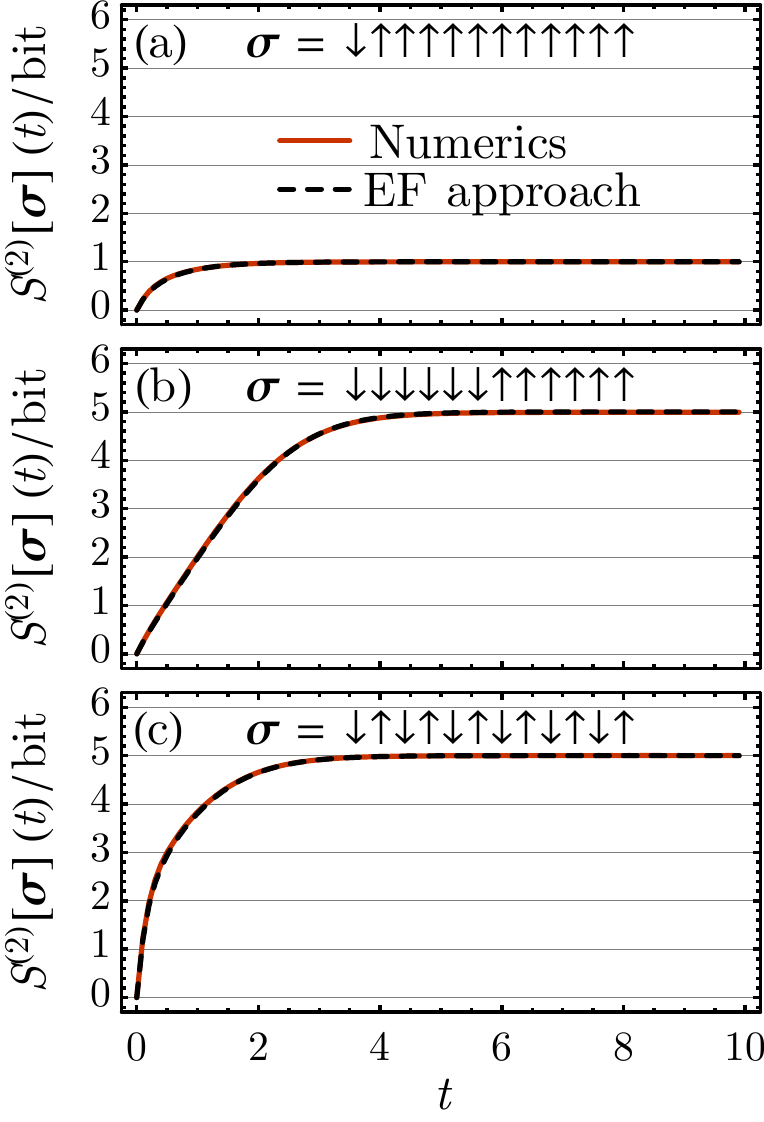}
\caption{The finial state EE of the locally scrambled Hamiltonian dynamics on a 12-site system for different choices of the entanglement regions: (a) single site, (b) half-system, (c) alternating\cite{Hsieh2014BESRQCWTS}. The qudit dimension is $d=2$ and the entropy is measured in unit of bit ($=\log 2$).}
\label{fig:Ising}
\end{center}
\end{figure}

We start with a fixed Hamiltonian on the one dimensional chain of qudits
\eq{\label{eq:H Ising}H=-\sum_{\langle ij\rangle}T_iT_j,}
where $T_i$ is one particular traceless Hermitian operator on site $i$ that squares to identity (i.e.~$\Tr T_i=0$ and $T_i^2=1$). For the qubit case ($d=2$), \eqnref{eq:H Ising} reduces to an Ising model. Note that there is no randomness in the Hamiltonian $H$. The randomness will be introduced by the local scramblers, when we use $H$ to generate the locally scrambled Hamiltonian dynamics following \eqnref{eq:U LSH}. The entanglement dynamics will be described by the following EF Hamiltonian
\eq{\label{eq:HEF Ising}\hat{H}_\text{EF}=\frac{2d^2}{d^2-1}\sum_{\langle ij\rangle}\frac{1-Z_iZ_j}{2}e^{-\delta(X_i+X_j)},}
which takes the same form as \eqnref{eq:HEF Brownian} but with a different parameter $g$. We can test the EF approach with numerical simulation on a 12-qubit system with the choice of $\epsilon=0.01$. We start with a product state $\ket{W_{\Psi_0}}=\ket{W_\text{prod}}$, evolve the EF state by \eqnref{eq:Sch Eq} and calculate the EE from \eqnref{eq:S def}. The result is shown in \figref{fig:Ising}. We can see that the averaged EE obtained from numerics matches well with the result of the EF approach over difference choices of the entanglement regions. These numerical evidences suggest that exchanging the order between taking ensemble average and taking logarithm does not seems to matter much, so the evolution equation we established for the EF in this work can provide reliable descriptions for the entanglement dynamics under locally scrambled quantum dynamics. Comparing \figref{fig:Ising} with \figref{fig:RUC}, one can see that the entanglement dynamics of the locally scrambled Hamiltonian dynamics closely resembles that of the Haar random unitary circuit. Thus the former can be considered as a continuum limit of the later.

\begin{figure}[htbp]
\begin{center}
\includegraphics[width=0.88\columnwidth]{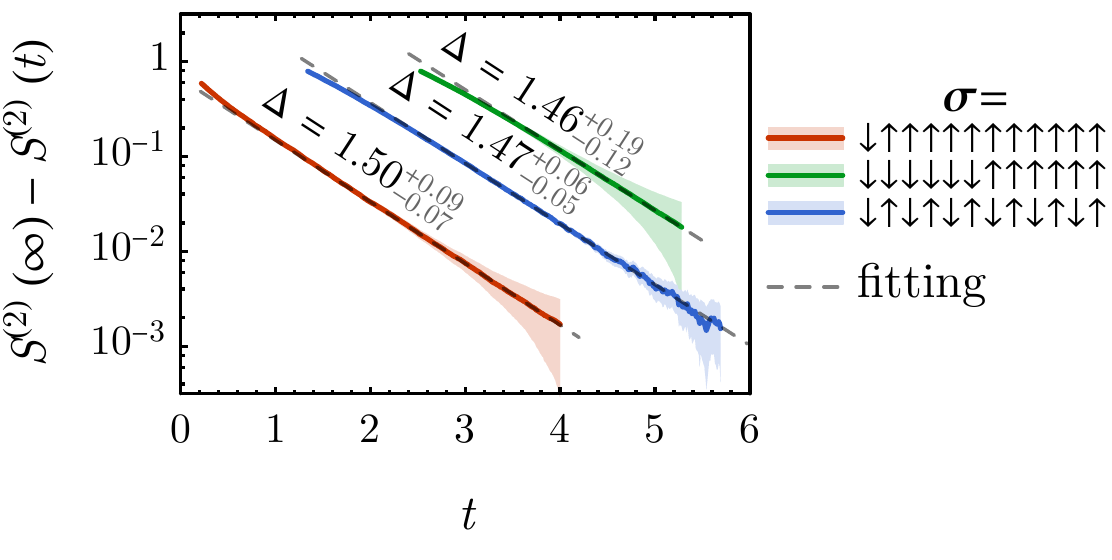}
\caption{The difference between the EE and its final saturation value, i.e.~$A[\vect{\sigma}]e^{-t/\tau}$, plot in the logarithmic scale vs time $t$. Different colors correspond to different choices of entanglement region (labeled by $\vect{\sigma}$). The shaded region indicates the error interval. The excitation gap $\Delta=\tau^{-1}$ is extracted by fitting the decay rate.}
\label{fig:gap}
\end{center}	
\end{figure}

We also notice that, in agreement with the imaginary time EF Schr\"odinger equation, the EE always approaches to its final thermalized value exponentially with the same relaxation time $\tau$ independent of the choice of the entanglement region,
\eq{\label{eq:tail}S^{(2)}[\vect{\sigma}](t)\to S^{(2)}[\vect{\sigma}](\infty)-A[\vect{\sigma}]e^{-t/\tau}.}
The relaxation time $\tau$ is intrinsically related to the excitation gap $\Delta$ of the EF Hamiltonian $\hat{H}_\text{EF}$, which can be estimated by \eqnref{eq:gap} in the thermodynamic limit,
\eq{\tau^{-1}=\Delta=2g\Big(1-\frac{1}{d}\Big)^2=\frac{4}{3},}
where the coupling $g$, according to \eqnref{eq:HEF Ising}, is given by $g=2d^2/(d^2-1)=8/3$ for qubits ($d=2$). To check this prediction, we fit the numerical simulation data using \eqnref{eq:tail} in the late-time regime to extract the excitation gap $\Delta$. As shown in \figref{fig:gap}, the EE approaches to the thermal value with the same rate (within error bars) regardless of the different choice of entanglement regions. The numerically fitted gap is around $\Delta=1.48$, which is close to the thermodynamic-limit analytic prediction $\Delta=4/3=1.33$. The small discrepancy mainly arises from the finite-size effect. If we use the finite-size gap formula $\Delta=0.56g$  based on the ED result in \figref{fig:gapED} at the system size $L=12$, we will obtain a better prediction of the gap $\Delta=1.49$, which matches the simulation result perfectly.

\section{Summary and Discussions}
\label{sec:summary}

In this work, we introduced the concept of locally scrambled quantum dynamics, where each step of the unitary evolution is randomized by local scramblers (on-site Haar random unitary gates). Surrounding each unitary gate in a quantum circuit by local scramblers effectively blocks the local-basis-specific quantum information from propagating in the circuit and  decouples the gates from each other under ensemble average. In this way, the average EF of the entire circuit can be constructed piece-by-piece from the EF of each gate, which makes the entanglement dynamics Markovian and enables us to write down the evolution equation for the EF of quantum states. The framework provides us the freedom to design the EF for each gate, such that we can go beyond the conventional Haar random gates and build the random circuit with more general random gates as long as their ensemble is local-basis-independent. This enables us to define and explore the continuum limit of locally scrambled quantum dynamics, under which the evolution of the EF state will be governed by an EF Hamiltonian. We obtained the general form of the EF Hamiltonian on symmetry ground and discussed the implication of its spectral properties on the entanglement dynamics. When the EF Hamiltonian is \emph{gapped}, the excitation modes in the EF state will decay exponentially in time $W_\ket{\Psi}=e^{-S}\sim e^{-t/\tau}$, which corresponds to a linear growth of EE in time, i.e.~$S\sim t/\tau$, as the system thermalizes. What has not been much discussed previously is the possibility that the EF Hamiltonian can become \emph{gapless} under fine-tuning, then the EF will decay in a power-law manner $W_\ket{\Psi}=e^{-S}\sim t^{-\alpha}$, which corresponds to a logarithmic growth of EE, i.e.~$S\sim \alpha\log t$. Such scenario could happen at  entanglement transitions,\cite{Vasseur2018Entanglement,Bao2019TPTRUCWM,Jian2019MCRQC} where the entanglement scaling of the long-time final state switches from volume-law to area-law. The transition can be interpreted as an order to disorder phase transition of the EF Hamiltonian. One interesting future direction is to explore different models of the EF Hamiltonian and to map out the phase diagram using analytical and numerical tools developed in quantum many-body physics.

Although we focused on the entanglement dynamics of pure states in this work, the EF formulation can be easily generalized to describe mixed state or operator entanglement. Following \eqnref{eq:WPsi}, it is straight forward to define the EF $W_{O}[\vect{\sigma}]$ for any many-body operator $O$ (including the density matrix $\rho$ as a special case),
\eq{W_{O}[\vect{\sigma}]=\Tr (\scX_{\vect{\sigma}}O^{\otimes 2}),}
and quantize the EF to a state $\ket{W_O}=\sum_{[\vect{\sigma}]}W_{O}[\vect{\sigma}]\ket{\vect{\sigma}}$. Suppose the operator evolves in time under a locally scrambled quantum dynamics $O'=UOU^\dagger$, the average EF will still be described by the same set of equation $\ket{W_{O'}}=\hat{W}_{U}\hat{W}_{\id}^{-1}\ket{W_{O}}$ as \eqnref{eq:EF relation}. Based on this, all the dynamic equation that we developed in this work applies directly, such that we do not need to derive a new set of equations for operator dynamics. The EF state $\ket{W_O}$ encodes the operator EE\cite{Zhou2017Operator} over all possible regions, which can be used to construct various quantities characterizing the operator size. To name a few, let us first assume $O$ to be a traceless Hermitian operator normalized to $\Tr O^2=d^L$. We can decompose the operator $O=\sum_{[a]}O_{[a]}T^{[a]}$ in the operator basis $T^{[a]}=\prod_i T_i^{a_i}$ (where $T_i^a$ denotes the $\SU(d)$ generator on the $i$th qudit), and define the operator weight $p_{[a]}=O_{[a]}^2$.\cite{Nahum2018Operator,Khemani2018Operator,Xu2018LQFS,Zhou2019ODBQC} The fraction of the operator in a subsystem $A$ then reads $p_A=\sum_{[a]\in A}p_{[a]},$\footnote{The notation $\sum_{[a]\in A}=\prod_{i\in A}\sum_{a_i=1}^{d^2-1}$ denotes the summation over the operator configuration $[a]$ which is non-trivial in region $A$.} which can be extracted from the EF state $\ket{W_O}$ by taking its inner product with a state $\ket{P_A}$ that labels the subsystem $A$:
\eq{p_A=\braket{P_A}{W_O},\quad\ket{P_A}=\frac{1}{d^{2L}}\prod_{i\in A}(dX_i-1)\ket{\Uparrow}.}
The fraction $p_A$ can be further used to characterize the average operator size $\ell_O=\sum_{A}p_{A}|A|$. The evolution equation for $p_A$ under Brownian dynamics is recently discussed in \refcite{Xu2018LQFS,Zhou2019ODBQC}. Another way to probe $\ket{W_O}$ is to consider the variance of the expectation value of $O$ on random mixed states $\rho$, i.e.~$\mathrm{var}_\rho\langle O\rangle_\rho=\dsE_{\rho}(\Tr \rho O)^2$. Suppose $\rho$ is drawn from a local-basis-independent ensemble characterized by its EF state $\ket{W_\rho}$, then the variance of operator expectation value is given by
\eq{\mathrm{var}_\rho\langle O\rangle_\rho=\bra{W_\rho}\hat{W}_{\id}^{-1}\ket{W_O}.}
It was recently pointed out by \refcite{Qi2019MOSGQQE} that $\mathrm{var}_\rho\langle O\rangle_\rho=\sum_{A}p_A (d+1)^{-|A|}$ can be expressed in terms of $p_A$, if $\rho$ is uniformly sampled from the ensemble of pure product states. The fact that $\mathrm{var}_\rho\langle O\rangle_\rho$ and $p_A$ are related to each other is less surprising in the EF formulation, because they are simply two different ways to probe the same EF state $\ket{W_O}$. The evolution equation of $\ket{W_O}$ under locally scrambled quantum dynamics is identical to that of $\ket{W_\Psi}$, from which the evolution equations of $p_A$, $\ell_O$ or $\mathrm{var}_\rho\langle O\rangle_\rho$ follow automatically. In this way, the EF formulation developed in our work provides a unified framework to discuss various aspects of the operator dynamics.

Another immediate generalization of the framework is to extend the unitary evolution to generic quantum channels allowing measurements to take place. The recent observation of measurement-induced entanglement transition in random unitary circuits \cite{Li2018QZEMET,Chan2019UED,Skinner2019MPTDE} has attracted much research interest.\cite{Li2019METHQC,Szyniszewski2019ETFVWM,Choi2019QECEPTRUCWPM,Bao2019TPTRUCWM,Jian2019MCRQC,Tang2019MPTCSNMDRGC} In these models, the quantum circuit is doped with local measurements (which can be either weak measurements or projective measurements happened with probability), and the final state EE is studied conditioned on the measurement outcome. If each measurements basis is randomly chosen each time, or if the local measurement take place only after the local basis has been sufficiently scrambled by the unitary evolution, the whole quantum channel still falls in the scope of locally scrambled quantum dynamics, which can be described by the EF approach developed in this work. In this case, each measurement,  described by the Kraus operator $M$, is also a local-basis-independent component in the quantum circuit, and has its own EF similar to \eqnref{eq:WU}
\eq{W_{M}[\vect{\sigma},\vect{\tau}]=\Tr\big(\scX_\vect{\sigma}M^{\otimes2}\scX_\vect{\tau}M^{\dagger\otimes2}\big),}
from which the EF operator 
\eq{\hat{W}_M=\sum_{\vect{\sigma},\vect{\tau}}\ket{\vect{\sigma}}W_M[\vect{\sigma},\vect{\tau}]\bra{\vect{\tau}}}
can be constructed. The EF state will evolve under measurement by $\ket{W_{\Psi'}}=\hat{W}_M\hat{W}_{\id}^{-1}\ket{W_\Psi}$, such that the approaches developed in this work seamlessly apply. The EF provides a finer resolution of the entanglement structure of a quantum many-body state beyond the single region scaling of EE, which turns out to be useful in diagnosing the error correction capacity\cite{Choi2019QECEPTRUCWPM} in the volume-law states prepared by the measurement-doped quantum circuits. We will leave this topic to future works\cite{Fan2019}. More generally, the EF formulation can be further generalized to locally scrambled tensor networks, which does not even need to have a preferential time direction. As long as each tensor in the tensor network is independently drawn from local-basis-independent ensembles, the entanglement structured of the random tensor network can be described by the EF approach. The freedom to design the EF for each separate tensor in the tensor networks opens up a large space of models to explore in the future.

There are also a few more challenging future directions that worth further investigation. The first direction is to generalize the 2nd R\'enyi EF to arbitrary R\'enyi index. As a consequence, the Ising variable on each site will be promoted to a permutation group element $\sigma_i\in S_n$. Such generalization will also allow us to access other measures of entanglement, such as R\'enyi negativity\cite{Kudler-Flam2019Quantum,Wu2019Entanglement,Lu2019Detecting}, as the moment of the partial transposed density matrix $\rho^{\intercal_A}$\cite{Calabrese2012Entanglement,Calabrese2013Entanglement,Alba2013Entanglement,Chung2014Entanglement,Alba2017Entanglement,Alba-Vincenzo2019Quantum} can be expressed in terms of the $n$th R\'enyi EF,
\eq{\Tr(\rho^{\intercal_A})^n=W_{\rho}[\vect{g}],\quad g_i=\left\{\begin{array}{cc}(n\cdots 21) & i\in A,\\(12\cdots n) & i\in\bar{A}.\end{array}\right.} 
The $n$th R\'enyi generalization of EF states $\ket{W_\Psi}$ and EF operators $\hat{W}_U$ can still be defined, but it will be more difficult to perform explicit calculations given that the number of group elements $n!$ grows quickly with $n$. Perhaps the most subtle issue is how to take the $n\to1$ replica limit systematically, which has been identified\cite{Vasseur2018Entanglement,Jian2019MCRQC,Bao2019TPTRUCWM} as an important step to understand the nature of entanglement transitions. The second direction is to include global symmetries and conservation laws\cite{Khemani2018Operator,Pai2019LFRC} into the discussion. This amounts to refining the generic local scramblers to symmetry-preserving local scramblers, which only performs basis transformations within each irreducible representations of the symmetry group. The formulation to describe the interference between the entanglement dynamics and the flow of symmetry representations in the quantum circuits still need to be developed. The third direction is to go beyond the locally scrambled quantum dynamics and to gradually introduce correlations among random gates in the spacetime. Can the current EF formulation serves as a good starting point to construct phenomenological descriptions for weakly correlated random gates? Can we eventually approach the limit of coherent quantum evolution for Hamiltonian or Floquet dynamics? There are many interesting open question awaiting us to explore.

\begin{acknowledgements}
We acknowledge the discussions with Adam Nahum, Xiao-Liang Qi, Chao-Ming Jian, Tarun Grover, John McGreevy, Ehud Altman, Shinsei Ryu, Sagar Vijay, Vedika Khemani, Yuri D. Lensky, and Tianci Zhou. In particular, we thank Ruihua Fan for introducing the swap gate circuit model to us. YZY thanks Yingfei Gu for early collaborations in developing the entanglement feature formulation. This research was done using resources provided by the Open Science Grid \cite{Pordes_2007,Sfiligoi_2009}, which is supported by the National Science Foundation and the U.S. Department of Energy's Office of Science.
\end{acknowledgements}

\bibliographystyle{apsrev4-1} 
\bibliography{EntanglementFeature}
\onecolumngrid

\newpage
\appendix

\section{Entanglement Feature of Page State}\label{app:Page}

The Page state can be considered as a single random tensor. According to \refcite{Hayden2016Holographic}, the 2nd R\'enyi entanglement feature of a random tensor network can be calculated as the partition function of an Ising model,
\eq{W_\text{RTN}[\vect{\sigma}]=\frac{1}{Z}\sum_{\vect{\tau}}e^{-E_\text{RTN}[\vect{\sigma},\vect{\tau}]},}
where each random tensor is mapped to an Ising spin $\tau_i$ coupled together via the network, and the boundary condition pinned by external Zeeman field along the direction specified by $\vect{\sigma}$. Applying this result to the Page state,
\eq{W_\text{Page}[\vect{\sigma}]=\frac{1}{Z}\sum_{\tau}e^{\eta\sum_{i=1}^L\sigma_i\tau},}
where there is only one Ising spin $\tau$ because the Page state is only a single random tensor. The $\tau$ spin couples to the boundary condition $\vect{\sigma}$ via uniform field strength $\eta=\frac{1}{2}\log d$, which is determined by the qudit Hilbert space dimension $d$ (see \refcite{Hayden2016Holographic} for derivation). Complete the summation over Ising spin $\tau$, we obtain
\eq{W_\text{Page}[\vect{\sigma}]=\frac{2}{Z}\cosh\Big(\eta\sum_{i=1}^L\sigma_i\Big).}
The normalization constant $Z$ is determined by the condition that $W_\text{Page}[\Uparrow]\equiv 1$, such that $Z=2\cosh(\eta L)$, hence
\eq{W_\text{Page}[\vect{\sigma}]=\frac{\cosh(\eta\sum_{i=1}^L\sigma_i)}{\cosh(\eta L)},}
which can be rewritten as the EF state $\ket{W_\text{Page}}$ in \eqnref{eq:WPage}.

\section{Entanglement Feature of Two-Qudit Haar Random Unitary}\label{app:EF Haar}

Here we derive the ensemble averaged EF operator for two-qudit Haar random unitary gate. We start with the definition
\eq{W_{U_{ij}}[\vect{\sigma},\vect{\tau}]=\Tr(\scX_{\vect{\sigma}}U_{ij}^{\otimes2}\scX_{\vect{\tau}}U_{ij}^{\dagger\otimes2}).}
$U_{ij}$ is a two-qudit gate acting on qudits labeled by $i$ and $j$. Focusing on these two qudits, the Ising variables $\vect{\sigma}=(\sigma_i,\sigma_j)$ and $\vect{\tau}=(\tau_i,\tau_j)$ both contain only two components. Consider averaging the EF $W_{U_{ij}}$ over unitary gates $U_{ij}$ in the Haar random unitary ensemble,
\eqs{\label{eq:EWU}
\dsE_{U_{ij}\in\text{Haar}}W_{U_{ij}}[\vect{\sigma},\vect{\tau}]&=\dsE_{U_{ij}\in\text{Haar}}\Tr(\scX_{\vect{\sigma}}U_{ij}^{\otimes2}\scX_{\vect{\tau}}U_{ij}^{\dagger\otimes2})\\
&=\sum_{g,h\in S_2}\mathsf{Wg}(g^{-1}h,d^2)\Tr(\scX_{g}\scX_{\sigma_1})\Tr(\scX_{g}\scX_{\sigma_2})\Tr(\scX_{h}\scX_{\tau_1})\Tr(\scX_{h}\scX_{\tau_2}),}
where $\mathsf{Wg}$ is the Weingarten function\cite{Weingarten1978Asymptotic,Collins2006Integration} and $g,h$ are $S_2$ group elements
\eq{\label{eq:Wg S2}
\mathsf{Wg}(g^{-1}h,d^2)=\left\{
\begin{array}{cc}
\frac{1}{d^4-1} & g^{-1}h=\dia{II}{11}{-2}\\
-\frac{1}{d^2(d^4-1)} & g^{-1}h=\dia{X}{11}{-2}\\
\end{array}\right..}
The cycle counting function $\Tr(\scX_{g}\scX_{h})$ follows
\eq{\label{eq:TrXX}
\Tr(\scX_{g}\scX_{h})=\left\{
\begin{array}{cc}
d^2 & g h=\dia{II}{11}{-2}\\
d & g h=\dia{X}{11}{-2}\\
\end{array}\right..}
Substitute \eqnref{eq:Wg S2} and \eqnref{eq:TrXX} into \eqnref{eq:EWU}, we can evaluate $\dsE_{U_{ij}\in\text{Haar}}W_{U_{ij}}[\vect{\sigma},\vect{\tau}]$ for all configurations of $\vect{\sigma},\vect{\tau}$. In terms of Ising variables (following the identification $\dia{II}{11}{-2}\Leftrightarrow\uparrow$ and $\dia{X}{11}{-2}\Leftrightarrow\downarrow$), we can summarize the result as the following matrix in the Ising basis $\vect{\sigma},\vect{\tau}=\uparrow\uparrow,\uparrow\downarrow,\downarrow\uparrow,\downarrow\downarrow$
\eq{\dsE_{U_{ij}\in\text{Haar}}\hat{W}_{U_{ij}}=\mat{
d^4 & d^3 & d^3 & d^2 \\
d^3 & \frac{2 d^4}{d^2+1} & \frac{2 d^4}{d^2+1} & d^3 \\
d^3 & \frac{2 d^4}{d^2+1} & \frac{2 d^4}{d^2+1} & d^3 \\
d^2 & d^3 & d^3 & d^4},}
which is also the matrix representation of the (ensemble averaged) EF operator $\hat{W}_{U_{ij}}$. The matrix can as well be written in terms of Pauli operators as
\eq{\dsE_{U_{ij}\in\text{Haar}}\hat{W}_{U_{ij}}=d^2(d+X_i)(d+X_j)-\frac{d^2(d^2-1)}{2(d^2+1)}(1-Z_iZ_j)(d^2-X_iX_j),}
as claimed in \eqnref{eq:WHaar}. For simplicity, we have omitted $\dsE_{U_{ij}\in\text{Haar}}$ notation in \eqnref{eq:WHaar}, with the understanding that the EF for an ensemble of unitaries is implicitly averaged.

\section{Relation Between State and Unitary Entanglement Features}
\label{app:diagram}

\begin{figure}[hbtp]
	\begin{center}
	\includegraphics[width=\textwidth]{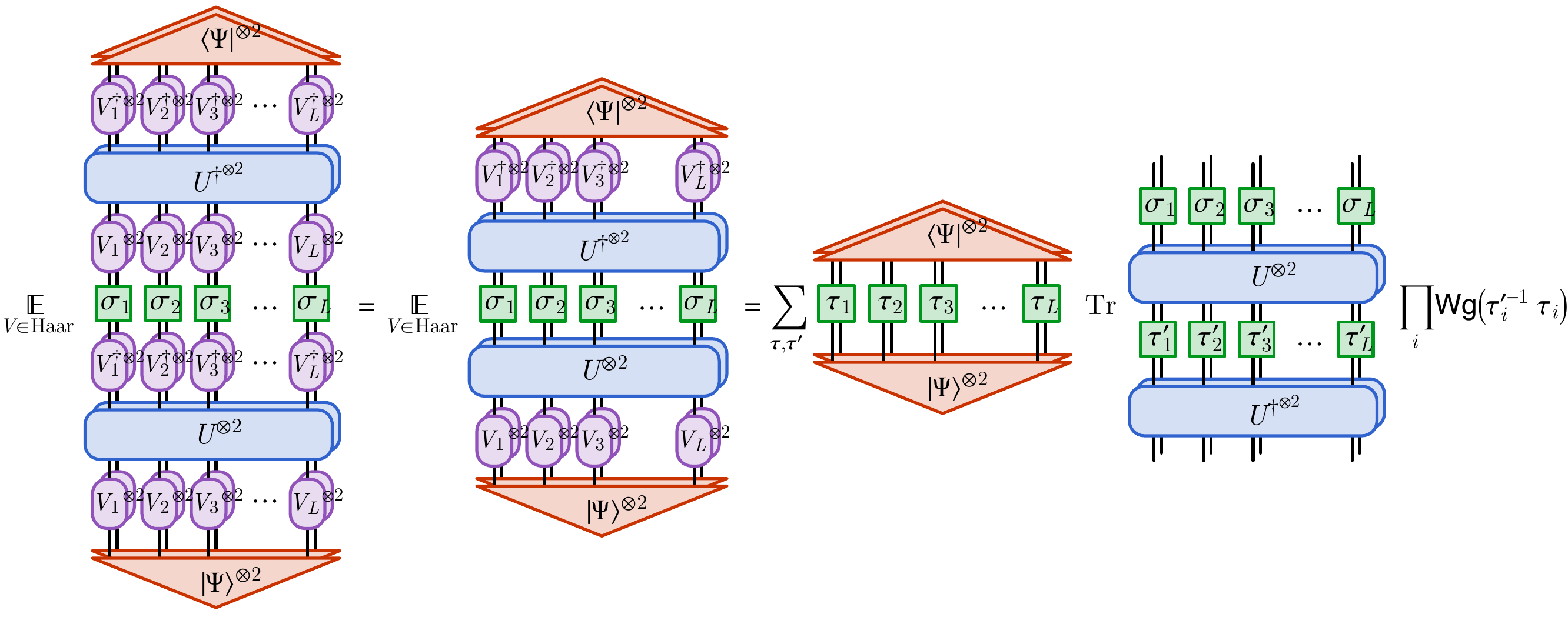}
	\end{center}
	\caption{Diagrammatic proof of \eqnref{eq:EF relation}}
	\label{fig:EF relation proof}
\end{figure}

Here we prove \eqnref{eq:EF relation}. Consider a many-body state (multi-qudit) state $\ket{\Psi}$ and an unitary operator $U_{t}$ supported in the same Hilbert space. Suppose that $\ket{\Psi'}=U_{t}\ket{\Psi}$, our goal is to derive the time evolution of the corresponding EF state. In general, this is not tractable since the unitary operator $U_{t}$ contains many non-universal features that are specific to the choice of local basis. Such features may affect the entanglement of the final state but such features are not captured in EF formalism. By this property, we instead consider an ensemble of unitary operator $U$,
\eq{\scE_U=\Big\{V^\dagger U V\Big|V=\bigotimes_{i=1}^{L}V_i, V_i\in\text{Haar}\Big\},}
where each $V_i$ independently follows Haar random unitary distribution on the $i$th qudit. Our goal is to compute $\mathop{\dsE}_{U'\in \scE_U}W_{U'\ket{\Psi}}[\vect{\sigma}]$,
\eqs{\label{eq:EF relation proof}\mathop{\dsE}_{U'\in \scE_U}W_{U'\ket{\Psi}}[\vect{\sigma}] &= \mathop{\dsE}_{U'\in \scE_U}\Tr[\scX_{\vect{\sigma}}(U'\ket{\Psi}\bra{\Psi}U'^{\dagger})^{\otimes 2}]\\
&= \mathop{\dsE}_{V\in \text{Haar}}\Tr[\scX_{\vect{\sigma}}(V^{\dagger}UV\ket{\Psi}\bra{\Psi}V^{\dagger}U^{\dagger}V)^{\otimes 2}]\\
&=\mathop{\dsE}_{V\in \text{Haar}}\bra{\Psi}^{\otimes 2}V^{\dagger\otimes 2}U^{\dagger\otimes 2}V^{\otimes 2}\scX_{\vect{\sigma}}V^{\dagger\otimes 2}U^{\otimes 2}V^{\otimes 2}\ket{\Psi}^{\otimes2 }\\
&=\mathop{\dsE}_{V\in \text{Haar}}\bra{\Psi}^{\otimes 2}V^{\dagger\otimes 2}U^{\dagger\otimes 2}\scX_{\vect{\sigma}}U^{\otimes 2}V^{\otimes 2}\ket{\Psi}^{\otimes2 }\\
&=\sum_{\vect{\tau},\vect{\tau}'}
\Tr(\scX_{\vect{\tau}}\ket{\Psi}\bra{\Psi}^{\otimes 2})\Tr(\scX_{\vect{\tau}'}U^{\dagger}(t)^{\otimes 2}\scX_{\vect{\sigma}}U_{t}^{\otimes 2})\prod_{i}\mathsf{Wg}(\tau'^{-1}_{i}\tau_{i},d)\\
&=\sum_{\vect{\tau},\vect{\tau}'}
W_\ket{\Psi}[\vect{\tau}]W_U[\vect{\sigma},\vect{\tau}']\prod_{i}\mathsf{Wg}(\tau'^{-1}_{i}\tau_{i},d),}
where $\mathsf{Wg}$ denotes the Weingarten function\cite{Weingarten1978Asymptotic,Collins2006Integration} originated from the Haar ensemble average of $V^{\dagger\otimes2}V^{\otimes2}$, and $\vect{\tau},\vect{\tau}'$ are new set of Ising variables. The derivation in \eqnref{eq:EF relation proof} can also be diagrammatically represented as \figref{fig:EF relation proof}. 

By definition, the Weingarten function, when viewed as a matrix indexed by $\vect{\tau}$ and $\vect{\tau}'$, is the inverse of the Gram matrix $\Tr\scX_\vect{\tau}\scX_{\vect{\tau}'}=\bra{\vect{\tau}'}\hat{W}_\id\ket{\vect{\tau}}$, which is simply the matrix representation of the EF operator $\hat{W}_\id$ of the identity operator. So the Weingarten function is given by the matrix element of $\hat{W}_{\id}^{-1}$ as
\eq{\prod_{i}\mathsf{Wg}(\tau'^{-1}_{i}\tau_{i},d)=\bra{\vect{\tau}'}\hat{W}_\id^{-1}\ket{\vect{\tau}}.}
Therefore, in operator form, we have
\eqs{\dsE_{U'\in\scE_U}\ket{W_{U'\Psi}}&=\sum_{\vect{\sigma}}W_{U'\ket{\Psi}}[\vect{\sigma}]\ket{\vect{\sigma}}\\
&=\sum_{\vect{\sigma},\vect{\tau}',\vect{\tau}}
\big(\ket{\vect{\sigma}}W_U[\vect{\sigma},\vect{\tau}']\bra{\vect{\tau}'}\big)\hat{W}_\id^{-1}\big(W_\ket{\Psi}[\vect{\tau}]\ket{\vect{\tau}}\big)\\
&=\hat{W}_U\hat{W}_\id^{-1}\ket{W_\Psi},}
as stated in \eqnref{eq:EF relation}.

\section{Spectral Properties of Entanglement Hamiltonian}
\label{app:HEF}

Let us start with the most general form of the EF Hamiltonian $\hat{H}_\text{EF}$ given in \eqnref{eq:HEF form} and investigate its spectral properties.
\eq{\label{eq:HEF=sumH}\hat{H}_\text{EF} = \sum_{i,j}\hat{H}_{ij},\qquad \hat{H}_{ij}=g_{ij}\frac{1-{Z}_{i}{Z}_{j}}{2}e^{-\beta_{ij}X_iX_j-\delta(X_i+X_j)},}
with $\coth\delta=d$. Our first goal is to show that $\hat{H}_\text{EF}$ is positive semi-definite. The trick is to first deform $\hat{H}_\text{EF}$ to a Hermitian version $\hat{H}'_\text{EF}$, given by
\eq{\label{eq:HEF'}\hat{H}'_\text{EF}=\hat{W}_{\id}^{-1/2}\hat{H}_\text{EF}\hat{W}_{\id}^{1/2}.}
Because $\hat{H}_\text{EF}$ and $\hat{H}'_\text{EF}$ are related by similar transformation, they share the same set of eigenvalues. So the positivity of the original EF Hamiltonian $\hat{H}_\text{EF}$ is equivalent to the positivity of the transformed Hermitian version $\hat{H}'_\text{EF}$. The later turns out to be easier to prove. By the way, to see that $\hat{H}'_\text{EF}$ is Hermitian (or real symmetric to be more precise), we use $\hat{W}_{\id}^\intercal=\hat{W}_{\id}$ and  \eqnref{eq:WH=HW} that $\hat{W}_{\id}\hat{H}_\text{EF}^\intercal=\hat{H}_\text{EF}\hat{W}_{\id}$, then
\eq{\hat{H}_\text{EF}^{\prime\intercal}
=\hat{W}_{\id}^{1/2}\hat{H}_\text{EF}^\intercal\hat{W}_{\id}^{-1/2}
=\hat{W}_{\id}^{-1/2}(\hat{W}_{\id}\hat{H}_\text{EF}^\intercal)\hat{W}_{\id}^{-1/2}
=\hat{W}_{\id}^{-1/2}(\hat{H}_\text{EF}\hat{W}_{\id})\hat{W}_{\id}^{-1/2}
=\hat{W}_{\id}^{-1/2}\hat{H}_\text{EF}\hat{W}_{\id}^{1/2}
=\hat{H}'_\text{EF},}
meaning that $\hat{H}'_\text{EF}$ is transpose symmetric. Moreover $\hat{H}'_\text{EF}$ is real by definition, so $\hat{H}'_\text{EF}$ is real and symmetric and therefore Hermitian. As a real symmetric operator, $\hat{H}'_\text{EF}$ admits the following spectral decomposition 
\eq{\hat{H}'_\text{EF}=\sum_{a}\ket{V_a}\lambda_a\bra{V_a},}
with $\ket{V_a}=\bra{V_a}^\intercal$ being the eigenvector corresponding to the eigenvalue $\lambda_a$. If we can show that the expectation value $\bra{V}\hat{H}'_\text{EF}\ket{V}\geq 0$ is non-negative on \emph{any} state $\ket{V}$ in the EF Hilbert space (including the eigenstates $\ket{V_a}$), we will be able to prove that all eigenvalues $\lambda_a=\bra{V_a}\hat{H}'_\text{EF}\ket{V_a}\geq 0$ are non-negative, hence $\hat{H}'_\text{EF}$ will be positive semi-definite.

We can show $\bra{V}\hat{H}'_\text{EF}\ket{V}\geq 0$ by finding the Cholesky decomposition for each terms in $\hat{H}'_\text{EF}$. A useful trick is to note that $d(d+X_i)=e^{\delta X_i}/(\tanh\delta\sinh\delta)$ given $d=\coth\delta$, so $\hat{W}_{\id}$ can be rewritten as
\eq{\hat{W}_{\id} =\prod_{i=1}^{L}d(d+X_i)=\prod_{i=1}^{L}\frac{e^{\delta X_i}}{\tanh\delta\sinh\delta}=\frac{1}{(\tanh\delta\sinh\delta)^L}\prod_{i=1}^Le^{\delta X_i},}
such that any $\hat{W}_{\id}^{\alpha}$ can be simply calculated,
\eq{\hat{W}_{\id}^{\alpha}=(\tanh\delta\sinh\delta)^{-\alpha L} \prod_{i=1}^Le^{\alpha\delta X_i}.}
With this, and substitute \eqnref{eq:HEF=sumH} in \eqnref{eq:HEF'}, we can show that 
\eqs{\hat{H}'_\text{EF} &= \sum_{i,j}\hat{H}'_{ij},\\
\hat{H}'_{ij}&=\hat{W}_{\id}^{-1/2}\hat{H}_{ij}\hat{W}_{\id}^{1/2}\\
&=\prod_{i=1}^Le^{-\frac{\delta}{2} X_i}\hat{H}_{ij}\prod_{i=1}^Le^{\frac{\delta}{2} X_i}\\
&=e^{-\frac{\delta}{2} (X_i+X_j)}\hat{H}_{ij}e^{\frac{\delta}{2} (X_i+X_j)}\\
&=g_{ij}e^{-\frac{\delta}{2}(X_i+X_j)}\frac{1-{Z}_{i}{Z}_{j}}{2}e^{-\beta_{ij}X_iX_j}e^{-\frac{\delta}{2}(X_i+X_j)}\\
&=g_{ij}e^{-\frac{\delta}{2}(X_i+X_j)}\frac{1-{Z}_{i}{Z}_{j}}{2}e^{-\beta_{ij}X_iX_j}\frac{1-{Z}_{i}{Z}_{j}}{2}e^{-\frac{\delta}{2}(X_i+X_j)}.}
In the last step, we use the fact that $\frac{1-{Z}_{i}{Z}_{j}}{2}$ is a projection operator, so $(\frac{1-{Z}_{i}{Z}_{j}}{2})^2=\frac{1-{Z}_{i}{Z}_{j}}{2}$. Also $\frac{1-{Z}_{i}{Z}_{j}}{2}$ and $e^{-\beta_{ij}X_iX_j}$ commute with each other, so we are free to move $e^{-\beta_{ij}X_iX_j}$ through $\frac{1-{Z}_{i}{Z}_{j}}{2}$. The final form of $\hat{H}'_{ij}$ admits the following Cholesky decomposition explicitly
\eq{\hat{H}'_{ij}=\hat{A}_{ij}^\intercal\hat{A}_{ij},\qquad \hat{A}_{ij}=g_{ij}^{1/2}e^{-\frac{\beta_{ij}}{2}X_iX_j}\frac{1-Z_iZ_j}{2}e^{-\frac{\delta}{2}(X_i+X_j)}.}
For any state $\ket{V}$ in EF Hilbert space, the expectation value $\bra{V}\hat{H}'_{ij}\ket{V}=\bra{V}\hat{A}_{ij}^\intercal\hat{A}_{ij}\ket{V}\geq 0$ is non-negative, therefore $\hat{H}'_{ij}$ is positive semi-definite. In consequence, the transformed EF Hamiltonian $\hat{H}'_\text{EF}=\sum_{i,j}\hat{H}'_{ij}$ is also positive semi-definite, as it is the sum of positive semi-definite terms $\hat{H}'_{ij}$. Recall that the similar transformation does not affect the eigenvalues, so $\hat{H}_\text{EF}=\hat{W}_{\id}^{1/2}\hat{H}'_\text{EF}\hat{W}_{\id}^{-1/2}$ is also positive semi-definite.

We can further show that $\hat{H}_\text{EF}$ always has two zero modes: one is even under $\dsZ_2$ Ising symmetry, and the other is odd. Using the left-null-state property $\bra{\Uparrow}\hat{H}_\text{EF}=0$ given in \eqnref{eq:left null}, it is ensured that $\bra{\Uparrow}$ is an left-eigenstate of $\hat{H}_\text{EF}$ with zero eigenvalue. Since $\hat{H}_\text{EF}$ is $\dsZ_2$ symmetric, the $\dsZ_2$ related state $\bra{\Downarrow}=\bra{\Uparrow}\prod_{i}X_i$ is also a left zero mode. So by explicit construction, we have shown that $\hat{H}_\text{EF}$ has at least two zero eigenvalues. The left zero mode subspace is spanned by $\bra{\Uparrow}$ and $\bra{\Downarrow}$. Using the relation between left- and right-eigenstate $\ket{R_a}\propto(\bra{L_a}\hat{W}_{\id})^\intercal$, the corresponding right zero mode subspace is spanned by $\hat{W}_{\id}\ket{\Uparrow}$ and $\hat{W}_{\id}\ket{\Downarrow}$. 

Since we are most interested about the EF of pure states, we should focus on the $\dsZ_2$ even state in the zero mode subspace. In that regard, the left and right zero modes in the $\dsZ_2$ even sector are given by
\eqs{\bra{L_0}&=\frac{\bra{\Uparrow}+\bra{\Downarrow}}{2},\\
\ket{R_0}&\propto\hat{W}_{\id}\frac{\ket{\Uparrow}+\ket{\Downarrow}}{2}\\
&=\frac{1}{2}\bigg(\prod_{i=1}^{L}d(d+X_i)\bigg)(\ket{\Uparrow}+\ket{\Downarrow})\\
&=\frac{1}{2}d^{3L/2}\bigg(\prod_{i=1}^{L}(e^{\eta}+e^{-\eta}X_i)\bigg)(\ket{\Uparrow}+\ket{\Downarrow})\qquad(\eta\equiv\tfrac{1}{2}\log d)\\
&=\frac{1}{2}d^{3L/2}\sum_{\vect{\sigma}}\bigg(\prod_{i=1}^Le^{\eta\sigma_i}\ket{\vect{\sigma}}+\prod_{i=1}^Le^{-\eta\sigma_i}\ket{\vect{\sigma}}\bigg)\\
&=d^{3L/2}\sum_{\vect{\sigma}}\cosh\Big(\eta\sum_{i=1}^L\sigma_i\Big)\ket{\vect{\sigma}}.}
The normalization of $\ket{R_0}$ is determined by the condition $\braket{L_0}{R_0}=1$, such that
\eq{\ket{R_0}=\sum_{\vect{\sigma}}\frac{\cosh\big(\eta\sum_{i=1}^L\sigma_i\big)}{\cosh(\eta L)}\ket{\vect{\sigma}}=\ket{W_\text{Page}}.}
In summary, we have shown that in the $\dsZ_2$ even sector, the EF Hamiltonian $\hat{H}_\text{EF}$ has at least one zero eigenvalue $\lambda_0=0$, whose left- and right-eigenstates are given by
\eq{\bra{L_0}=\frac{\bra{\Uparrow}+\bra{\Downarrow}}{2},\qquad \ket{R_0}=\ket{W_\text{Page}},}
as claimed in \eqnref{eq:L0R0}.

\section{Derivation of the Dispersion Relation for two-domain-wall Ansatz  }\label{app:dispersiontwodomain}

Here, we show the derivation of \eqnref{eq:dispersion}. Our goal here is to obtain the analytical expression of excited state energy, namely dispersion relation, $\omega(k).$ Note that the left and right eigenstates are \textit{not} simply each other's conjugate transpose due to the non-hermitian nature of EF Hamiltonian (\eqnref{eq:HEF form}). For simplicity, we focus on the excitation of left eigenstates and construct the corresponding right eigenstate with $\ket{R} = (\bra{L}\hat{W}_\id)^\intercal$. From the discussion in \secref{sec:universal}, the  universal left ground state for any parameters $g_{ij},\beta$ is the linear combination of all spin-up and spin-down state,
\eq{\bra{L_{0}}=\frac{\bra{\Uparrow}+\bra{\Downarrow}}{2}.} 
Based on our ED result in \figref{fig:weightdis}, the low energy left excited state mainly consists of two-domain-wall states. The generic form of two-domain-wall state can be expressed as
\eq{\bra{k}\equiv C_{k}\sum_{n,a}\bra{k_{n},a} \equiv C_{k}\sum_{n,a} e^{ikn}\phi^{*}(a)\bra{\Uparrow}\prod_{i=n}^{n+a}X_{i}}
where $C_{k}$ is the normalization constant and $\bra{k_{n},a}$ represents the two-domain-wall state ranging from $n$ to $n+a$.

First, we start from deriving the normalization constant. 
\eq{\label{app:normalization}\braket{k}{k} = |C_{k}|^{2}\sum_{n,m,a,b}e^{ik(n-m)}\phi^{*}(a)\phi(b)\bra{\Uparrow}\prod^{n+a}_{i=n}X_{i}\hat{W}_\id\prod^{m+b}_{i=m}X_{i}\ket{\Uparrow}=1.} Next step is to evaluate $\bra{\Uparrow}\prod^{n+a}_{i=n}X_{i}\hat{W}_\id\prod^{m+b}_{j=m}X_{j}\ket{\Uparrow}.$ The physical meaning is the transition amplitude between two Bethe string states with evolution as $\hat{W}_\id$. There are two possibilities for each site. When both Bethe strings have/do not have excitation at site $i$, the answer would be $\bra{\Uparrow}(\hat{W}_\id)_{i}\ket{\Uparrow}=d^{2}$. When either Bethe string has excitation on site $i$, the result becomes $\bra{\Uparrow}X_{i}(\hat{W}_\id)_{i}\ket{\Uparrow}=d$. To evaluate this quantity, we perform perturbative expansion as $1/d$ series. To obtain analytical expression of $|C_{k}|^{2}$, we also approximate $\phi(a)$ as plane wave $\sim e^{-ika/2}$. Physical intuition is that we assume these domain walls have little interaction with each other and thus they can move through each other almost freely. Consequently, plane wave solution is assumed and $a/2$ represents the center location of domain wall.  Let's start evaluating the normalization constant up to the order of $1/d^{2}$,
\eqs{
	& \bra{\Uparrow}\prod^{n+a}_{i=n}X_{i}\hat{W}_\id\prod^{m+b}_{j=m}X_{j}\ket{\Uparrow} = \delta_{n,m}\delta_{a,b}d^{2N} + (\delta_{n,m}\delta_{a,b+1}+\delta_{n,m}\delta_{a,b-1}+\delta_{n,m+1}\delta_{a,b-1}+\delta_{n,m-1}\delta_{a,b+1} )d^{2N-1}\\
	&+(\delta_{n,m}\delta_{a,b+2}+\delta_{n,m}\delta_{a,b-2}+\delta_{n,m-2}\delta_{a,b+2}+\delta_{n,m+2}\delta_{a,b-2}+\delta_{n,m+1}\delta_{a,b}+\delta_{n,m+1}\delta_{a,b-2}+\delta_{n,m-1}\delta_{a,b}+\delta_{n,m-1}\delta_{a,b+2})d^{2N-2}\\
	&+\mathcal{O}(d^{2N-3})\\}

For $b=0,1$, we would have different terms,

\eqs{
 \bra{\Uparrow}\prod^{n+a}_{i=n}X_{i}\hat{W}_\id X_{m}\ket{\Uparrow} &= \delta_{n,m}\delta_{a,0}d^{2N} + (\delta_{n,m}\delta_{a,1}+\delta_{n,m-1}\delta_{a,1} )d^{2N-1}\\
	&+(\delta_{n,m}\delta_{a,2}+\delta_{n,m-2}\delta_{a,2}+\delta_{n,m+1}\delta_{a,0}+\delta_{n,m-1}\delta_{a,0}+\delta_{n,m-1}\delta_{a,2})d^{2N-2}+\mathcal{O}(d^{2N-3})\\}

\eqs{
	 \bra{\Uparrow}\prod^{n+a}_{i=n}X_{i}\hat{W}_\id X_{m}X_{m+1}\ket{\Uparrow} &= \delta_{n,m}\delta_{a,1}d^{2N} + (\delta_{n,m}\delta_{a,2}+\delta_{n,m}\delta_{a,0}+\delta_{n,m+1}\delta_{a,0}+\delta_{n,m-1}\delta_{a,2} )d^{2N-1}\\
	&+(\delta_{n,m}\delta_{a,3}+\delta_{n,m-2}\delta_{a,3}+\delta_{n,m+1}\delta_{a,1}+\delta_{n,m-1}\delta_{a,1}+\delta_{n,m-1}\delta_{a,3})d^{2N-2}+\mathcal{O}(d^{2N-3})\\}

Put them back to \eqnref{app:normalization} and we can obtain
\eqs{&|C_{k}|^{2}d^{2N}N^{2}\{\frac{N-2}{N}[1+\frac{4}{d}\cos\frac{k}{2} + \frac{1}{d^{2}}( 2 + 6\cos k)]+\frac{1}{N}[1+\frac{2}{d}\cos\frac{k}{2}+\frac{1}{d^{2}}(1+4\cos{k})]\\
&+\frac{1}{N}[1+\frac{4}{d}\cos\frac{k}{2}+\frac{1}{d^{2}}(1+4\cos k)]+\scO(d^{-3})\}=1\\}

To simplify the whole calculation, the thermodynamics limit is taken $N\rightarrow \infty$. The main effect of thermodynamics limit is that the contribution from short two domain wall states (e.g. single-site or two-site excitations) is fully suppressed. Thus, up to $\scO(\frac{1}{d^{2}})$, we have
\eq{|C_{k}|^{2}d^{2N}N^{2}[1+\frac{4}{d}\cos\frac{k}{2} + \frac{1}{d^{2}}( 2 + 6\cos k)]=1}

Now, we are ready to evaluate the energy expectation value of our two-domain-wall state, $\bra{k}H_{EF}\ket{k}$. For simplicity, we assume $g_{ij}= 1,\beta_{ij}=\beta$ and reorganize the EF Hamiltonian 
\eqs{\label{HEFexpression}\hat{H}_\text{EF}& = \sum_{i,j}\frac{1-{Z}_{i}{Z}_{j}}{2}e^{-\beta X_iX_j-\delta(X_i+X_j)}\\
&=\sum_{i,j}\frac{1-Z_{i}Z_{j}}{2}\frac{d^{2}}{d^{2}-1}[\cosh\beta-\sinh\beta X_{i}X_{j}][1-\frac{1}{d}(X_{i}+X_{j})+\frac{1}{d^{2}}X_{i}X_{j}]
\\
&=\sum_{i,j}\frac{1-Z_{i}Z_{j}}{2}\frac{d^{2}}{d^{2}-1}[\cosh\beta-\frac{\sinh\beta}{d^{2}}-\frac{1}{d}(\cosh\beta-\sinh\beta)(X_{i}+X_{j})+(\frac{\cosh\beta}{d^{2}}-\sinh\beta)X_{i}X_{j}]\\
&\equiv \sum_{i} \frac{1-Z_{i}Z_{i+1}}{2}\big[ a(\beta,d) + b(\beta,d)(X_{i}+X_{i+1}) + c(\beta,d)X_{i}X_{i+1} \big] \\}
where \eqs{&a(\beta,d) = \frac{d^{2}}{d^{2}-1}(\cosh\beta -\frac{\sinh\beta}{d^{2}}) =\cosh\beta +\frac{\cosh\beta-\sinh\beta}{d^{2}}+\scO(\frac{1}{d^{4}})\\
	&b(\beta,d)=-\frac{d}{d^{2}-1}(\cosh\beta-\sinh\beta)=-\frac{1}{d}(\cosh\beta-\sinh\beta)+\scO(\frac{1}{d^{3}})\\ &c(\beta,d)=\frac{d^{2}}{d^{2}-1}(\frac{\cosh\beta}{d^{2}}-\sinh\beta)=\frac{\cosh\beta-\sinh\beta}{d^{2}}-\sinh\beta+\scO(\frac{1}{d^{4}}).\\}
The first term in $\bra{k}H_{EF}\ket{k}$ is 
\eqs{\label{app:first}&|C_{k}|^{2}a(\beta,d)\sum_{n,m,a,b}e^{ik(n-m)}e^{ik(a-b)/2}\bra{\Uparrow}\prod^{n+a}_{i=n}X_{i}\sum_{l}\frac{1-Z_{l}Z_{l+1}}{2}\hat{W}_\id\prod^{m+b}_{j=m}X_{j}\ket{\Uparrow}\\
	&= 2a(\beta,d)
	|C_{k}|^{2}\sum_{n,m,a,b}e^{ik(n-m)}e^{ik(a-b)/2}\bra{\Uparrow}\prod^{n+a}_{i=n}X_{i}\hat{W}_\id\prod^{m+b}_{j=m}X_{j}\ket{\Uparrow}=2a(\beta,d)\\}
As for the second term, since $b(\beta,d)$ already contains $1/d$ power, we just compute the terms up to $1/d$ order and the result is 

\eqs{&|C_{k}|^{2}b(\beta,d)\sum_{n,m,a,b}e^{ik(n-m)}e^{ik(a-b)/2}\bra{\Uparrow}\prod^{n+a}_{i=n}X_{i}\sum_{l}\frac{1-Z_{l}Z_{l+1}}{2}(X_{l}+X_{l+1})\hat{W}_\id\prod^{m+b}_{j=m}X_{j}\ket{\Uparrow}\\
	&=|C_{k}|^{2}b(\beta,d)\sum_{n,m,a,b}e^{ik(n-m)}e^{ik(a-b)/2}\\
	&\times[h(n-1,m,a+1,b)+h(n+1,m,a-1,b)+h(n,m,a+1,b)+h(n,m,a-1,b)]\\} 
where 
\eq{h(n,m,a,b)=\bra{\Uparrow}\prod^{n+a}_{i=n}X_{i}\hat{W}_\id\prod^{m+b}_{j=m}X_{j}\ket{\Uparrow}.}
For each $h(n,m,a,b)$, the boundary terms would have different result. For example, the results of $h(n-1,m,a+1,b)$ are as follows
\eqs{
&h(n-1,m,a+1,0) =  (\delta_{n,m+1}\delta_{a,0}+\delta_{n,m}\delta_{a,0} )d^{2N-1}+\mathcal{O}(d^{2N-2})\\
&h(n-1,m,a+1,1) = \delta_{n,m+1}\delta_{a,0}d^{2N} + (\delta_{n,m+1}\delta_{a,1}+\delta_{n,m}\delta_{a,1} )d^{2N-1}+\mathcal{O}(d^{2N-2})\\
&h(n-1,m,a+1,2) = \delta_{n,m+1}\delta_{a,1}d^{2N} + (\delta_{n,m+1}\delta_{a,2}+\delta_{n,m+1}\delta_{a,0}+\delta_{n,m+2}\delta_{a,0}+\delta_{n,m}\delta_{a,2} )d^{2N-1}+\mathcal{O}(d^{2N-2})\\
&h(n-1,m,a+1,b\neq 0,1,2) = \delta_{n,m+1}\delta_{a,b-1}d^{2N} + (\delta_{n,m+1}\delta_{a,b}+\delta_{n,m+1}\delta_{a,b-2}+\delta_{n,m+2}\delta_{a,b-2}+\delta_{n,m}\delta_{a,b} )d^{2N-1}+\mathcal{O}(d^{2N-2}).\\}
Since the thermodynamics limit would be taken ($N\rightarrow\infty$), the "boundary effect" from short two-domain-wall states would be suppressed. Consequently, we only keep the last term in our calculation. Combine these four terms and compute the sum with thermodynamic limit,
\eq{\label{app:second}|C_{k}|^{2}b(\beta,d)d^{2N}N^{2}[4\cos \frac{k}{2}+\frac{8}{d}(1+\cos k)]+\scO(\frac{1}{d^{2}})=b(\beta,d)[4\cos \frac{k}{2}-\frac{16}{d}\cos^{2}\frac{k}{2}+\frac{4}{d}(2+2\cos k)]+\mathcal{O}(\frac{1}{d^{2}})}

For the third term, the following quantity is computed
\eq{|C_{k}|^{2}c(\beta,d)\sum_{n,m,a,b}e^{ik(n-m)}e^{ik(a-b)/2}\bra{\Uparrow}\prod^{n+a}_{i=n}X_{i}\sum_{l}\frac{1-Z_{l}Z_{l+1}}{2}X_{l}X_{l+1}\hat{W}_{I}\prod^{m+b}_{j=m}X_{j}\ket{\Uparrow}.\\}

The EF Hamiltonian would give extra $X_{i}X_{i+1}$ term. In most two-domain-wall states (length $>1$), the two-domain-wall structure would destroyed. However, for single site excitation, this  $X_{i}X_{i+1}$ term would only shift the position of excitation with one site. Due to the suppression of thermodynamic limit, we would also drop this term. Eventually, we can have

\eqs{\label{app:third}
	&c(\beta,d)\big[\frac{4}{d}\cos\frac{k}{2}+\frac{1}{d^{2}}(8+8\cos k)\big](1-\frac{4}{d}\cos \frac{k}{2})+\mathcal{O}(\frac{1}{d^{3}})=c(\beta,d)\big[\frac{4}{d}\cos\frac{k}{2}+\frac{1}{d^{2}}(8+8\cos k)-\frac{16}{d^{2}}\cos^{2}\frac{k}{2} \big]+\mathcal{O}(\frac{1}{d^{3}}).\\}
Combining \eqnref{app:first}, \eqnref{app:second} and \eqnref{app:third} and keeping terms up to $\scO(\frac{1}{d^{3}})$, $\bra{k}H_{EF}\ket{k}$ would be 
\eqs{\bra{k}H_{EF}\ket{k} &= 2[\cosh \beta + \frac{\cosh \beta -\sinh\beta}{d^{2}}]-\frac{1}{d}(\cosh\beta-\sinh\beta)[4\cos \frac{k}{2}-\frac{16}{d}\cos^{2}\frac{k}{2}+\frac{4}{d}(2+2\cos k)]\\
&-\sinh\beta\big[\frac{4}{d}\cos\frac{k}{2}+\frac{1}{d^{2}}(8+8\cos k)-\frac{16}{d^{2}}\cos^{2}\frac{k}{2} \big]+\scO(\frac{1}{d^{3}})\\}

\section{Derivation of the Dispersion Relation for Single-Site Excitation ansatz }\label{app:dispersionSMA}

This appendix is similar with the calculation in \appref{app:dispersiontwodomain}. The only difference is the ansatz state we use. The single-site excitation ansatz is defined as
\eq{\bra{k} =C_{k}\bra{\Uparrow}\sum_{n}X_{n}e^{ikn}, \ket{k} = \hat{W}_\id \sum_{n}X_{n}e^{-ikn}\ket{\Uparrow}.}

First, we start from the normalization condition $\braket{k}{k}=1,$
\eqs{\braket{k}{k} = 1 &= C_{k}\bra{\Uparrow}\sum_{n,m}e^{ik(n-m)}X_{n}\hat{W}_\id X_{m}\ket{\Uparrow}=C_{k}[Nd^{2(N-1)}(d^{2}-1)+N^{2}\delta_{k,0}d^{2(N-1)}].\\}

Following the expression in \eqnref{HEFexpression},
\eqs{\label{HEFexpression}\hat{H}_\text{EF}=
	 \sum_{i} \frac{1-Z_{i}Z_{i+1}}{2}\big[ a(\beta,d) + b(\beta,d)(X_{i}+X_{i+1}) + c(\beta,d)X_{i}X_{i+1} \big] \\}
where \eq{a(\beta,d) = \frac{d^{2}}{d^{2}-1}(\cosh\beta -\frac{\sinh\beta}{d^{2}}), b(\beta,d)=-\frac{d}{d^{2}-1}(\cosh\beta-\sinh\beta),c(\beta,d)=\frac{d^{2}}{d^{2}-1}(\frac{\cosh\beta}{d^{2}}-\sinh\beta).}

The first term is
\eqs{a(\beta,d)\bra{k}\sum_{l}\frac{1-Z_{l}Z_{l+1}}{2}W_{I}\ket{k} &= a(\beta,d)C_{k}\sum_{m,n}e^{ik(n-m)}\bra{n}\sum_{l}\frac{1-Z_{l}Z_{l+1}}{2}\hat{W}_\id\ket{m}\\
	&=a(\beta,d)C_{k}\sum_{m,n}e^{ik(n-m)}\bra{n}\sum_{l}(\delta_{l,n-1}+\delta_{l,n})\hat{W}_\id\ket{m}\\
	&=2a(\beta,d)C_{k}\sum_{m,n}e^{ik(n-m)}\bra{n}\hat{W}_\id\ket{m}\\
	&=2a(\beta,d)C_{k}\sum_{m,n}e^{ik(n-m)}[\delta_{n,m}d^{2N}+(1-\delta_{n,m})d^{2(N-1)}]\\
	&=2a(\beta,d)C_{k}[Nd^{2(N-1)}(d^{2}-1)+N^{2}\delta_{k,0}d^{2(N-1)}]=2a(\beta,d).\\}
The second term is 

\eqs{&b(\beta,d)\bra{k}\sum_{l}\frac{1-Z_{l}Z_{l+1}}{2}(X_{l}+X_{l+1})W_{I}\ket{k} =b(\beta,d) C_{k}\sum_{m,n}e^{ik(n-m)}\bra{n}\sum_{l}\frac{1-Z_{l}Z_{l+1}(X_{l}+X_{l+1})}{2}\hat{W}_\id\ket{m}\\
	&=b(\beta,d) C_{k}\sum_{m,n}e^{ik(n-m)}\bra{n}(X_{n-1}+2X_{n}+X_{n+1})\hat{W}_\id\ket{m}\\
	&=b(\beta,d) C_{k}\sum_{m,n}e^{ik(n-m)}(\bra{n,n-1}W_{I}\ket{m}+\bra{n,n+1}\hat{W}_\id\ket{m}+2\bra{\Uparrow}\hat{W}_\id\ket{m})\\
	&=b(\beta,d) C_{k}\sum_{m,n}e^{ik(n-m)}[d^{2N-3}(1-\delta_{m,n})(1-\delta_{m,n-1})+d^{2N-3}(1-\delta_{m,n})(1-\delta_{m,n+1})\\
	&+d^{2N-1}(\delta_{m,n}+\delta_{m,n-1})+d^{2N-1}(\delta_{m,n}+\delta_{m,n+1})+2d^{2N-1}]\\
	&=b(\beta,d) C_{k}\sum_{m,n}e^{ik(n-m)}[2d^{2N-3}+2d^{2N-1}+(d^{2N-1}-d^{2N-3})(\delta_{m,n-1}+2\delta_{m,n}+\delta_{m,n+1})]\\
	&=b(\beta,d) C_{k}[N^{2}\delta_{k,0}2d^{2N-3}(d^{2}+1)+2d^{2N-3}(d^{2}-1)N(1+\cos k)].\\}

For single-site excitation, we focus on the region which $k\neq 0$. The result would be \eq{2b(\beta,d)\frac{1+\cos k}{d}\times C_{k}[Nd^{2(N-2)}(d^{2}-1)]=\frac{2b(\beta,d)}{d}(1+\cos k).}

The third term is 
\eqs{c(\beta,d)\bra{k}\sum_{l}\frac{1-Z_{l}Z_{l+1}}{2}X_{l}X_{l+1}\hat{W}_\id\ket{k} &=c(\beta,d)C_{k}\sum_{m,n}e^{ik(n-m)}\bra{n}\sum_{l}\frac{(1-Z_{l}Z_{l+1})X_{l}X_{l+1}}{2}\hat{W}_\id\ket{m}\\
	&=c(\beta,d)C_{k}\sum_{m,n}e^{ik(n-m)}\bra{n}\sum_{l}(\delta_{l,n-1}+\delta_{l,n})X_{l}X_{l+1}\hat{W}_\id\ket{m}\\
	&=c(\beta,d)C_{k}\sum_{m,n}e^{ik(n-m)}(\bra{n-1}\hat{W}_{\id}\ket{m}+\bra{n+1}\hat{W}_\id\ket{m})\\
	&=c(\beta,d)C_{k}\sum_{m,n}e^{ik(n-m)}(2d^{2(N-1)}+(\delta_{m,n-1}+\delta_{m,n+1})(d^{2N}-d^{2(N-1)})\\
	&=2c(\beta,d)C_{k}[Nd^{2(N-1)}(d^{2}-1)\cos k+N^{2}d^{2(N-1)}\delta_{k,0}]=2c(\beta,d)\cos k\\}

The overall result would be 
\eqs{&\bra{k}\hat{H}_{EF}\ket{k}= 2a(\beta,d) +\frac{2b(\beta,d)}{d}(1+\cos k) + 2c(\beta,d)\cos k=2a(\beta,d)+\frac{2b(\beta,d)}{d}+\cos k[2c(\beta,d)+\frac{2b(\beta,d)}{d}]\\
&= \frac{2d^{2}}{d^{2}-1}(\cosh\beta -\frac{\sinh\beta}{d^{2}})-\frac{2}{d^{2}-1}(\cosh\beta-\sinh\beta)+\cos k [\frac{2d^{2}}{d^{2}-1}(\frac{\cosh\beta}{d^{2}}-\sinh\beta)-\frac{2}{d^{2}-1}(\cosh\beta-\sinh\beta)]\\}
\section{Diagrammatic Expansion of Entanglement Feature Hamiltonian}\label{app:epsilon expansion}

In this appendix, we derive the EF Hamiltonian for the locally scrambled Hamiltonian dynamics. We start from the definition of the EF for $e^{-\ii \epsilon H}$ following \eqnref{eq:WU},
\eq{W_{e^{-\ii \epsilon H}}[\vect{\sigma},\vect{\tau}]=\Tr\big(\scX_\vect{\sigma}(e^{-\ii\epsilon H})^{\otimes 2}\scX_\vect{\tau}(e^{\ii\epsilon H})^{\otimes 2}\big)=\Tr\big(\scX_\vect{\sigma}e^{-\ii\epsilon \dsH}\scX_\vect{\tau}e^{\ii\epsilon\dsH}\big),}
where we have introduced $\dsH=H\otimes\id+\id\otimes H$ to denote the double Hamiltonian. Given the locality of $H=\sum_x H_x$, the double Hamiltonian $\dsH$ is also a sum of local terms $\dsH=\sum_{x}\dsH_{x}$ with $\dsH_{x}=H_{x}\otimes\id+\id\otimes H_{x}$ being the doubled version of $H_{x}$. Expand around $\epsilon\to 0$ to the order of $\epsilon^2$, we obtain
\eqs{\label{eq:WU GUE expand}
W_{e^{-\ii \epsilon H}}[\vect{\sigma},\vect{\tau}]&=\Tr(\scX_\vect{\sigma}\scX_\vect{\tau})-\epsilon^2\Tr(\scX_\vect{\sigma}\scX_\vect{\tau}\dsH^2-\scX_\vect{\sigma}\dsH\scX_\vect{\tau}\dsH)+\scO(\epsilon^4),\\
&=W_{\id}[\vect{\sigma},\vect{\tau}]-\epsilon^2\sum_{x,x'}\Tr(\scX_\vect{\sigma}\scX_\vect{\tau}\dsH_{x}\dsH_{x'}-\scX_\vect{\sigma}\dsH_{x}\scX_\vect{\tau}\dsH_{x'})+\scO(\epsilon^4),\\
&=W_{\id}[\vect{\sigma},\vect{\tau}]-\epsilon^2\sum_{x}\Tr(\scX_\vect{\sigma}\scX_\vect{\tau}\dsH_{x}^2-\scX_\vect{\sigma}\dsH_{x}\scX_\vect{\tau}\dsH_{x})+\scO(\epsilon^4)}
where the first order term in $\epsilon$ vanishes by the cyclic identity of trace, confirming the argument in \secref{sec:Hamiltonian} that $W_{U(\epsilon)}$ will be even in $\epsilon$. The last equality in \eqnref{eq:WU GUE expand} relies on the fact that $\Tr(\scX_\vect{\sigma}\scX_\vect{\tau}\dsH_{x}\dsH_{x'}-\scX_\vect{\sigma}\dsH_{x}\scX_\vect{\tau}\dsH_{x'})=0$ as long as $x\neq x'$. To prove this, we first consider the case when $x=\langle ij\rangle$ and $x'=\langle kl\rangle$ do not overlap,
\eqs{\Tr\scX_\vect{\sigma}\dsH_{\langle ij\rangle}\scX_\vect{\tau}\dsH_{\langle kl\rangle}&=\Tr\scX_\vect{\sigma}\dsH_{\langle ij\rangle}\Big(\scX_{\tau_i}\scX_{\tau_j}\scX_{\tau_k}\scX_{\tau_l}\bigotimes_{m\neq i,j,k,l}\scX_{\tau_m}\Big)\dsH_{\langle kl\rangle}\\
&=\Tr\scX_\vect{\sigma}\Big(\scX_{\tau_k}\scX_{\tau_l}\bigotimes_{m\neq i,j,k,l}\scX_{\tau_m}\Big)\dsH_{\langle ij\rangle}\dsH_{\langle kl\rangle}(\scX_{\tau_i}\scX_{\tau_j})\\
&=\Tr(\scX_{\tau_i}\scX_{\tau_j})\scX_\vect{\sigma}\Big(\scX_{\tau_k}\scX_{\tau_l}\bigotimes_{m\neq i,j,k,l}\scX_{\tau_m}\Big)\dsH_{\langle ij\rangle}\dsH_{\langle kl\rangle}\\
&=\Tr\scX_\vect{\sigma}\Big(\scX_{\tau_i}\scX_{\tau_j}\scX_{\tau_k}\scX_{\tau_l}\bigotimes_{m\neq i,j,k,l}\scX_{\tau_m}\Big)\dsH_{\langle ij\rangle}\dsH_{\langle kl\rangle}\\
&=\Tr\scX_\vect{\sigma}\scX_\vect{\tau}\dsH_{\langle ij\rangle}\dsH_{\langle kl\rangle},}
where we have used the fact that $[\dsH_{\langle ij\rangle},\scX_{\tau_k}\scX_{\tau_l}]=0$ for $i,j\neq k,l$, and $[\scX_{\sigma_i},\scX_{\tau_j}]=0$ for any $i,j$ as the $S_2$ group is Abelian. We then consider the case when $x=\langle ij\rangle$ and $x'=\langle jk\rangle$ overlaps on a single site $j$,
\eqs{\Tr\scX_\vect{\sigma}\dsH_{\langle ij\rangle}\scX_\vect{\tau}\dsH_{\langle jk\rangle}&=\Tr\scX_\vect{\sigma}\dsH_{\langle ij\rangle}\Big(\scX_{\tau_i}\scX_{\tau_j}\scX_{\tau_k}\bigotimes_{m\neq i,j,k}\scX_{\tau_m}\Big)\dsH_{\langle jk\rangle}\\
&=\Tr\scX_\vect{\sigma}\Big(\scX_{\tau_k}\bigotimes_{m\neq i,j,k}\scX_{\tau_m}\Big)\dsH_{\langle ij\rangle}\scX_{\tau_j}\dsH_{\langle jk\rangle}\scX_{\tau_i}.\\}
At this point, it seems that $\scX_{\tau_j}$ is caught between $\dsH_{\langle ij\rangle}$ and $\dsH_{\langle jk \rangle}$. The solution is to make use of the property that $\dsH_{\langle jk \rangle}=\scX_{\alpha_j}^{-1}\scX_{\alpha_k}^{-1}\dsH_{\langle jk \rangle}\scX_{\alpha_k}\scX_{\alpha_j}$ for any $\alpha_j=\alpha_k\in S_2$, due to the permutation symmetry to exchange the two replicas of the double Hamiltonian. Now we choose $\alpha_j=\alpha_k=\tau_j$, such that $\scX_{\tau_j}\scX_{\alpha_j}^{-1}=1$, then
\eqs{\Tr\scX_\vect{\sigma}\dsH_{\langle ij\rangle}\scX_\vect{\tau}\dsH_{\langle jk\rangle}&=\Tr\scX_\vect{\sigma}\Big(\scX_{\tau_k}\bigotimes_{m\neq i,j,k}\scX_{\tau_m}\Big)\dsH_{\langle ij\rangle}\scX_{\tau_j}\scX_{\alpha_j}^{-1}\scX_{\alpha_k}^{-1}\dsH_{\langle jk \rangle}\scX_{\alpha_k}\scX_{\alpha_j}\scX_{\tau_i}\\
&=\Tr\scX_\vect{\sigma}\Big(\scX_{\tau_k}\bigotimes_{m\neq i,j,k}\scX_{\tau_m}\Big)\dsH_{\langle ij\rangle}\scX_{\alpha_k}^{-1}\dsH_{\langle jk \rangle}\scX_{\alpha_k}\scX_{\alpha_j}\scX_{\tau_i}\\
&=\Tr\scX_\vect{\sigma}\Big(\scX_{\tau_i}\scX_{\alpha_j}\scX_{\alpha_k}\scX_{\tau_k}\scX_{\alpha_k}^{-1}\bigotimes_{m\neq i,j,k}\scX_{\tau_m}\Big)\dsH_{\langle ij\rangle}\dsH_{\langle jk \rangle}\\
&=\Tr\scX_\vect{\sigma}\Big(\scX_{\tau_i}\scX_{\tau_j}\scX_{\tau_k}\bigotimes_{m\neq i,j,k}\scX_{\tau_m}\Big)\dsH_{\langle ij\rangle}\dsH_{\langle jk \rangle}\\
&=\Tr\scX_\vect{\sigma}\scX_{\vect{\tau}}\dsH_{\langle ij\rangle}\dsH_{\langle jk \rangle}.\\}
Hence we have shown that $\Tr\scX_\vect{\sigma}\dsH_{\langle ij\rangle}\scX_\vect{\tau}\dsH_{\langle kl\rangle}=\Tr\scX_\vect{\sigma}\scX_\vect{\tau}\dsH_{\langle ij\rangle}\dsH_{\langle kl\rangle}$ as long as $\langle ij\rangle\neq\langle kl\rangle$, meaning that $\Tr(\scX_\vect{\sigma}\scX_\vect{\tau}\dsH_{x}\dsH_{x'}-\scX_\vect{\sigma}\dsH_{x}\scX_\vect{\tau}\dsH_{x'})=\delta_{xx'}\Tr(\scX_\vect{\sigma}\scX_\vect{\tau}\dsH_{x}^2-\scX_\vect{\sigma}\dsH_{x}\scX_\vect{\tau}\dsH_{x})$. Thus the derivation of \eqnref{eq:WU GUE expand} is justified.

If we consider the difference between $W_{e^{-\ii \epsilon H}}$ and $W_\id$, denoted as $\delta W$,
\eq{\delta W[\vect{\sigma},\vect{\tau}]\equiv W_{e^{-\ii \epsilon H}}[\vect{\sigma},\vect{\tau}]-W_\id[\vect{\sigma},\vect{\tau}]=-\epsilon^2\sum_{x}\Tr(\scX_\vect{\sigma}\scX_\vect{\tau}\dsH_{x}^2-\scX_\vect{\sigma}\dsH_{x}\scX_\vect{\tau}\dsH_{x})+\scO(\epsilon^4).}
$\delta W[\vect{\sigma},\vect{\tau}]=\sum_x\delta W_x[\vect{\sigma},\vect{\tau}]W_{\id_{\bar{x}}}[\vect{\sigma},\vect{\tau}]$ can be expressed as a sum of terms on each bond $x$ (at least to the order of $\epsilon^2$). To carry out the $\epsilon$ expansion more systematically, we choose to focus on a single bond, and define the EF difference
\eq{\delta W_x[\vect{\sigma},\vect{\tau}]\equiv W_{e^{-\ii \epsilon H_x}}[\vect{\sigma},\vect{\tau}]-W_{\id_x}[\vect{\sigma},\vect{\tau}]=\Tr\big(\scX_\vect{\sigma}e^{-\ii\epsilon \dsH_x}\scX_\vect{\tau}e^{\ii\epsilon\dsH_x}\big)-\Tr\big(\scX_\vect{\sigma}\scX_\vect{\tau}\big),}
where $\vect{\sigma}=(\sigma_i,\sigma_j)$ is restricted to the two sites $i, j$ connected by the bond $x$ and similarly for $\vect{\tau}$. $\delta W_x[\vect{\sigma},\vect{\tau}]=0$ vanishes as long as $\sigma_i=\sigma_j$ or $\tau_i=\tau_j$, because in that case, $\scX_\vect{\sigma}$ or $\scX_\vect{\tau}$ will commute with $\dsH_x$ and hence the two traces will cancel with each other. Therefore there are only two independent non-trivial components of $\delta W[\vect{\sigma},\vect{\tau}]$, which we denote as $u$ and $v$:
\eqs{u&=\delta W[\dia{II}{11}{-2}_i\dia{X}{11}{-2}_j,\dia{II}{11}{-2}_i\dia{X}{11}{-2}_j]=\delta W[\dia{X}{11}{-2}_i\dia{II}{11}{-2}_j,\dia{X}{11}{-2}_i\dia{II}{11}{-2}_j],\\
v&=\delta W[\dia{II}{11}{-2}_i\dia{X}{11}{-2}_j,\dia{X}{11}{-2}_i\dia{II}{11}{-2}_j]=\delta W[\dia{X}{11}{-2}_i\dia{II}{11}{-2}_j,\dia{II}{11}{-2}_i\dia{X}{11}{-2}_j].}
So we only need to focus on these terms and perform the $\epsilon$ expansion following the definition
\eqs{\delta W_x[\vect{\sigma},\vect{\tau}]&=\Tr\big(\scX_\vect{\sigma}e^{-\ii\epsilon H_x}\otimes e^{-\ii\epsilon H_x}\scX_\vect{\tau}e^{\ii\epsilon H_x}\otimes e^{\ii\epsilon H_x}\big)-\Tr\big(\scX_\vect{\sigma}\scX_\vect{\tau}\big)\\
&=\sum_{k=1}^\infty\epsilon^{2k}\sum_{n_1+n_2+n_3+n_4=2k}\frac{\ii^{-n_1-n_2+n_3+n_4}}{n_1!n_2!n_3!n_4!}\Tr\big(\scX_\vect{\sigma}H_x^{n_1}\otimes H_x^{n_2}\scX_\vect{\tau} H_x^{n_3}\otimes H_x^{n_4}\big).}
The $\epsilon$ odd power terms must vanish because $\delta W_x[\vect{\sigma},\vect{\tau}]$ must be real. To the $\epsilon^4$ order, we found
\eqs{u&=\epsilon^2\Big(-2\dia{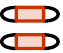}{11}{-2}+\big(2d\dia{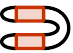}{11}{-2}+2d\dia{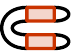}{11}{-2}\big)-\tfrac{1}{2!}\big(4d^2\dia{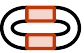}{11}{-2}\big)\Big)\\
&\phantom{=\;}+\epsilon^4\bigg(\dia{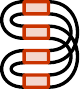}{20}{-8}+\tfrac{1}{2!}\Big(4\dia{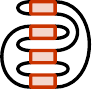}{20}{-8}-4\dia{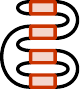}{20}{-8}-4\dia{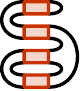}{20}{-8}\Big)+\tfrac{1}{(2!)^2}\Big(2\dia{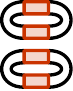}{20}{-8}+2d\dia{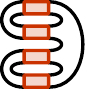}{20}{-8}+2d\dia{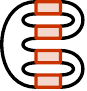}{20}{-8}\Big)\\
&\phantom{=+\epsilon^4\bigg(}+\tfrac{1}{3!}\Big(4\dia{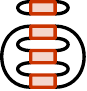}{20}{-8}-4d\dia{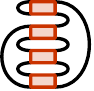}{20}{-8}-4d\dia{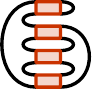}{20}{-8}\Big)+\tfrac{1}{4!}\Big(4d^2\dia{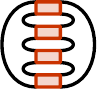}{20}{-8}\Big)\bigg)+\scO(\epsilon^6)\\
&=-\epsilon^2(2\R{(1)(2)}{(1)(2)}-2d\big(\R{(1)(2)}{(12)}+\R{(12)}{(1)(2)}\big)+2d^2\R{(12)}{(12)})\\
&\phantom{=\;}+\epsilon^4(\R{(12)(34)}{(13)(24)}+2\big(\R{(123)(4)}{(124)(3)}-\R{(123)(4)}{(12)(34)}-\R{(12)(34)}{(123)(4)}\big)+\tfrac{1}{2}\big(\R{(12)(34)}{(12)(34)}+d\R{(12)(34)}{(1234)}+d\R{(1234)}{(12)(34)}\big)\\
&\phantom{=+\epsilon^4(}+\tfrac{2}{3}\big(\R{(123)(4)}{(123)(4)}-d\R{(123)(4)}{(1234)}-d\R{(1234)}{(123)(4)}\big)+\tfrac{1}{6}d^2\R{(1234)}{(1234)})+\scO(\epsilon^6),\\
v&=\epsilon^4\bigg(\dia{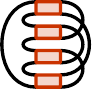}{20}{-8}+\tfrac{4-8}{2!}\dia{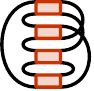}{20}{-8}+\Big(\tfrac{6}{(2!)^2}+\tfrac{4-8}{3!}+\tfrac{4}{4!}\Big)\dia{R_1234_1234}{20}{-8}\bigg)+\scO(\epsilon^6)\\
&=\epsilon^4\big(\R{(1234)}{(1432)}-2\R{(1234)}{(1243)}+\R{(1234)}{(1234)}\big)+\scO(\epsilon^6).}
In the diagrams, each small red block represents a copy of the bond Hamiltonian $H_x$. Their legs are contracted according to the assignment of the permutations $\vect{\sigma}$ and $\vect{\tau}$. The result can be expressed in terms of the generalized spectral form factor $R_{g_i}^{g_j}$, labeled by two permutations $g_i,g_j\in S_n$ acting separately on sites $i$ and $j$,
\eq{\label{eq:R def}
R_{g_i}^{g_j}=\Tr(H_{ij}^{\otimes n} \scX_{g_ig_j}).}
where $\scX_{g_ig_j}=\scX_{g_i}\scX_{g_j}$ is the representation of $g_i$ and $g_j$ in the $n$-replicated Hilbert space. For example, $\R{(1)(2)}{(1)(2)}=(\Tr H_{ij})^2$, $\R{(1)(2)}{(12)}=\Tr_j(\Tr_i H_{ij})^2$ (where $\Tr_i$ denotes the partial trace over site $i$), and $\R{(12)}{(12)}=\Tr (H_{ij}^2)$. 

Given the components $u$ and $v$, we can rewrite $\delta W_x[\vect{\sigma},\vect{\tau}]$ in the operator form
\eq{\delta\hat{W}_{x}=\frac{1-Z_iZ_j}{2}(u+v X_iX_j)\frac{1-Z_iZ_j}{2},}
therefore the EF operator reads
\eq{\hat{W}_{e^{-\ii\epsilon H}}=\hat{W}_\id+\sum_{x}\delta\hat{W}_x\otimes\hat{W}_{\id_{\bar{x}}}=\hat{W}_\id+\sum_{ij}\frac{1-Z_iZ_j}{2}(u+v X_iX_j)\frac{1-Z_iZ_j}{2}\otimes\hat{W}_{\id_{\bar{ij}}}.}
The EF Hamiltonian is therefore given by
\eqs{\label{eq:HEF1}\hat{H}_\text{EF}&=\frac{1}{\epsilon^2}(\id-\hat{W}_{e^{-\ii\epsilon H}}\hat{W}_{\id}^{-1})\\
&=-\frac{1}{\epsilon^{2}}\sum_{ij}\frac{1-Z_iZ_j}{2}(u+v X_iX_j)\frac{1-Z_iZ_j}{2}\hat{W}_{\id_{ij}}^{-1}\\
&=-\frac{1}{\epsilon^{2}}\sum_{ij}\frac{1-Z_iZ_j}{2}(u+v X_iX_j)\frac{1-Z_iZ_j}{2}\frac{1}{d^2(d^2-1)}e^{-\delta(X_i+X_j)}\\
&=-\sum_{ij}\frac{1-Z_iZ_j}{2}\frac{u+v X_iX_j}{\epsilon^{2}d^2(d^2-1)}e^{-\delta(X_i+X_j)}}
Therefore the EF Hamiltonian generally take the form of
\eq{\label{eq:HEF2}\hat{H}_\text{EF} = \sum_{ ij}g\frac{1-{Z}_{i}{Z}_{j}}{2}e^{-\beta X_iX_j-\delta(X_i+X_j)},}
consistent with the general form in \eqnref{eq:HEF form}. Comparing \eqnref{eq:HEF1} with \eqnref{eq:HEF2}, we should identify
\eq{g e^{-\beta X_iX_j}=-\frac{u+v X_iX_j}{\epsilon^{2}d^2(d^2-1)},}
which indicates
\eqs{\label{eq:g beta} g \cosh\beta&=-\frac{u}{\epsilon^2d^2(d^2-1)}=\frac{1}{d^2(d^2-1)}(u_2-u_4\epsilon^2+\scO(\epsilon^4))\\
g \sinh\beta&=\frac{v}{\epsilon^2d^2(d^2-1)}=\frac{1}{d^2(d^2-1)}(v_4\epsilon^2+\scO(\epsilon^4)),}
where the coefficients $u_2,u_4,v_4$ are defined in terms of generalized spectral form factors $R_{g_i}^{g_j}$ as
\eqs{\label{eq:uv}
u_2&=2\R{(1)(2)}{(1)(2)}-2d\big(\R{(1)(2)}{(12)}+\R{(12)}{(1)(2)}\big)+2d^2\R{(12)}{(12)},\\
u_4&=\R{(12)(34)}{(13)(24)}+2\big(\R{(123)(4)}{(124)(3)}-\R{(123)(4)}{(12)(34)}-\R{(12)(34)}{(123)(4)}\big)+\tfrac{1}{2}\big(\R{(12)(34)}{(12)(34)}+d\R{(12)(34)}{(1234)}+d\R{(1234)}{(12)(34)}\big)\\
&+\tfrac{2}{3}\big(\R{(123)(4)}{(123)(4)}-d\R{(123)(4)}{(1234)}-d\R{(1234)}{(123)(4)}\big)+\tfrac{1}{6}d^2\R{(1234)}{(1234)},\\
v_4&=\R{(1234)}{(1432)}-2\R{(1234)}{(1243)}+\R{(1234)}{(1234)}.}
For specific model of $H_{ij}$, we can evaluate the generalized spectral form factors, then we can determined the parameters $g$ and $\beta$ as well as the EF Hamiltonian. In the following, we will perform the calculation for random $\U(d)$ spin model and the locally scrambled Ising model.

For two-qudit GUE Hamiltonians, the generalized spectral form factors, defined in \eqnref{eq:R def}, can be evaluated under the GUE average using the basic property that
\eqs{\mathop{\dsE}_\text{GUE}H_{ij}^{\otimes2}&\equiv\mathop{\dsE}_\text{GUE}\dia{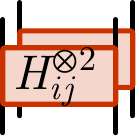}{28}{-12}\\
&=\frac{1}{d^2}\dia{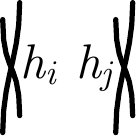}{28}{-12}\equiv\frac{1}{d^2}\scX_{(12)_i(12)_j},}
the GUE average of $n$-replicated Hamiltonian $H_{ij}$ can be obtained by summing over Wick contractions
\eq{\mathop{\dsE}_\text{GUE}H_{ij}^{\otimes n}=\left\{\begin{array}{ll}d^{-n}\sum_{h_i=h_j\in P_n}\scX_{h_ih_j}&n\in\text{even},\\0&n\in\text{odd},\end{array}\right.}
where $P_n$ denotes all possible pair-wise exchange of $n$ replicas. Then the generalized spectral form factor reads
\eq{\label{eq:R GUE}\mathop{\dsE}_\text{GUE}R_{g_i}^{g_j}=\frac{1}{d^n}\sum_{h\in P_n}\Tr(\scX_{g_i}\scX_h)\Tr(\scX_{g_j}\scX_h),}
whose results are enumerated in \tabref{tab:R}. Substitute these results to \eqnref{eq:uv}, we find $u_2=2(d^2-1)^2$, $u_4=\frac{11}{6}(d^2-1)^2$, and $v_4=2(d^2-1)^2/d^2$. By solving \eqnref{eq:g beta}, we can determine the parameters $g$ and $\beta$ to the order of $\epsilon^2$,
\eqs{\label{eq:g beta randH}
g&=2(1-d^{-2})\big(1-\tfrac{11}{12}\epsilon^2+\scO(\epsilon^4)\big),\\
\beta&=\epsilon^2/d^2+\scO(\epsilon^4).}
In conclusion, as we consider the locally scrambled quantum dynamics by alternatively applying the small unitary $e^{-\ii\epsilon H}$ and the local scramblers, the evolution of the corresponding EF state will be governed by $\partial_t\ket{W_{\Psi_t}}=-\hat{H}_\text{EF}\ket{W_{\Psi_t}}$, with the EF Hamiltonian $\hat{H}_\text{EF}$ given by \eqnref{eq:HEF2}. The random $\U(d)$ spin model $H$ in \eqnref{eq:randH} corresponds to the set of parameters in \eqnref{eq:g beta randH} for $\hat{H}_\text{EF}$.

\begin{table}[htp]
\caption{Spectral form factor of two-qudit GUE Hamiltonian}
\begin{center}
\begin{tabular}{|ccc||ccc||ccc||ccc|}
\hline
$R_{(1)(2)}^{(1)(2)}$ & \dia{R_1_2_1_2}{11}{-2} & $1$ & $R_{(1)(2)}^{(12)}$ & \dia{R_1_2_12}{11}{-2} & $d$ & $R_{(12)}^{(1)(2)}$ & \dia{R_12_1_2}{11}{-2} & $d$ & $R_{(12)}^{(12)}$ & \dia{R_12_12}{11}{-2} & $d^2$\\
\hline
$R_{(123)(4)}^{(123)(4)}$ & \dia{R_123_4_123_4}{20}{-8} & $3$ & $R_{(123)(4)}^{(124)(3)}$ & \dia{R_123_4_124_3}{20}{-8} & $3$ & $R_{(123)(4)}^{(1234)}$ & \dia{R_123_4_1234}{20}{-8} & $2 d+\frac{1}{d}$ & $R_{(1234)}^{(123)(4)}$ & \dia{R_1234_123_4}{20}{-8} & $2 d+\frac{1}{d}$\\
\hline
$R_{(123)(4)}^{(12)(34)}$ & \dia{R_123_4_12_34}{20}{-8} & $d^2+2$ & $R_{(1234)}^{(1243)}$ & \dia{R_1234_1243}{20}{-8} & $d^2+2$ & $R_{(12)(34)}^{(123)(4)}$ & \dia{R_12_34_123_4}{20}{-8} & $d^2+2$ & $R_{(12)(34)}^{(13)(24)}$ & \dia{R_12_34_13_24}{20}{-8} & $2 d^2+1$\\
\hline
$R_{(1234)}^{(1234)}$ & \dia{R_1234_1234}{20}{-8} & $2 d^2+\frac{1}{d^2}$ & $R_{(1234)}^{(1432)}$ & \dia{R_1234_1432}{20}{-8} & $2 d^2+\frac{1}{d^2}$ & $R_{(1234)}^{(12)(34)}$ & \dia{R_1234_12_34}{20}{-8} & $d^3+d+\frac{1}{d}$ & $R_{(12)(34)}^{(1234)}$ & \dia{R_12_34_1234}{20}{-8} & $d^3+d+\frac{1}{d}$\\
\hline
$R_{(12)(34)}^{(12)(34)}$ & \dia{R_12_34_12_34}{20}{-8} & $d^4+2$\\
\cline{1-3}
\end{tabular}
\end{center}
\label{tab:R}
\end{table}%

\end{document}